\documentclass[twocolumn,astrosymb,tighten,times,trackchanges]{aastex701}
\usepackage{adjustbox}

\begin{document}

\title{Small and Complex II: Characterizing the Disk and Stellar Envelope of Edge-on $z \sim 0$ Massive Compact Galaxies}

\author[orcid=0000-0002-0953-8224,gname=Allan,sname='Schnorr-M\"uller']{A. Schnorr-M\"uller}
\affiliation{Universidade Federal do Rio Grande do Sul – Departamento de Astronomia – 91501-970, Porto Alegre-RS, Brazil}
\email[show]{allan.schnorr@ufrgs.br}  

\author[gname=Katia, sname='Slodkowski Clerici']{K. Slodkowski Clerici} 
\affiliation{Universidade Federal do Rio Grande do Sul – Departamento de Astronomia – 91501-970, Porto Alegre-RS, Brazil}
\email{clericikatia@gmail.com} 

\author[gname=Francesca,sname=Pinna]{F. Pinna}
\affiliation{Instituto de Astrof\'isica de Canarias, calle Vía L\'actea s/n, E-38205 La Laguna, Tenerife, Spain}
\affiliation{Departamento de Astrof\'isica, Universidad de La Laguna, Avenida Astrof\'isico Francisco S\'anchez s/n, E-38206 La Laguna, Spain}
\email{francesca.pinna@iac.es} 

\author[gname=Rafael,sname=Merib-Dias]{R. Merib-Dias}
\affiliation{Universidade Federal do Rio Grande do Sul – Departamento de Astronomia – 91501-970, Porto Alegre-RS, Brazil}
\email{rafameribdias@hotmail.com} 

\author[gname=Marina,sname=Trevisan]{M. Trevisan}
\affiliation{Universidade Federal do Rio Grande do Sul – Departamento de Astronomia – 91501-970, Porto Alegre-RS, Brazil}
\email{marina.trevisan@ufrgs.br} 

\author[sname=Vecchi Ricci,gname=Tiago]{T. V. Ricci}
\affiliation{Universidade Federal da Fronteira Sul – Campus Cerro Largo – 97900-000, Cerro Largo-RS, Brazil}
\email{tiago.ricci@uffs.edu.br} 

\author[sname=Ferrari,gname=Fabricio]{F. Ferrari}
\affiliation{Universidade Federal do Rio Grande - Instituto de Matemática, Estatística e Física - 96203-900, Rio Grande-RS, Brazil}
\email{fabricio.ferrari@gmail.com} 

\author[sname=Santiago-Menezes,gname=Ana Carolina]{A. C. Santiago-Menezes}
\affiliation{Universidade Federal do Rio Grande do Sul – Departamento de Astronomia – 91501-970, Porto Alegre-RS, Brazil}
\affiliation{European Southern Observatory, Alonso de Córdova 3107, Vitacura, Región Metropolitana, Chile}
\email{carolsantiago020@gmail.com}

\author[sname=Linck Becker,gname=Weslley]{W. L. Becker}
\affiliation{Universidade Federal do Rio Grande do Sul – Departamento de Astronomia – 91501-970, Porto Alegre-RS, Brazil}
\email{weslleylinck0@gmail.com} 

\author[sname=Palacios,gname=Felícia]{F. Palacios}
\affiliation{Universidade Federal do Rio Grande do Sul – Departamento de Astronomia – 91501-970, Porto Alegre-RS, Brazil}
\email{felicia.palacios@gmail.com}
\begin{abstract}

We present multi-component photometric decompositions of $r$-band Hyper Suprime-Cam images for a sample of 75 edge-on massive compact galaxies (MCGs) at $z < 0.1$, selected as $+2\sigma$ outliers in the stellar mass--velocity dispersion relation and $-2\sigma$ outliers in the velocity dispersion--size relation. MCGs are composed of compact bulges and disks embedded within stellar envelopes of unclear physical nature. Comparing MCGs to a mass- and redshift-matched control sample of non-compact edge-on S0 galaxies with a similar three-component structure, we find that the smaller sizes of MCGs are not driven by a single component. MCGs host more compact bulges and envelopes ($R_\mathrm{e,bulge} \sim 0.3$ versus $0.5$\,kpc; $R_\mathrm{e,env} \sim 4.4$ versus $5.7$\,kpc), as well as shorter and thicker disks ($h_R \sim 1.1$ versus $1.7$\,kpc; $h_R/z_0 \sim 3.9$ versus $5.3$). The sizes of the structural components are coupled, suggesting their formation processes are linked. Median bulge- and disk-to-total flux fractions are similar in both samples, with $B/T \sim 0.3$ and $D/T \sim 0.4$. Envelope ellipticities span $\epsilon_\mathrm{env} \sim 0$--$0.7$, with MCGs exhibiting rounder envelopes. Low- and high-ellipticity envelopes are broadly consistent with stellar halos and thick disks, respectively. However, the nature of intermediate ellipticity envelopes remains ambiguous from photometry alone. The coupling between component sizes, together with the survival of a substantial disk component, argues against dry minor mergers as the dominant envelope-building mechanism. A comparison with 8 relic galaxies reveals that MCGs and relics share similar bulge-disk-envelope structures and follow the same component size--mass relations, consistent with belonging to a common structural family.
\end{abstract}


\section{Introduction} 

In the local universe, galaxies are broadly divided into two main populations: the blue cloud, composed of star-forming galaxies in which stellar mass and star formation rate are tightly correlated \citep{brinchmann04}, and the red sequence, populated by quiescent galaxies whose star formation has been quenched. Observations at high redshift have shown that this color bimodality was already in place by $z \sim 3$, and that by $z \sim 2$, quiescent galaxies dominated the high-mass end of the stellar mass function \citep{davidzon17}. Compared to their local counterparts, these early quiescent systems were significantly more compact and often exhibited stellar disks \citep{vanderwel14,bundy10,davari17}. Similarly compact quiescent galaxies are rare in the local universe, implying substantial structural evolution of the quiescent population over the past $\sim$10\,Gyr.

Two main mechanisms have been proposed to explain this evolution. The first is a two-phase scenario, in which individual compact quiescent galaxies grow through dry mergers \citep{naab09,oser10,barro13}. The second is progenitor bias, whereby newly quenched galaxies are systematically larger because the sizes of star-forming progenitors increase with cosmic time \citep{carollo13}, thereby shifting the average properties of the quiescent population. Observations indicate that both effects contribute, with dry-merger growth becoming dominant at $\log (M_\star/M_\odot) \gtrsim 11.0$, while progenitor bias plays an increasingly important role at lower masses \citep{fagioli16,williams17,damjanov19}. As a result, although the evolutionary path of the most massive compact quiescent galaxies is relatively well understood, that of intermediate- and lower-mass systems ($\log (M_\star/M_\odot) \lesssim 11.0$) remains less clear, since most low-redshift quiescent galaxies in this mass range quenched at later times, and only a small fraction are likely direct descendants of systems already in place at $z \sim 2$.

Studies of local ($z \lesssim 0.2$) compact quiescent galaxies have revealed a mix of young and old systems \citep{ferre-mateu12,fangzhou12,llasker13,yildirim17,buitrago18,spiniello21,schnorr21}. They are typically fast rotators, consistent with dissipative formation followed by limited dry merging \citep{yildirim17,schnorr21}. The oldest among them ($\gtrsim 10$\,Gyr), which exhibit kinematic and morphological properties consistent with passive evolution since $z \sim 2$, are classified as relic galaxies \citep{trujillo14,ferre-mateu15}. However, searches for relic galaxies have traditionally focused on high-mass ($\log M_\star/M_\odot \gtrsim 10.8$), ultra-compact ($R_\mathrm{e} \lesssim 2$\,kpc) systems, thereby probing only a small part of the region occupied by high-redshift quiescent galaxies in the $\log M_\star$--$\log R_\mathrm{e}$ plane. The intermediate-mass regime ($10.0 \lesssim \log M_\star/M_\odot \lesssim 10.8$) therefore remains largely unexplored.

To probe this less-studied regime, \citet{schnorr21} selected compact quiescent galaxies using a combination of stellar mass, velocity dispersion, and effective radius, adopting more relaxed criteria than the strict mass and size thresholds typically used in relic searches. Their sample included somewhat younger galaxies (median age $\sim 8$\,Gyr), but still metal-rich and $\alpha$-enhanced, suggesting that star formation was quenched after an intense early burst. Their kinematics revealed signatures of embedded stellar disks, indicating that quenching was followed by a merger-quiet phase. Crucially, they found that, at fixed stellar mass, older galaxies exhibit higher velocity dispersions, suggesting that the combination of compactness and high central velocity dispersion may be a more effective tracer of early quenching than compactness alone. Building on this result, \citet{slodkowski_clerici24} selected compact quiescent galaxies as $+2\sigma$ outliers from the local stellar mass–velocity dispersion relation. These objects lie below the local size–mass relation—though not all are extremely compact—and are very old ($\gtrsim 10$\,Gyr), metal-rich, and $\alpha$-enhanced, with stellar population properties comparable to those of relic galaxies.

While the stellar population properties of relics, and compact quiescent galaxies in general, have been extensively studied, their morphological properties remain poorly constrained. Until recently, detailed morphological analyses were largely limited to a few case studies \citep{fangzhou12,llasker13,yildirim15}, which hinted at complex stellar structures not well captured by simple bulge--disk decompositions and requiring additional structural components. Building on these works, in the first paper of our series on the morphology of massive compact galaxies (\citealt{clerici26}, hereafter Paper I), we presented the results of a multi-component morphological decomposition of 245 massive compact galaxies, corresponding to a representative subset of the \citet{slodkowski_clerici24} sample. Our main finding was that massive compact galaxies are typically three-component systems, with compact bulges and disks embedded within an outer, lower-ellipticity envelope. In contrast, the majority of galaxies in a mass-matched control sample of early-type systems are well described by a bulge+disk model, with only a small fraction requiring an analogous three-component structure. The nature of the envelope remains unclear, but given its broad ellipticity range ($\epsilon_\mathrm{Env} \sim 0$--$0.7$), it may correspond to a thick disk in some galaxies and to a stellar halo in others. In the absence of kinematic data, distinguishing between these possibilities requires constraining the intrinsic envelope ellipticity, which can only be done reliably in edge-on systems.

In this second paper of a series on the morphology of massive compact quiescent galaxies, we analyze a subset of 75 edge-on galaxies from the parent sample of \citet{slodkowski_clerici24} observed by the Hyper Suprime-Cam Subaru Strategic Program (HSC-SSP) imaging survey. Our goals are threefold: (i) to constrain the intrinsic envelope ellipticity and test the hypothesis that massive compact galaxies host thick disks and stellar haloes; (ii) to compare the structural properties of massive compact galaxies with those of three-component S0 galaxies of similar stellar mass, in order to identify how compact galaxies resemble and differ from their non-compact counterparts; and (iii) to test whether relic galaxies and massive compact galaxies can be considered part of the same structural family.

This paper is organized as follows. In Sec.\,\ref{sec:data}, we describe the data and sample selection criteria. In Sec.\,\ref{sec:methods}, we present our multi-component decomposition strategy. Our results are shown and interpreted in Sec.\,\ref{sec:results}, with further discussion in Sec.\,\ref{subsec:discuss}. Finally, we summarize our findings and present our conclusions in Sec.\,\ref{sec:conclusions}. Throughout this work, we adopt a simplified $\Lambda$CDM cosmology with $\Omega_{\rm M} = 0.3$, $\Omega_\Lambda = 0.7$, and $H_0 = 70\,\mathrm{km\,s^{-1}\,Mpc^{-1}}$.

\section{Data and Sample Selection}\label{sec:data}

\begin{figure}
\centering
 \includegraphics[width=\columnwidth, trim=35 10 80 0, clip]{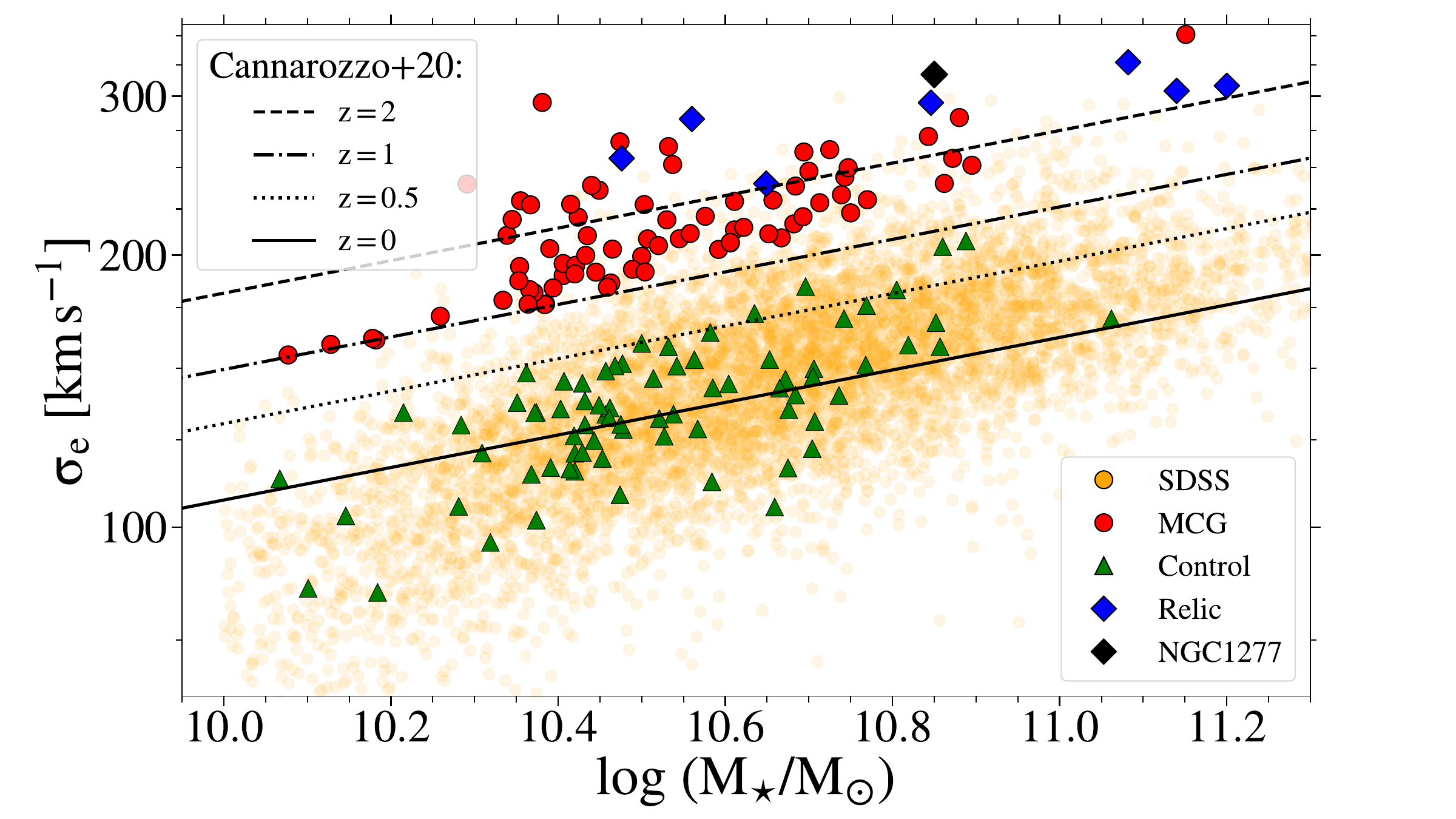}
 \includegraphics[width=\columnwidth, trim=35 10 80 0, clip]{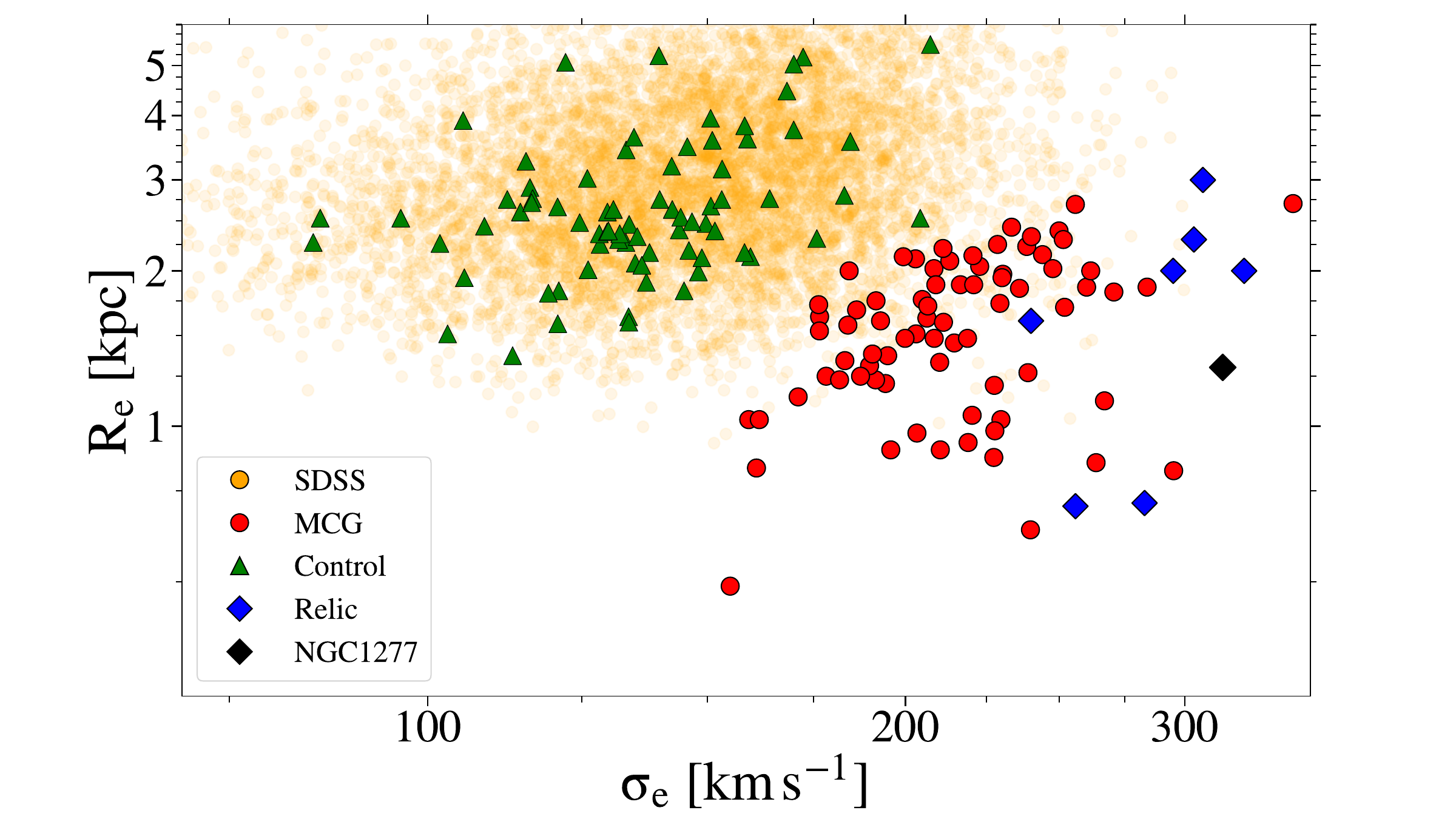}
 \includegraphics[width=\columnwidth, trim=35 10 80 0, clip]{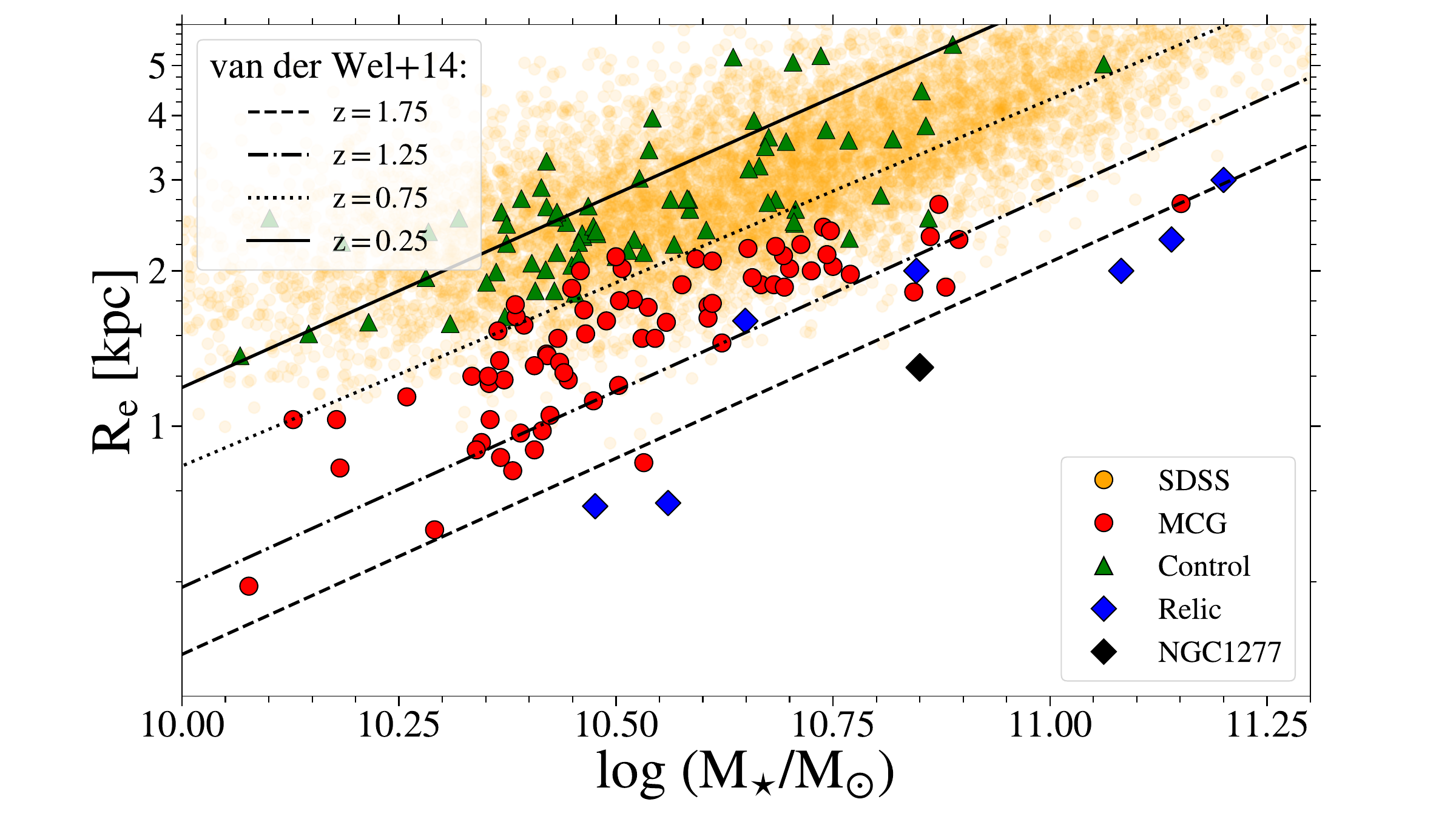}
    \caption{Sample selection. MCGs are shown as red circles, CSGs as green triangles, relic galaxies as blue diamonds (with the prototypical relic galaxy NGC\,1277 highlighted as a black diamond) and SDSS quiescent edge-on S0 galaxies as orange circles. The top panel displays the stellar mass–velocity dispersion diagram, the middle panel shows the velocity dispersion–effective radius diagram, and the bottom panel presents the stellar mass–effective radius diagram. The top and bottom panels also include the stellar mass–velocity dispersion and size–mass relations at different redshifts.}
    \label{fig:MCGs_CSGs_SDSS}
\end{figure}

\subsection{Massive Compact Galaxy Sample}

The massive compact galaxy (hereafter MCG) sample studied in this work is drawn from the parent sample of \citet{slodkowski_clerici24}. We briefly summarize its selection here and refer the reader to \citet{slodkowski_clerici24} for a detailed description. The parent sample was constructed from massive quiescent galaxies in the Sloan Digital Sky Survey (SDSS) Data Release 14 \citep{abolfathi18}, selected to have $\log{(M_\star/M_\odot)} > 10$ and $\log (\mathrm{sSFR}/\mathrm{yr}^{-1}) < -11$. Linear fits were then performed to the stellar mass--velocity dispersion and velocity dispersion--effective radius relations. MCGs were defined as galaxies lying more than $+2\sigma$ above the stellar mass--velocity dispersion relation and more than $-2\sigma$ below the velocity dispersion--effective radius relation, yielding a sample of 1858 galaxies at $z < 0.1$.

Stellar masses ($M_\star$) and star formation rates (SFR) were taken from the GALEX-SDSS-WISE Legacy Catalog (GSWLC; \citealt{salim18}). Effective radii ($R_\mathrm{e}$), defined as the semi-major axis of the half-light ellipse, were extracted from the catalog of \citet{simard11}. Stellar velocity dispersions were retrieved from the SDSS spectroscopic catalog and converted to effective velocity dispersions ($\sigma_\mathrm{e}$) using the aperture correction relation $\sigma_{\mathrm{ap}} = \sigma_\mathrm{e} [R_{\mathrm{ap}}/R_\mathrm{e}]^{-0.066}$ \citep{cappellari06}, where $R_{\mathrm{ap}}$ is the SDSS aperture radius.

Given their small sizes, characterizing the structure of MCGs requires imaging with higher spatial resolution than that provided by SDSS. The Hyper Suprime-Cam Subaru Strategic Program (HSC-SSP), a multi-band (\textit{grizy}) imaging survey conducted with the Hyper Suprime-Cam (HSC) on the 8.2-meter Subaru Telescope \citep{aihara18}, provides an attractive alternative. Its wide layer covers $\sim 1400\,\mathrm{deg}^2$ to a depth of $r \sim 26$\,mag, with a median seeing of $0.75\arcsec$ in the \textit{r} band, and largely overlaps the SDSS footprint. A cross-match between the parent sample and the HSC-SSP source catalog yielded 245 matches. In the present study, we focus on the subset of HSC-observed MCGs with edge-on orientations, while an analysis of the full sample is presented in Paper I.

We identified edge-on MCGs from the multi-component decompositions of the HSC \textit{r}-band images presented in Paper I. First, we selected galaxies for which a three-component model (Sérsic bulge + exponential disk + Sérsic envelope) was preferred over a two-component one, yielding a sample of 185 MCGs. We then visually inspected the HSC images to identify galaxies exhibiting disky isophotes, which are generally interpreted as signatures of highly inclined disks. Since the intrinsic thickness of the disk is unknown, the observed ellipticity cannot be directly converted into an inclination angle. Rather than restricting the sample exclusively to galaxies with clearly visible disky isophotes, we used these systems to determine an ellipticity threshold for highly inclined galaxies. This is because the prominence of disky isophotes depends not only on inclination, but also on the disk-to-total flux ratio, while point-spread function smearing can make compact flattened disks appear less elongated. As a result, some highly inclined systems are expected to exhibit only weakly disky isophotes and could therefore be missed in a purely visual classification. We found that galaxies exhibiting disky isophotes consistently have fitted disk ellipticities $\geq 0.6$, and therefore adopted this value as the threshold for selecting edge-on candidates. Applying this criterion yielded 89 galaxies. Of these, two were excluded due to heavy blending with nearby sources, two due to strong isophotal twists, one due to a strongly warped disk, and two due to prominent outer star-forming rings. An additional seven galaxies were removed because the edge-on disk models provided poor fits. The final sample therefore consists of 75 edge-on MCGs.

\subsection{Control Sample}

To highlight how MCGs differ from non-compact S0 galaxies, we built a control sample of edge-on lenticular galaxies matched in stellar mass and redshift. In Paper I, we showed that central and satellite MCGs exhibit consistent morphological properties. We therefore restrict the control sample to central galaxies in order to exclude systems whose morphologies may have been significantly shaped by environmental processes acting exclusively on satellites.

We begin by selecting massive, quiescent lenticular galaxies ($\log{(M_\star/M_\odot)} > 10$, $\log (\mathrm{sSFR}/\mathrm{yr}^{-1}) < -11$, $\mathrm{T\mbox{-}Type} < 0$, $\mathrm{P_{S0}} > 0.7$) from a cross-match of the GSWLC, \citet{simard11}, and \citet{dominguez-sanchez18} catalogs. Highly inclined candidates were identified using the probability of being edge-on ($\mathrm{P_{Edge-On}} > 0.7$). Galaxies were classified as centrals or satellites using the \citet{lim17} group catalog. After removing satellite galaxies, as well as $+2\sigma$ outliers from the mass–velocity dispersion relation and $-2\sigma$ outliers from the velocity dispersion–effective radius relation, we obtained a sample of 563 highly inclined S0 galaxies at $z < 0.1$ with HSC imaging.

We then refined the sample by applying criteria analogous to those used for the MCGs. Two- and three-component models were fitted to all candidate galaxies, and the preferred model was selected using the same criteria adopted in Paper I. Galaxies for which a two-component model was preferred, as well as those with fitted disk ellipticities $< 0.6$, were excluded. It is worth noting that, unlike MCGs, many highly inclined S0s are adequately described by a two-component model, as shown in Paper I. For this reason we restrict the control sample to the subset of highly inclined S0s for which a three-component model is preferred. We further excluded galaxies with saturated centers, severe blending with nearby sources, strong isophotal twists, warped disks, or poor fits with the edge-on three-component model. This yielded a final pool of 155 highly inclined S0s. MCGs were then matched to these control galaxies in stellar mass and redshift using the {\scshape MatchIt} R package \citep{ho11} and the Propensity Score Matching (PSM) technique \citep{rosenbaum83}, resulting in a final matched sample of 75 edge-on S0s.

Figure\,\ref{fig:MCGs_CSGs_SDSS} shows the distribution of MCGs and control sample galaxies (CSGs) in the stellar mass–velocity dispersion, velocity dispersion–effective radius, and stellar mass–effective radius diagrams. The top and bottom panels also include the stellar mass–velocity dispersion \citep{cannarozzo20} and size–mass relations at different redshifts \citep{vanderwel14}. Some overlap exists in the stellar mass–effective radius plane, but excluding the overlapping galaxies does not significantly impact our results. 

\begin{figure*}
\centering
\includegraphics[width=0.9\linewidth, trim=0 0 0 0,clip]{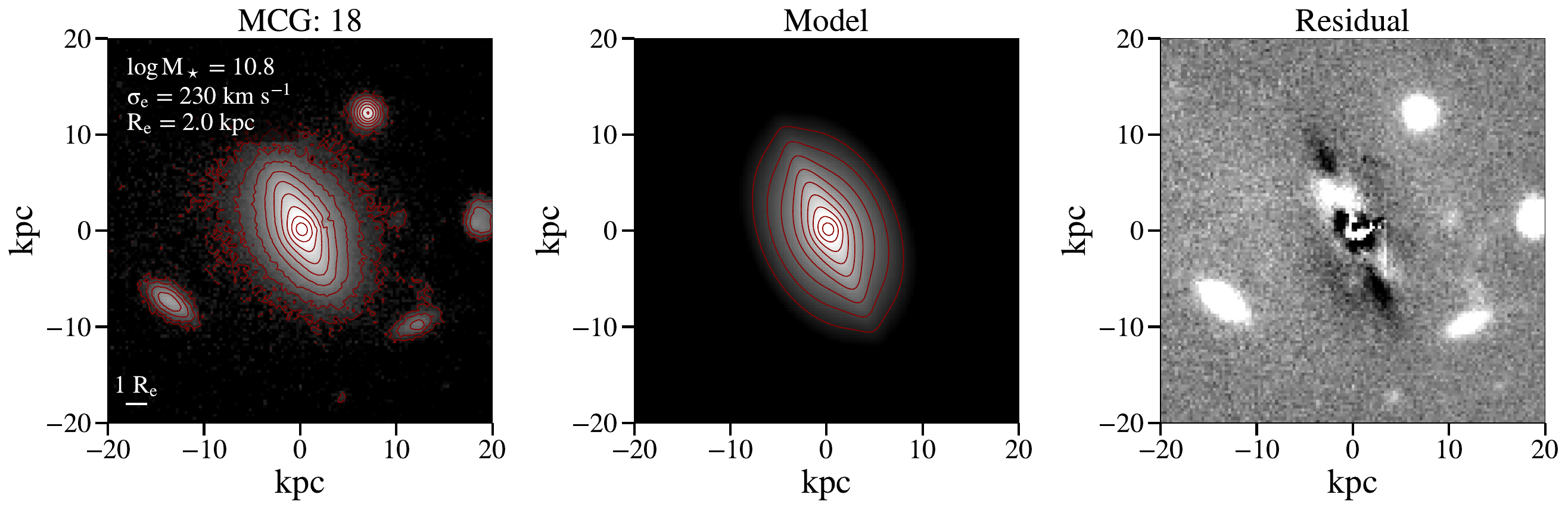}
\includegraphics[width=0.9\linewidth, trim=0 0 0 0,clip]{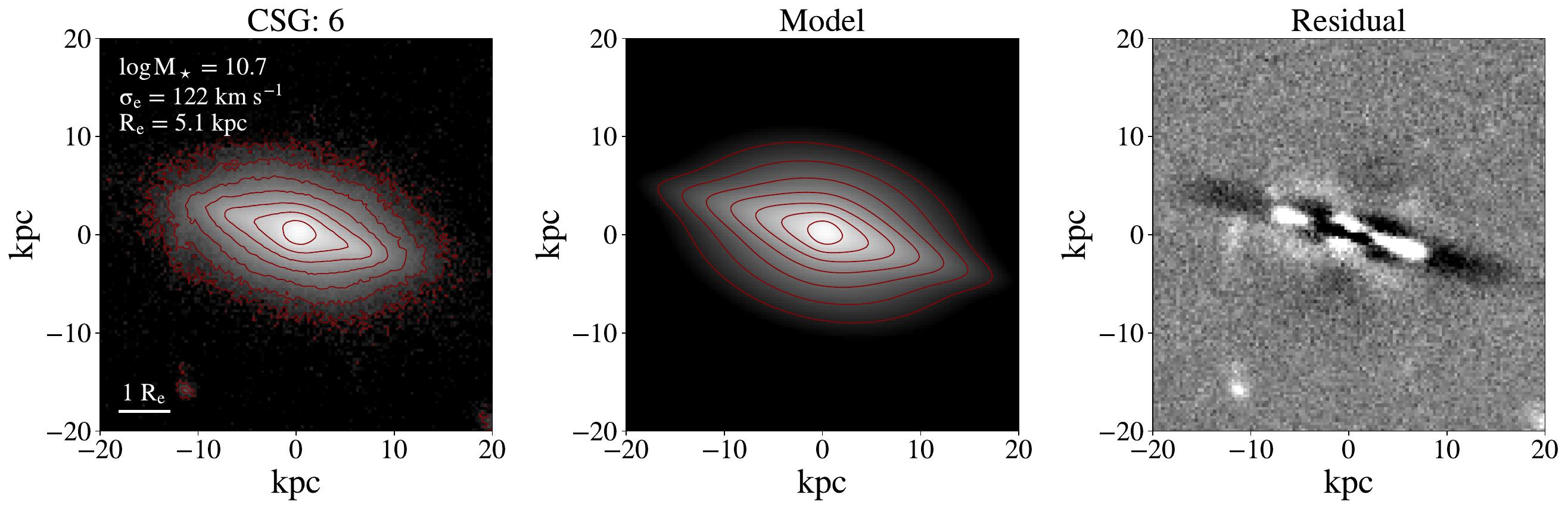}
\caption{$r$-band image, best-fitting model, and residuals for an MCG (top) and a CSG (bottom). The image and model panels share the same logarithmic flux scale. The residuals are scaled separately to emphasize structural features.}
\label{fig:imfit}
\end{figure*}

\begin{table}
\caption{Properties of the relic galaxy sample. Asterisks denote stellar masses from Y17 converted to a Kroupa IMF.}
\hspace*{-1.1cm}
\begin{tabular}{lccccc}
\hline
Name & $\log{(M_\star/M_\odot)}$ & $\sigma_{e}$ & $R_e$ & Inc.  \\
     &                           &  [km\,s$^{-1}$]      & [kpc] & [$^\circ$] \\
\hline
Mrk\,1216     & 11.2$^*$ & 308 & 3.0 & 70 \\
NGC\,1270    & 11.1     & 327 & 2.0 & 80 \\
NGC\,1271    & 10.8     & 295 & 2.0 & 83 \\
NGC\,1277    & 10.8$^*$ & 317 & 1.3 & 75 \\
NGC\,1281    & 10.6     & 240 & 1.6 & 70 \\
PGC\,12557   & 10.6     & 283 & 0.7 & 75\\
PGC\,12562   & 10.5     & 256 & 0.7 & 80 \\
PGC\,32873   & 11.1$^*$ & 304 & 2.3 & 84 \\
\hline
\label{tab:relics}
\end{tabular}
\end{table}

\subsection{Relic Galaxy Sample}

We define a second comparison sample composed of nearby galaxies previously classified as relics and for which archival \textit{Hubble Space Telescope} (HST) imaging is available. Relic classifications were taken from \citet{trujillo14}, \citet{ferre-mateu15}, and \citet{ferre_mateu17}. Nine relic galaxies have suitable HST imaging, however NGC\,2767 was excluded because it exhibits a strong central position angle twist.  For NGC\,1277, we use  F814W imaging from program GO:15235 (PI: W. E. Harris). For PGC\,12557, we use F814W-WFC3 (Wide Field Camera 3) imaging from GO:14924 (PI: A.~Seth). The remaining galaxies were observed in the near-infrared $H$ band (F160W filter) with WFC3 as part of program GO:13050 (PI: R.~van den Bosch). 

The stellar masses, effective velocity dispersions, and effective radii of the relic galaxies are listed in Table\,\ref{tab:relics}. Stellar masses were taken from the GSWLC catalog when available, while the remaining galaxies have dynamically determined stellar masses from \citet{yildirim17}. Since GSWLC stellar masses assume a Kroupa initial mass function (IMF), we converted the \citet{yildirim17} stellar masses using their published mass-to-light ratios to ensure consistency between the datasets. Effective radii and velocity dispersions were adopted from \citet{yildirim17}, except for PGC\,12557, for which we use the values reported by \citet{ferre-mateu15}. Inclinations are also listed in Table\,\ref{tab:relics}. \citet{yildirim17} provided inclination measurements for all galaxies except PGC\,12557. Since a nuclear disk is clearly visible in the HST image of PGC\,12557, we estimated its inclination from the observed disk geometry assuming an infinitely thin disk, obtaining $75^\circ$. Although not all relic galaxies in the sample are viewed exactly edge-on, all have inclinations $> 70^\circ$. This allows a structural analysis broadly comparable to that employed for the MCGs, requiring only minor adjustments to the fitting strategy, which are described in detail in the following section.

Relic galaxies are massive, compact, high-velocity-dispersion early-type systems. Their velocity fields exhibit regular rotation, and they host very old stellar populations ($\gtrsim 13$\,Gyr) out to a few $R_e$. In Fig.\,\ref{fig:MCGs_CSGs_SDSS}, relic galaxies are shown as blue diamonds, while the prototypical relic galaxy NGC\,1277 is highlighted as a black diamond. As can be seen, relic galaxies satisfy the stellar mass, size, and velocity-dispersion criteria used to select MCGs. 

\section{Multi-Component Decomposition Strategy} \label{sec:methods}

We modeled the surface brightness distributions of the galaxies using \textsc{imfit} \citep{erwin15}, which includes both standard image functions, such as the Sérsic profile, and 3D luminosity-density models integrated along the line of sight, particularly useful for fitting highly inclined disks. The required inputs for \textsc{imfit} are a galaxy image, an associated noise map, a point-spread function (PSF), a mask image, and a configuration file specifying the model components, together with the free and fixed parameters, their bounds, and their initial values.

\textsc{imfit} provides three different $\chi^2$ minimization methods: the Levenberg--Marquardt algorithm, the Nelder--Mead simplex method, and differential evolution. We adopted the differential evolution algorithm because it does not require initial guesses for the fitted parameters, relying instead on user-defined lower and upper bounds, and is less prone to becoming trapped in local minima than the alternative optimizers available in \textsc{imfit}. It has also been shown to yield robust results with minimal user intervention (see \citealt{gadotti26} for a detailed discussion). Parameter bounds were defined following the strategy described in Paper~I.

For the MCG and control samples, the PSF and variance maps were provided by the HSC pipeline \citep{bosch18}. For the relic galaxy sample, we constructed empirical effective point-spread functions (ePSFs), which account for both the telescope PSF and detector sampling, using bright, isolated stars in the corresponding HST fields. The ePSFs were generated with the \textsc{EPSFBuilder} module from the \texttt{astropy} Python package, following the methodology described by \citet{anderson00} and \citet{anderson16}. Mask images were created from segmentation maps produced with \textsc{SExtractor} \citep{bertin96}, with all pixels not associated with the target galaxy masked out.

\subsection{MCG and CSG Samples}

Each galaxy was modeled with three components: an inner Sérsic bulge, an edge-on disk, and a Sérsic envelope. The projected intensity distribution of the edge-on disk component is defined in cylindrical coordinates as:
\begin{equation}
    I(r,z)=\mu_0\frac{r}{h_R}K_1\left(\frac{r}{h_R}\right)\mathrm{sech}^{2/\alpha}\left(\frac{z}{2z_0}\right),
\end{equation}
where $h_R$ is the scale length, $z_0$ is the scale height, $K_1$ is the modified Bessel function of the second kind, $\mu_0$ is the central surface brightness, and $\alpha$ controls the shape of the vertical profile. This expression corresponds to the projected intensity distribution of a perfectly edge-on disk with an exponential radial profile, using the Bessel-function solution of \citet{vanderKruit81} for the radial dependence. We adopt a $\mathrm{sech}^2$ vertical profile (i.e., $\alpha = 1$), corresponding to the solution for a self-gravitating isothermal sheet \citep{spitzer42}.

The disk model includes four free parameters: position angle ($PA_{\mathrm{disk}}$), central surface brightness ($\mu_0$), radial scale length ($h_R$), and vertical scale height ($z_0$). The bulge and envelope are each modeled with a Sérsic profile defined by their position angle, ellipticity, effective radius, and intensity at the effective radius. The Sérsic index $n$ is treated as a free parameter for the bulge, but fixed to $n=1$ for the envelope in order to limit model complexity, since allowing it to vary yields comparable $\chi^2$ values and variations of typically only a few percent in the envelope parameters. For most galaxies, the bulge effective radius is smaller than half the PSF full width at half maximum (FWHM). We therefore tested replacing the bulge Sérsic profile with a point-source model, implemented as a scaled version of the user-supplied PSF. This substitution consistently produced poorer fits, reinforcing our decision to model the bulge with a Sérsic profile despite the limited spatial resolution. 

Nine MCGs required an additional outer low-ellipticity ($\epsilon \sim 0$) exponential component to adequately reproduce both the surface-brightness distribution and the radial variation of ellipticity in their outskirts. In these cases, the residual maps of the three-component fits showed significant large-scale residual structures with a rounder spatial distribution. Allowing the envelope Sérsic index to vary freely was insufficient to reproduce these features, requiring the inclusion of an additional low-ellipticity exponential component. Since it is unclear whether the envelope component in these galaxies has the same physical nature as the envelope identified in galaxies well described by a three-component model, these systems were excluded from all envelope-related plots and from the computation of median envelope properties.

All galaxies were fit using 40\,kpc\,$\times$\,40\,kpc image cutouts. An example fit is shown in Fig.\,\ref{fig:imfit}. The overlaid isophotes demonstrate that our model successfully reproduces the disky isophotes near the mid-plane as well as the more elliptical isophotes at larger vertical distances. However, the modeled disky isophotes extend to larger radial distances than observed. This discrepancy arises because the disk flares, and our modeling approach does not account for this effect.

Finally, one potential source of systematic uncertainty is the comparison between the optical $r$-band HSC images of MCGs and the near-infrared $H$-band HST images of relic galaxies, since morphological parameters are known to vary with wavelength. However, multi-component morphological decompositions of $z$-band HSC images yield changes of only a few percent in the derived parameters compared to the $r$-band images adopted as the default. This indicates that any wavelength-dependent bias in our analysis is negligible.

The full table containing the galaxy properties and structural parameters derived from our multi-component photometric decompositions of the MCG and CSG samples will be made available in machine-readable format through VizieR upon publication of this paper.

\subsection{Relic Galaxy Sample}

Fitting the relic galaxies required minor adjustments to our fitting strategy. First, since inclination estimates are available for each relic galaxy, we replaced the perfectly edge-on disk model with a 3D disk luminosity-density model projected onto the plane of the sky at the adopted inclination. The 3D disk luminosity density is given by
\begin{equation}
    j(r,z) = J_0 \exp\left(-\frac{r}{h_R}\right) \mathrm{sech}^{2/\alpha}\left(\frac{z}{2 z_0}\right),
\end{equation}
where $r$ and $z$ are cylindrical coordinates aligned with the disk. The free parameters of the 3D disk model are the position angle ($PA_{\mathrm{disk}}$), central luminosity density ($J_0$), radial scale length ($h_R$), and vertical scale height ($z_0$). As in the edge-on disk model, we adopt $\alpha = 1$, corresponding to a $\mathrm{sech}^2$ vertical profile.

Second, Mrk\,1216, NGC\,1270, NGC\,1271, and PGC\,32873 exhibit boxy envelopes surrounding their disks. For these galaxies, we modeled the envelope using the \textit{SérsicGenEllipse} function within \textsc{imfit}, which generalizes the standard Sérsic profile by allowing non-elliptical isophotes through an additional parameter, $c_0$ ($c_0 > 0$ for boxy, $c_0 < 0$ for disky, and $c_0 = 0$ for purely elliptical isophotes).

Third, the same four galaxies required an additional outer round component ($\epsilon \sim 0$) to achieve satisfactory fits. These structures were modeled with an exponential component.

Fourth, owing to their small distances and the high spatial resolution of the HST images, most relic galaxies exhibit substructures within the photometric bulge, such as nuclear stellar and/or dust disks and bright nuclear point sources. Modeling or masking these structures does not significantly affect the bulge ellipticity or effective radius measurements, although $n_\mathrm{bulge}$ is substantially impacted. To identify such components, we visually inspected both the galaxy images and the residual maps from the three- (or four-) component fits. Bright nuclear point sources are present in NGC\,1270, NGC\,1281, PGC\,12562, and PGC\,32873, and were modeled using a PointSource function. We also repeated the fitting process while masking the nuclear disks in NGC\,1277, NGC\,1281, PGC\,12557, PGC\,12562, and PGC\,32873. NGC\,1270 exhibits an edge-on nuclear disk but no large-scale edge-on disk; in this case, we treat the nuclear disk as the main galaxy disk. Accounting for these substructures decreases $n_\mathrm{bulge}$ by up to $\sim 20\%$ relative to fits in which they are ignored.

\begin{figure*}
\centering
\vspace{0.05cm}
\makebox[\textwidth][c]{\textbf{Bulge Properties}}
\vspace{0.05cm}
\begin{minipage}[t]{0.265\textwidth}
\includegraphics[width=\linewidth, trim=30 30 30 30,clip]{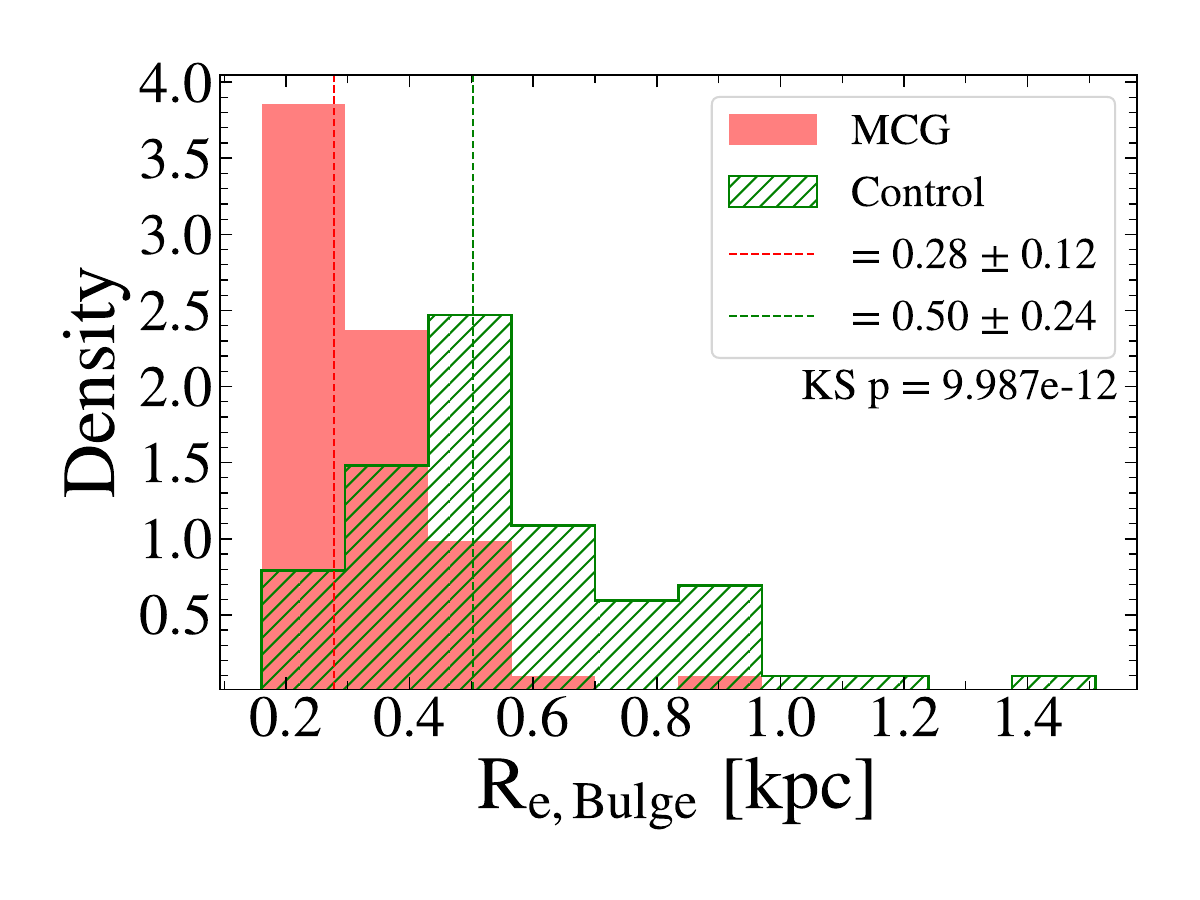}
\end{minipage}
\begin{minipage}[t]{0.265\textwidth}
\includegraphics[width=\linewidth, trim=30 30 30 30,clip]{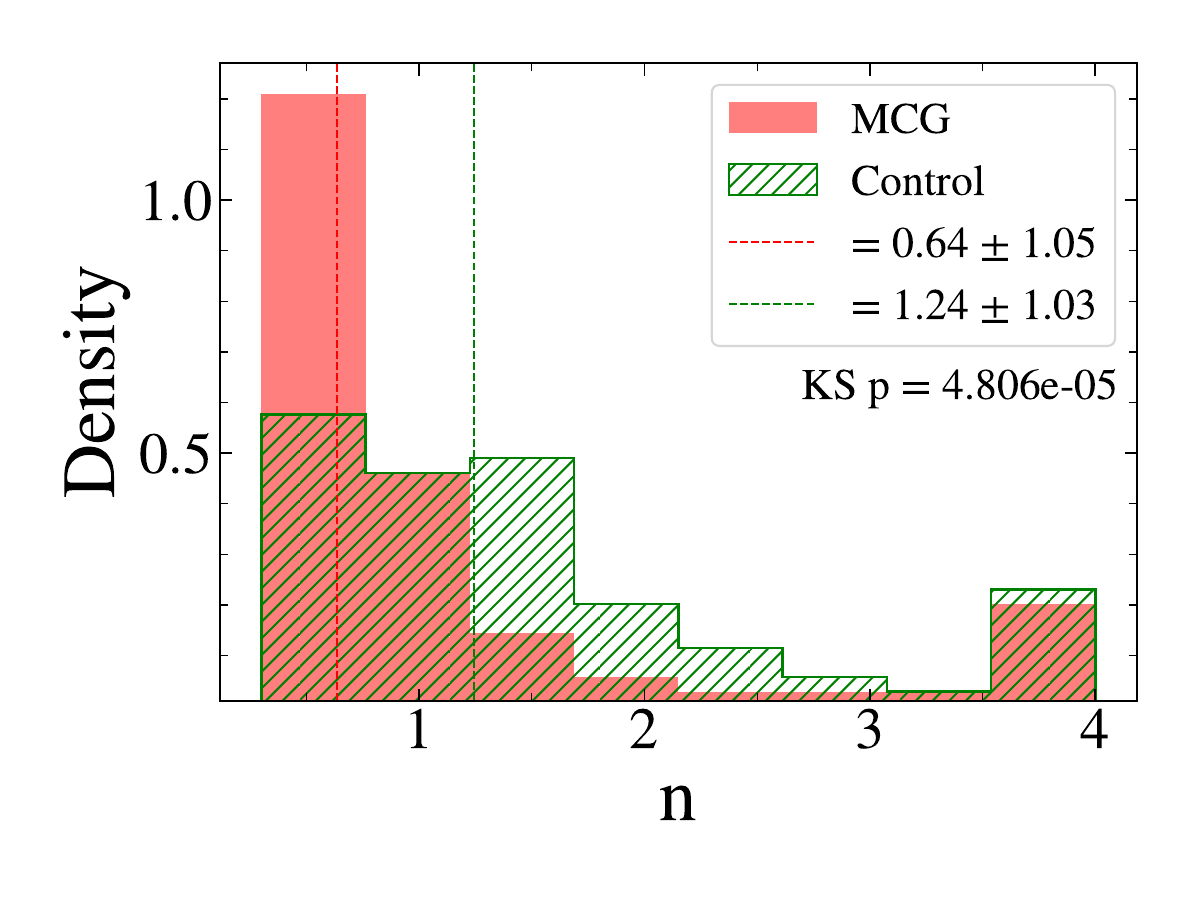}
\end{minipage}
\begin{minipage}[t]{0.265\textwidth}
\includegraphics[width=\linewidth, trim=30 30 30 30,clip]{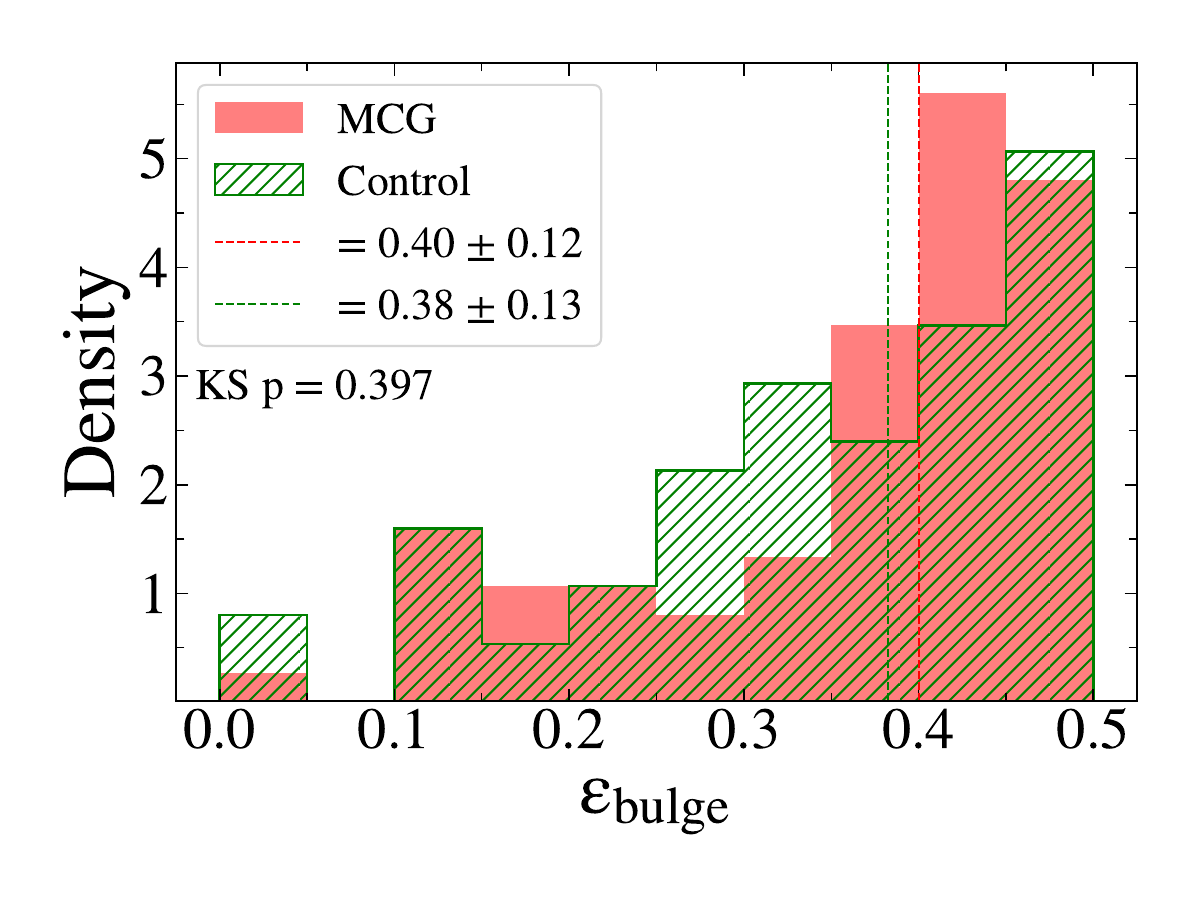}
\end{minipage}
\vspace{0.05cm}
\makebox[\textwidth][c]{\textbf{Disk Properties}}
\vspace{0.05cm}
\begin{minipage}[t]{0.265\textwidth}
\includegraphics[width=\linewidth, trim=30 30 30 30,clip]{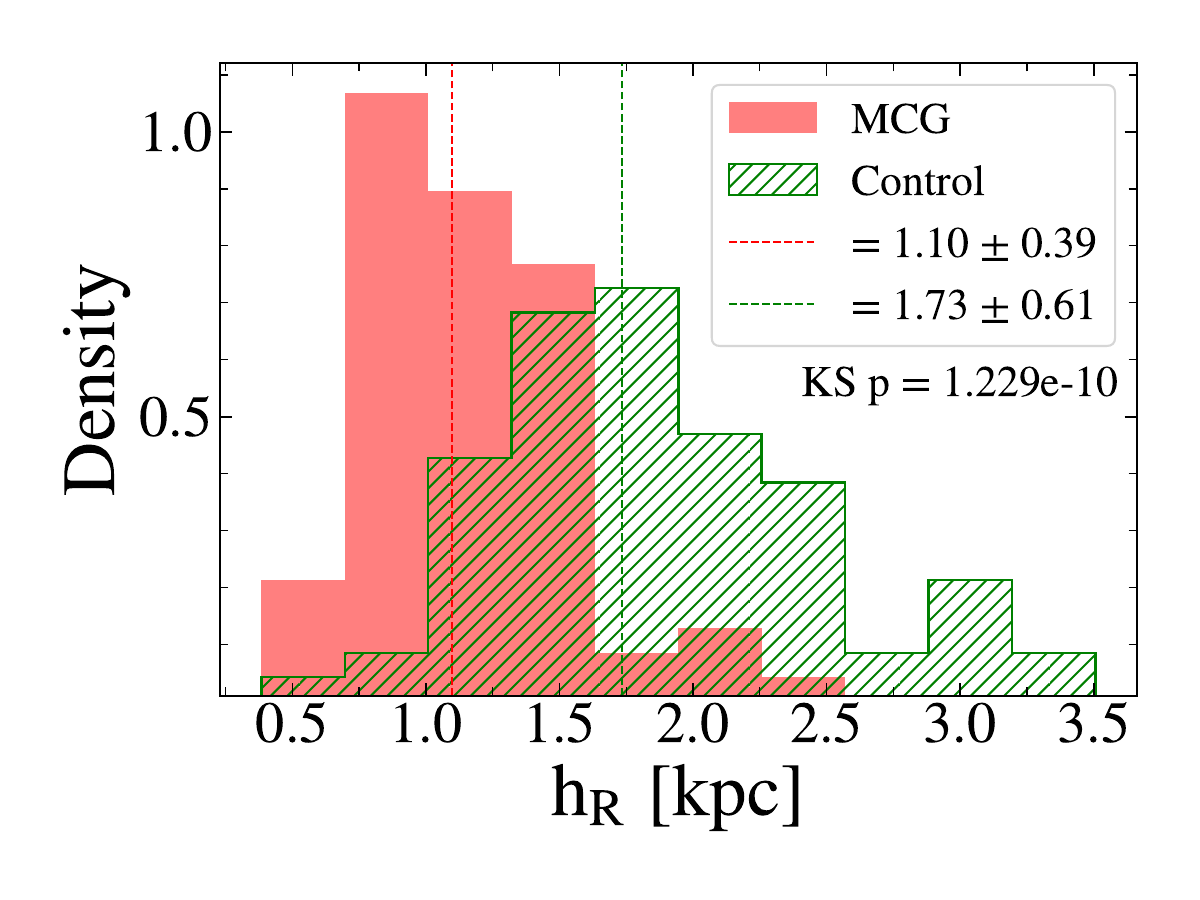}
\end{minipage}
\begin{minipage}[t]{0.265\textwidth}
\includegraphics[width=\linewidth, trim=30 30 30 30,clip]{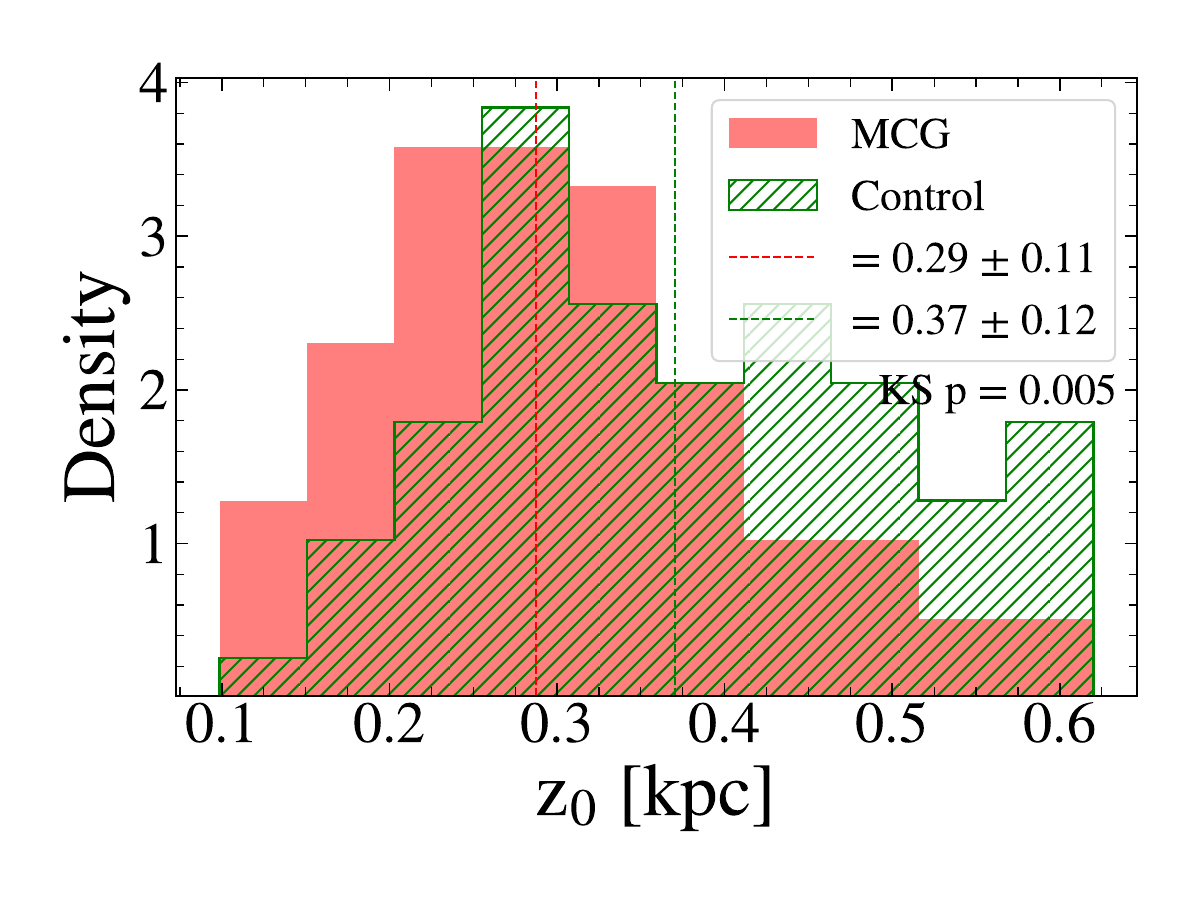}
\end{minipage}
\begin{minipage}[t]{0.265\textwidth}
\includegraphics[width=\linewidth, trim=30 30 30 30,clip]{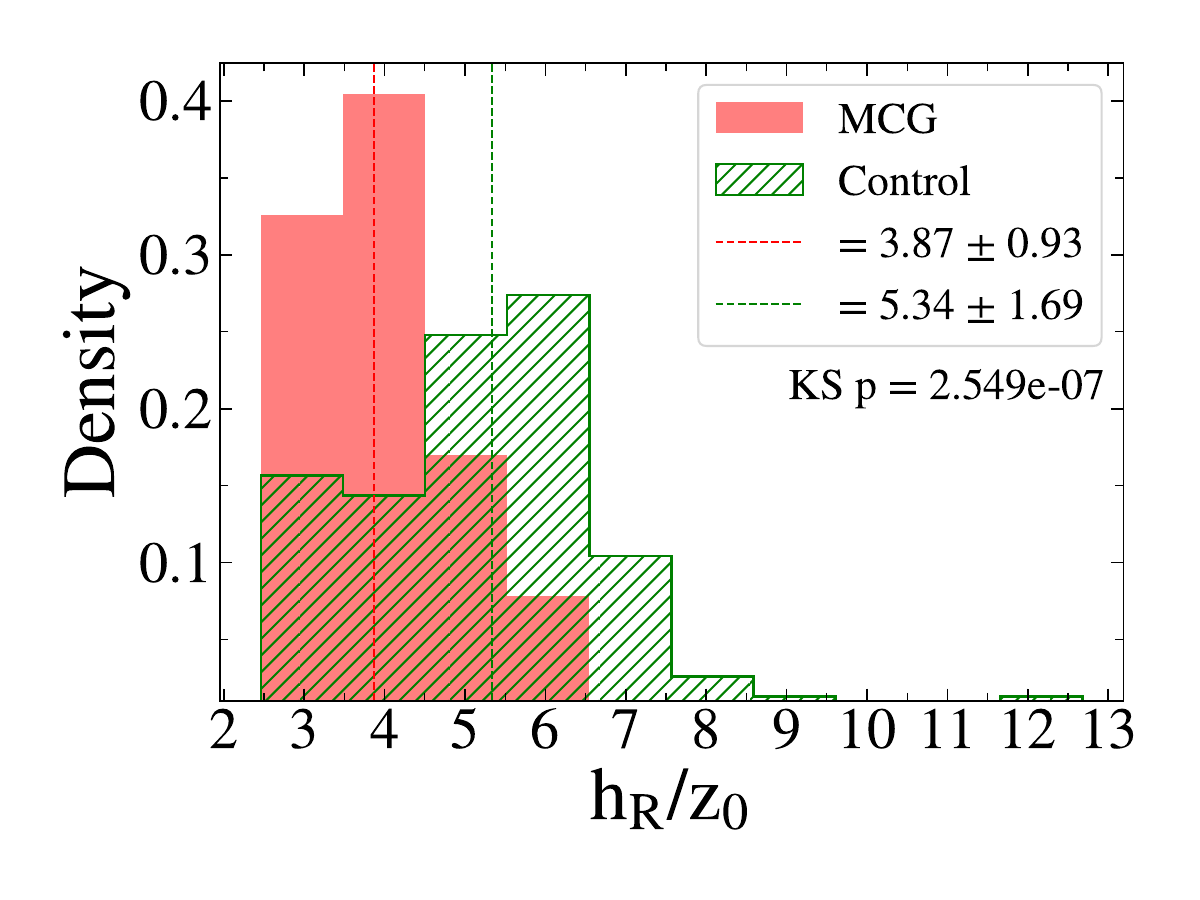}
\end{minipage}
\vspace{0.05cm}
\makebox[\textwidth][c]{\textbf{Envelope Properties}}
\vspace{0.05cm}
\makebox[\textwidth][c]{%
  \begin{minipage}[t]{0.265\textwidth}
  \includegraphics[width=\linewidth, trim=30 30 30 30,clip]{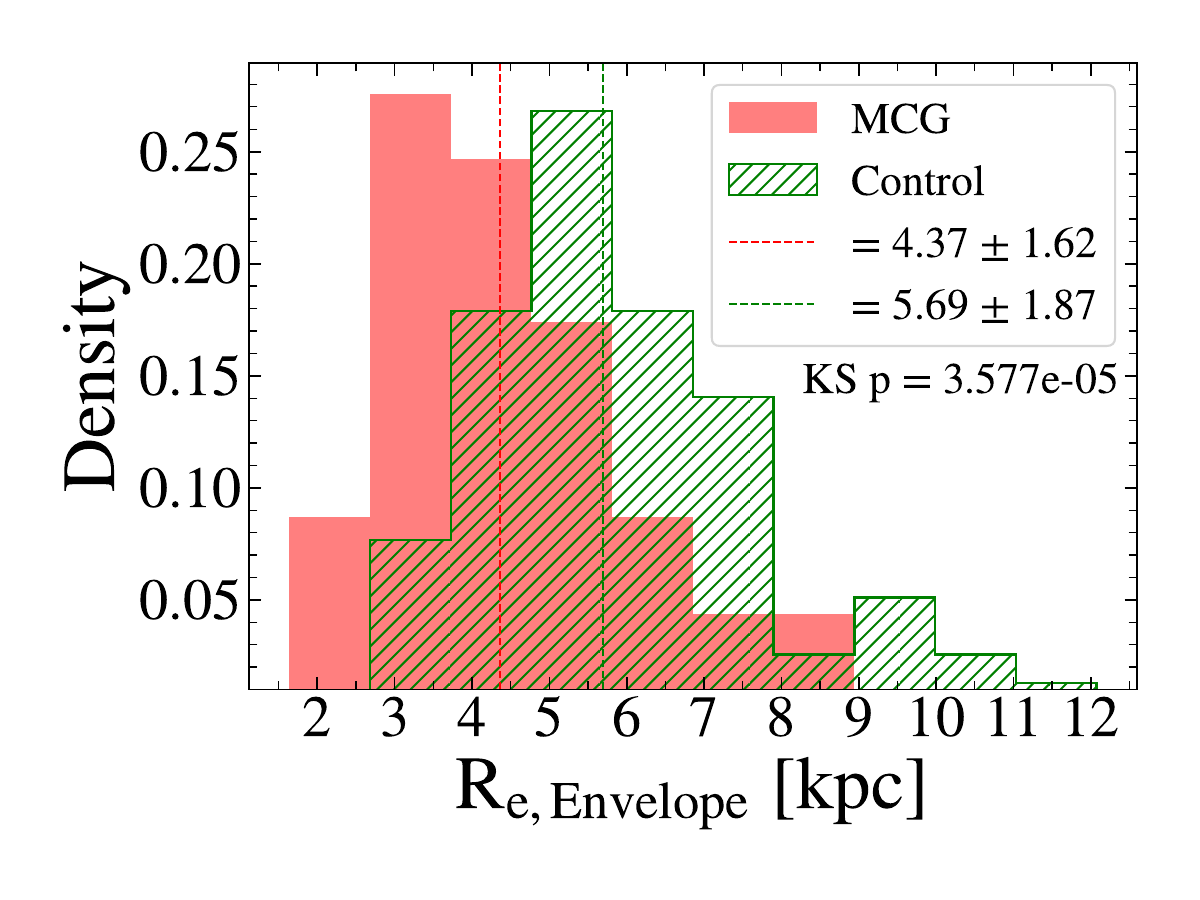}
  \end{minipage}%
  \hspace{0.01\textwidth}%
  \begin{minipage}[t]{0.265\textwidth}
  \includegraphics[width=\linewidth, trim=30 30 30 30,clip]{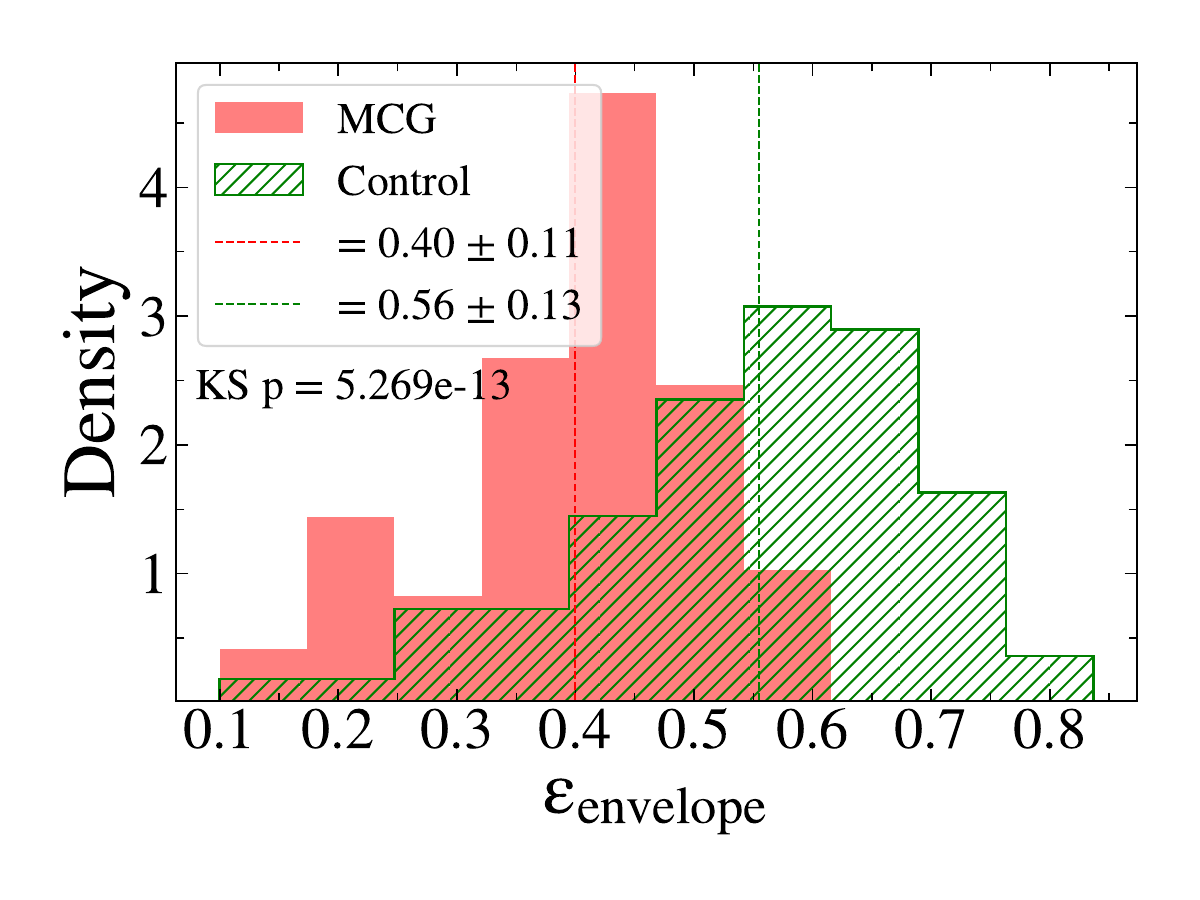}
  \end{minipage}%
}
\vspace{0.05cm}
\makebox[\textwidth][c]{\textbf{Flux-to-total ratio}}
\vspace{0.05cm}
\begin{minipage}[t]{0.265\textwidth}
\includegraphics[width=\linewidth, trim=30 30 30 30,clip]{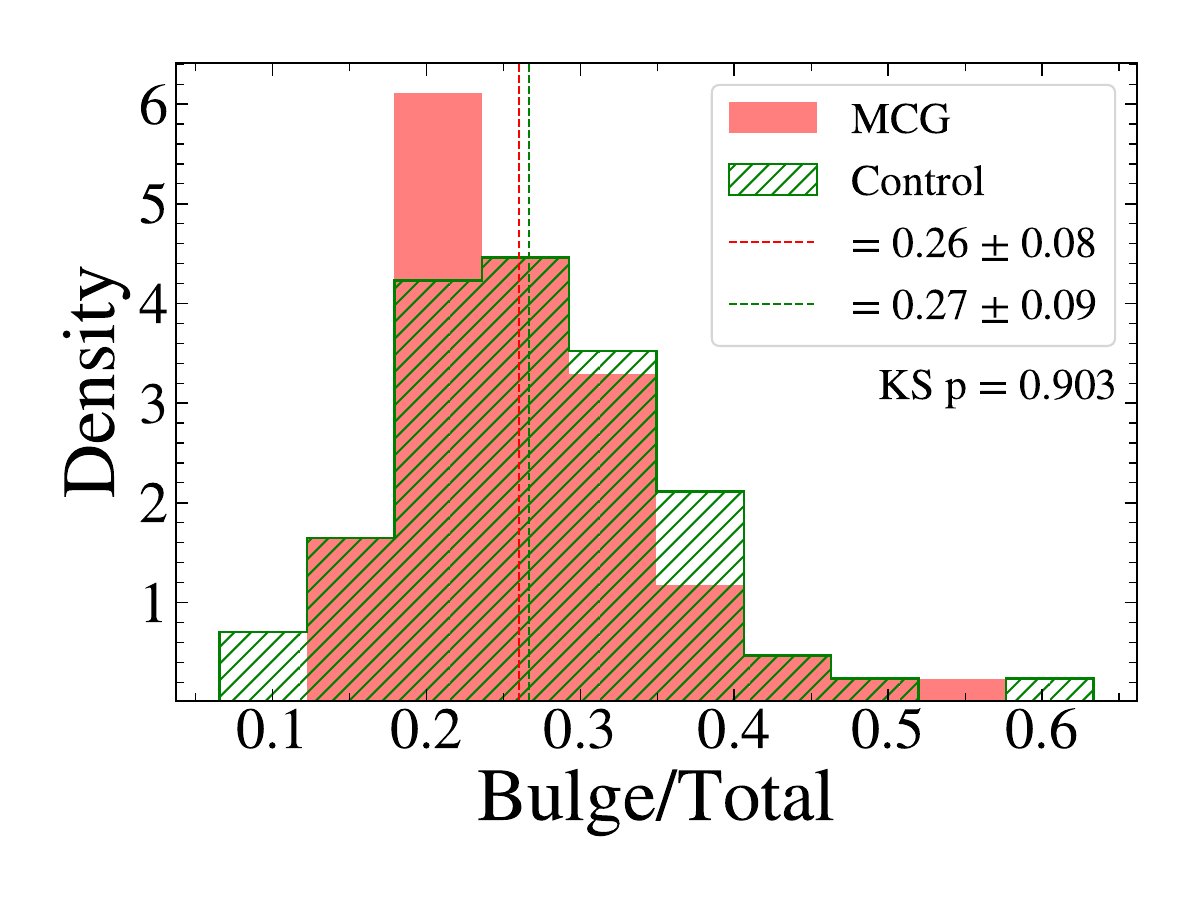}
\end{minipage}
\begin{minipage}[t]{0.265\textwidth}
\includegraphics[width=\linewidth, trim=30 30 30 30,clip]{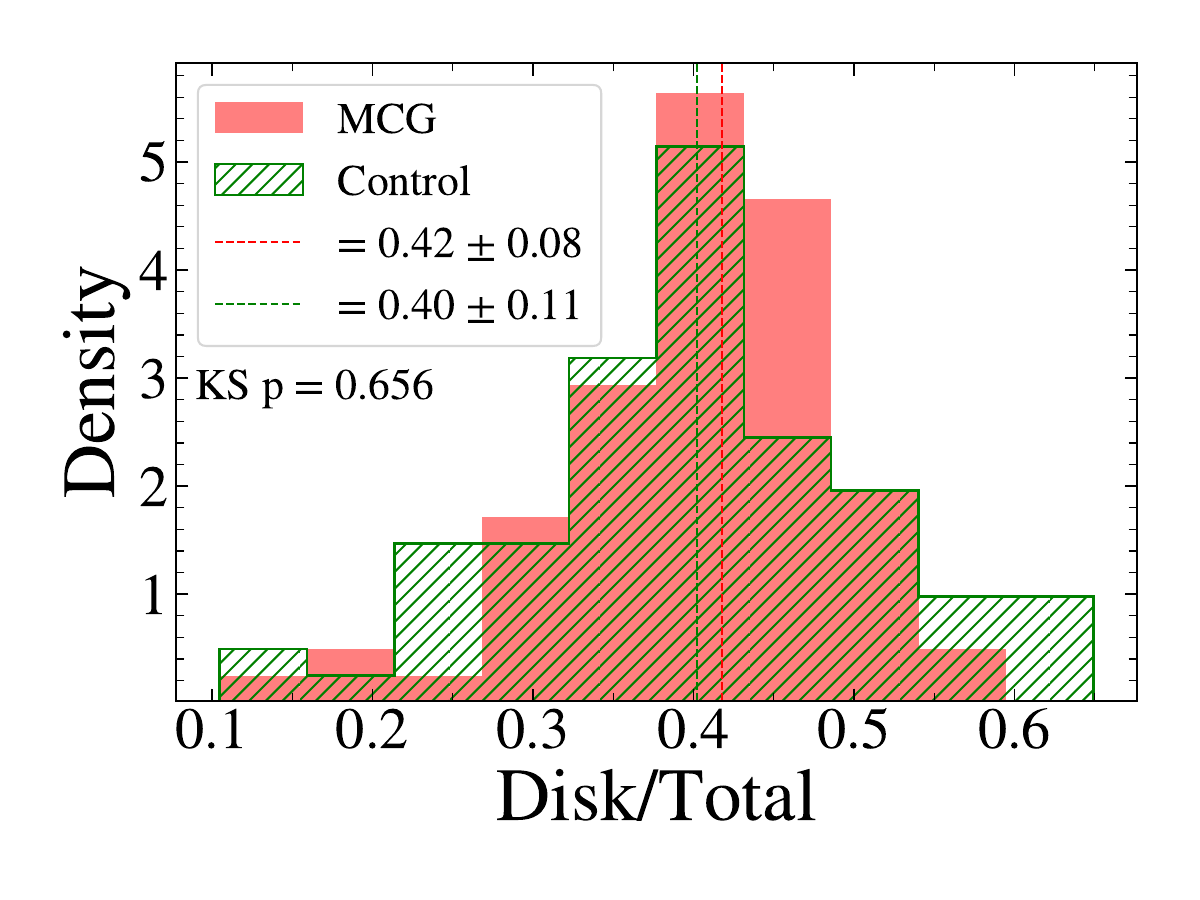}
\end{minipage}
\begin{minipage}[t]{0.265\textwidth}
\includegraphics[width=\linewidth, trim=30 30 30 30,clip]{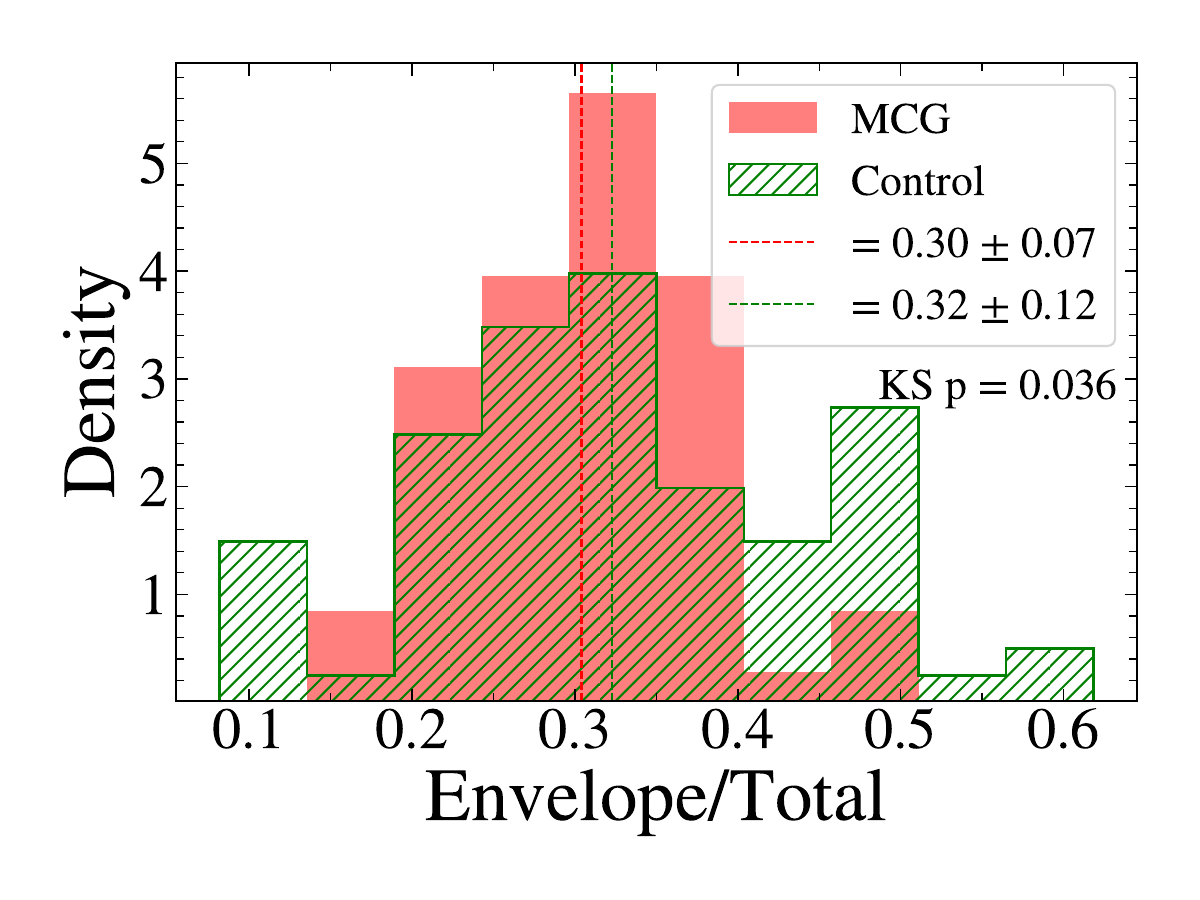}
\end{minipage}
\vspace{0.05cm}

\caption{Structural properties of MCGs and CSGs. The first row shows the effective radius, Sérsic index and ellipticity of the bulge component. The second row presents the disk scale length, scale height, and the ratio between them. In the third row, we show the effective radius and ellipticity of the stellar envelope, modeled with a fixed Sérsic index of $n = 1$. The fourth row displays the flux-to-total ratio of each component. Dashed lines indicate the median values for each sample. Galaxies best fitted with a four-component model were excluded from the histograms of the envelope properties.}
\label{fig:morph}
\end{figure*}

\section{Results}\label{sec:results}

\begin{figure*}
\centering
\includegraphics[width=0.32\textwidth, trim=8 20 30 20,clip]{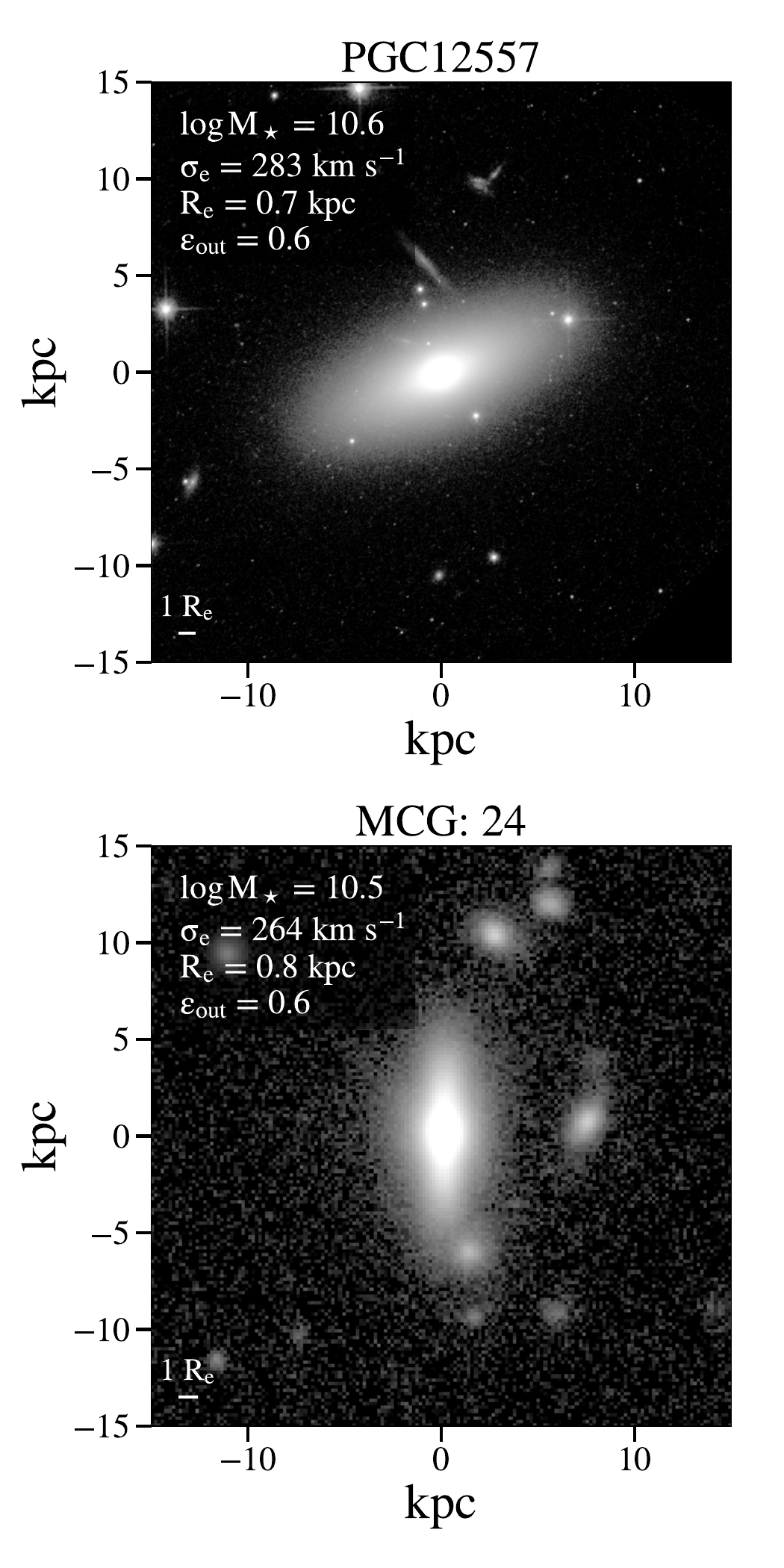}
\includegraphics[width=0.32\textwidth, trim=8 20 30 20,clip]{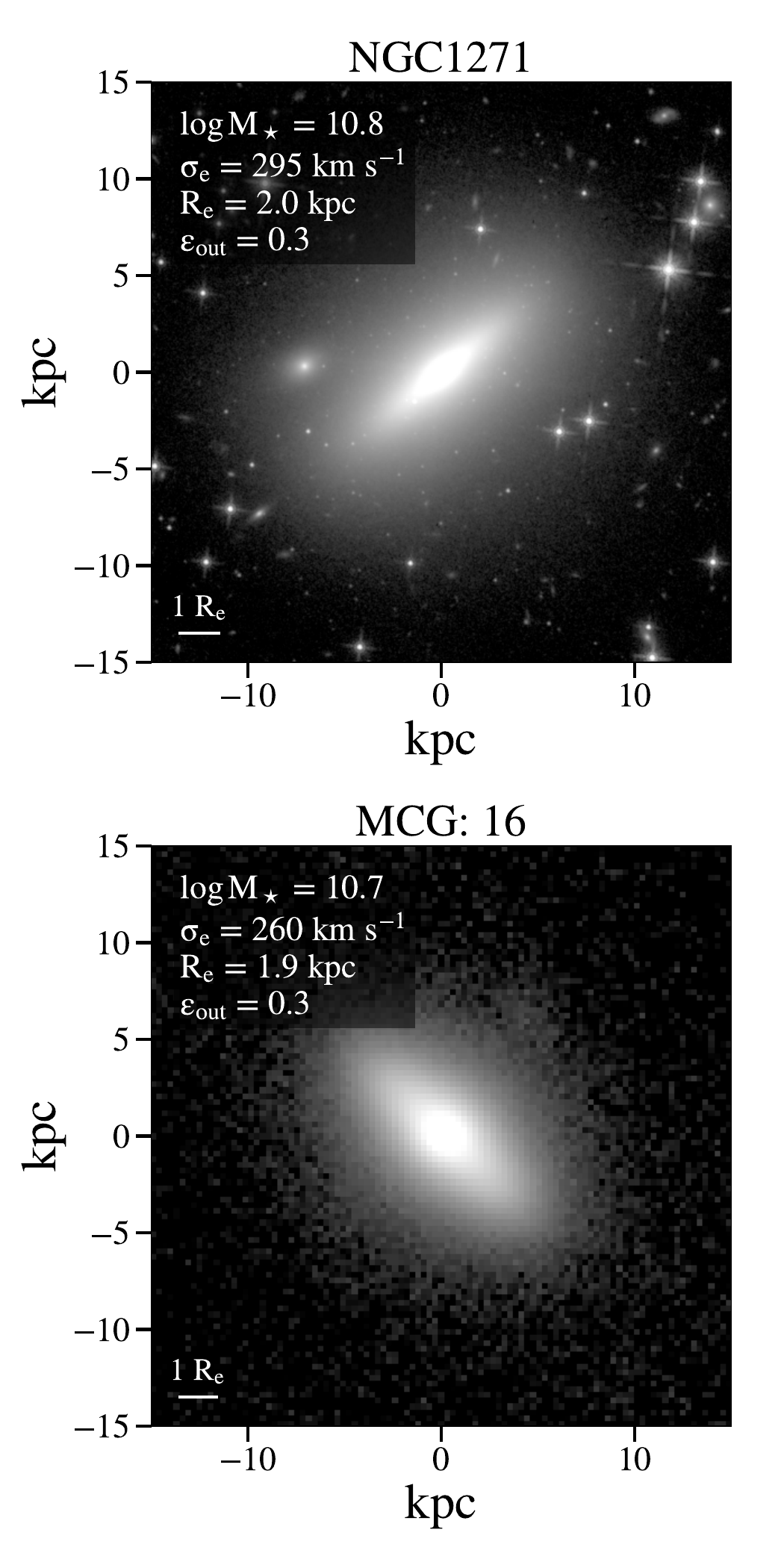}
\includegraphics[width=0.32\textwidth, trim=8 20 30 20,clip]{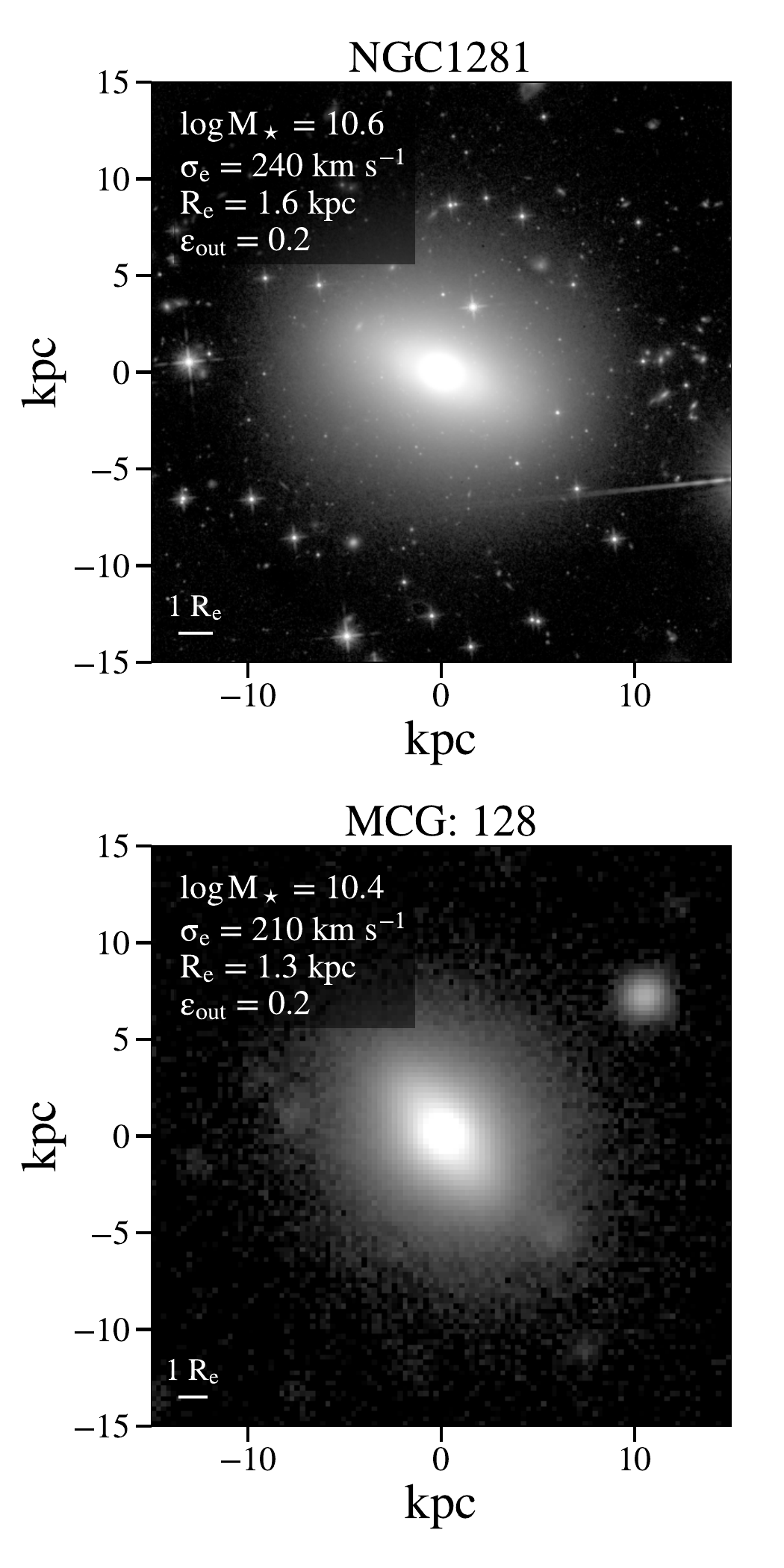}
\caption{A comparison between relic galaxies and MCGs. The top panels displays HST images of three relic galaxies, while the bottom panel shows MCGs  with similar stellar mass, velocity dispersion, effective radius, and ellipticity of the outermost component. The morphological similarity between relics and MCGs is clear.}
\label{fig:mcg_vs_relics}
\end{figure*}

\begin{table*}
\centering
\hspace*{-1.95cm}
\scriptsize
\caption{Structural parameters and flux-to-total ratios for relic galaxies. Median values and standard deviations are given for relics and MCGs for comparison. Standard deviations for the halo component are not shown due to small number of objects requiring a fourth component. Columns: bulge ellipticity ($\epsilon_{\mathrm{bulge}}$), Sérsic index ($n_{\mathrm{bulge}}$), bulge effective radius ($R_{e,\mathrm{bulge}}$ [kpc]), bulge-to-total ratio (B/T), disk radial scale length ($h_R$ [kpc]), vertical scale height ($z_0$ [kpc]), scale length-to-height ratio ($h_R/z_0$), disk-to-total ratio (D/T), envelope ellipticity ($\epsilon_{\mathrm{env}}$), envelope effective radius ($R_{e,\mathrm{env}}$ [kpc]), envelope-to-total ratio (Env/T), halo ellipticity  ($\epsilon_{halo}$), halo effective radius ($R_{e,\mathrm{halo}}$ [kpc]), and halo-to-total ratio (Halo/T). Ellipticities were converted from observed to intrinsic assuming each component is well described by an axisymmetric oblate spheroid. Galaxies best fitted with a four-component model were excluded from the computation of the median and standard deviation of envelope structural parameters.}

\begin{tabular}{lccccccccccccccc}
\hline
Name & $\epsilon_{\mathrm{bulge}}$ & $n_{\mathrm{bulge}}$ & $R_{e,\mathrm{bulge}}$ & B/T & $h_R$ & $z_0$ & $h_R/z_0$ & D/T & $\epsilon_{\mathrm{env}}$ & $R_{e,\mathrm{env}}$ & Env/T & $\epsilon_\mathrm{halo}$ & $R_{e,\mathrm{halo}}$ & Halo/T \\
 &  &  & [kpc] &  & [kpc] & [kpc] &  &  &  & [kpc] &  &  & [kpc] & \\
\hline
Mrk1216   & 0.4 & 2.3 & 0.42 & 0.25 & 0.7 & 0.16 &  4.5 & 0.08 & 0.6 & 3.1 & 0.32 & 0.0 & 8.2 & 0.35 \\
NGC1270   & 0.3 & 1.7 & 0.87 & 0.39 & 0.1 & 0.06 &  2.3 & 0.05 & 0.4 & 3.3 & 0.19 & 0.0 & 5.2 & 0.37 \\
NGC1271   & 0.4 & 2.2 & 0.25 & 0.25 & 0.9 & 0.13 &  7.0 & 0.12 & 0.7 & 2.5 & 0.38 & 0.3 & 4.5 & 0.25 \\
NGC1277   & 0.4 & 1.5 & 0.41 & 0.35 & 1.0 & 0.28 &  3.6 & 0.20 & 0.6 & 3.1 & 0.44 & -- & -- & -- \\
NGC1281   & 0.4 & 1.2 & 0.37 & 0.29 & 1.3 & 0.33 &  3.8 & 0.32 & 0.2 & 3.2 & 0.38 & -- & -- & -- \\
PGC12557  & 0.2 & 1.9 & 0.34 & 0.41 & 0.6 & 0.12 &  5.3 & 0.16 & 0.6 & 2.7 & 0.44 & -- & -- & -- \\
PGC12562  & 0.3 & 1.7 & 0.37 & 0.49 & 1.1 & 0.34 &  3.4 & 0.31 & 0.6 & 3.8 & 0.20 & -- & -- & -- \\
PGC32873  & 0.3 & 2.0 & 0.68 & 0.42 & 2.0 & 0.45 &  4.4 & 0.10 & 0.6 & 4.9 & 0.32 & 0.3 & 13.9 & 0.16 \\
\hline
Median (Relics) & 0.35 & 1.8 & 0.39 & 0.37 & 0.9 & 0.20 & 4.1 & 0.14 & 0.58 & 3.2 & 0.41 & 0.15 & 6.7 & 0.30 \\
\scriptsize Std dev & 0.06 & 0.3 & 0.21 & 0.08 & 0.5 & 0.14 & 2.8 & 0.10 & -- & -- & -- & -- & -- & -- \\
Median (MCGs)   & -- & -- & 0.28 & 0.26 & 1.1 & 0.29 & 3.9 & 0.42 & 0.40 & 4.4 & 0.30 & 0.15 & 7.0 & 0.23 \\
\scriptsize Std dev & -- & -- & 0.12 & 0.08 & 0.4 & 0.11 & 0.9 & 0.08 & 0.11 & 1.6 & 0.07 & -- & -- & -- \\
\hline
\end{tabular}
\label{tab:relics_2}
\end{table*}

\subsection{Morphological Properties of MCGs and CSGs}\label{sec:results_mcg_csg}

Figure\,\ref{fig:morph} presents the morphological properties and flux-to-total ratios of the different structural components. Throughout this paper, we consider differences between distributions to be statistically significant when the Kolmogorov–Smirnov test yields $p < 0.05$. The first row shows the effective radius, Sérsic index, and ellipticity of the bulges. A comparison of the effective radius distributions ($R_{e,\mathrm{Bulge}}$) shows that the bulges of MCGs are significantly more compact, with $65\%$ having $R_{e,\mathrm{Bulge}} < 400$\,pc, compared to only $24\%$ in the control sample, although this result should be interpreted with caution given the resolution limit of the HSC data; for all MCGs, $R_{e,\mathrm{Bulge}}$ is smaller than the full width at half maximum of the PSF and should therefore be regarded as an upper limit. This limitation is also reflected in the Sérsic index and ellipticity distributions: MCGs cluster near the boundaries of the allowed parameter space—toward the lower bound in $n$ and the upper bound in $\epsilon_\mathrm{Bulge}$—indicating that these parameters are driven by the fitting limits and are thus poorly constrained.

To assess the reliability of our $R_{e,\mathrm{Bulge}}$ measurements, we searched for \textit{HST} archival observations of the parent MCG sample. Of the 1858 MCGs in the \citet{slodkowski_clerici24} sample, 15 have available data, four of which satisfy our edge-on criteria. Details of these observations and their multi-component decompositions are provided in Appendix\,\ref{app:mcg_hst}. Briefly, for one galaxy the bulge is best fit by a point source, consistent with $R_{e,\mathrm{Bulge}} \lesssim 200$\,pc, while the other three have $R_{e,\mathrm{Bulge}} \sim 0.15$--0.35\,kpc, $\epsilon_{\mathrm{Bulge}} \sim 0.2$--0.5, and $n_{\mathrm{Bulge}} \sim 2$. Overall, the \textit{HST}-based sizes are comparable to those measured from the HSC images, suggesting that the $R_{e,\mathrm{Bulge}}$ distribution is broadly reliable, even though individual measurements remain uncertain. Given the above, we adopt $R_{e,\mathrm{Bulge}}$ as the primary bulge structural parameter throughout the remainder of this paper, and defer the analysis of Sérsic indices and bulge ellipticities to future work with higher-resolution data.

The second row of Fig.\,\ref{fig:morph} displays the scale length, scale height, and scale-length-to-height ratio ($h_r/z_0$) of the edge-on disk component. MCG disks are generally more compact, with $77\%$ having radial scale lengths $h_r < 1.5$\,kpc, compared to only $31\%$ of control S0s. They also tend to have smaller vertical scale heights, with median values of $z_0 \sim 0.29$\,kpc for MCGs and $z_0 \sim 0.38$\,kpc for CSGs, although this difference is less pronounced than that in scale length and the distributions overlap significantly. The difference between the distributions is primarily driven by the high-$z_0$ tail: $40\%$ of control S0s have $z_0 > 0.4$\,kpc, whereas only $17\%$ of MCGs exceed this value. The $h_r/z_0$ distribution further highlights the structural differences, with MCGs showing systematically lower ratios: $84\%$ have $h_r/z_0 < 5$, compared to $44\%$ of control galaxies. This indicates that, although MCGs have smaller scale heights on average, their even shorter radial scale lengths result in disks that are proportionally thicker.

In the third row of Fig.\,\ref{fig:morph}, we present the distributions of envelope ellipticity and effective radius. As seen for the bulge and disk components, MCG envelopes are typically more compact. Although the ellipticity ranges overlap, their distributions differ markedly: MCGs preferentially host rounder envelopes, with $72\%$ having $0.2 < \epsilon_{\mathrm{Envelope}} < 0.5$, compared to only $25\%$ of CSGs. Conversely, $73\%$ of CSGs have $\epsilon_{\mathrm{Envelope}} > 0.5$, versus only $19\%$ among MCGs.

The fourth row of Fig.\,\ref{fig:morph} displays the flux-to-total ratios of the bulge, disk, and envelope components. The median $B/T$ ($\sim 26\%$), $D/T$ ($\sim 42\%$), and $Env/T$ ($\sim 30\%$) values are similar between the samples. In the majority of galaxies the disk contributes the largest fraction of the total flux ($69\%$ of MCGs and $60\%$ of CSGs). In contrast, the fraction of galaxies in which the bulge contributes the largest fraction of the total flux is low, at $\sim 10\%$ in both samples.

\subsection{Morphological properties of Relic Galaxies}

Table\,\ref{tab:relics_2} lists the morphological parameters of the eight relic galaxies analyzed in this work, along with the median values for MCGs for comparison. Relic galaxies share several structural characteristics with MCGs: they host compact bulges ($R_{e,\mathrm{Bulge}} \sim 0.3$--$0.9$\,kpc), and their disks are short ($h_R \sim 0.5$--$2$\,kpc), thick ($h_R/z_0 \sim 2-7$), and embedded within flattened stellar envelopes ($\epsilon_{\mathrm{Envelope}} \sim 0.2$--$0.7$). Four relic galaxies required an additional outer component to achieve satisfactory fits. Given its low ellipticity ($\epsilon_h \sim 0$--$0.3$), we interpret this fourth component as a stellar halo.  We note that only nine MCGs required a similar fourth component. It remains unclear whether such four-component structures are intrinsically rarer among MCGs or simply go undetected due to the depth limitations of the HSC data.

In Fig.\,\ref{fig:mcg_vs_relics}, we compare images of relic galaxies and MCGs. The top panels show three relic galaxies, while the bottom panels present three MCGs selected to have similar stellar masses, effective radii, and velocity dispersions. It is evident from the images that MCGs and relic galaxies are morphologically very similar. However, inspection of the structural parameters in Table\,\ref{tab:relics_2} reveals a few significant differences. Notably, relic galaxies have larger B/T ratios. Additionally, five out of eight relics have $D/T \lesssim 0.15$. Conversely, no MCGs show similarly low $D/T$ values. 

Although the small sample size precludes firm conclusions, these differences raise some possibilities. One is that our sample is biased toward systems with higher $D/T$ ratios due to the selection criteria: galaxies with low $D/T$ may not display sufficiently disky isophotes to be well described by our three-component model and would therefore be excluded from the sample. Alternatively, galaxies with lower $D/T$ ratios may simply become more common at $\log(M_\star/M_\odot) \gtrsim 11$, a stellar-mass regime represented by only one MCG in our current sample.

\subsection{Size-mass Relations of the Bulge, Disk, and Envelope}\label{sec:size_mass}

\begin{figure*}
\centering
\begin{minipage}{0.445\linewidth}
\includegraphics[width=\linewidth, trim=50 8 90 0, clip]{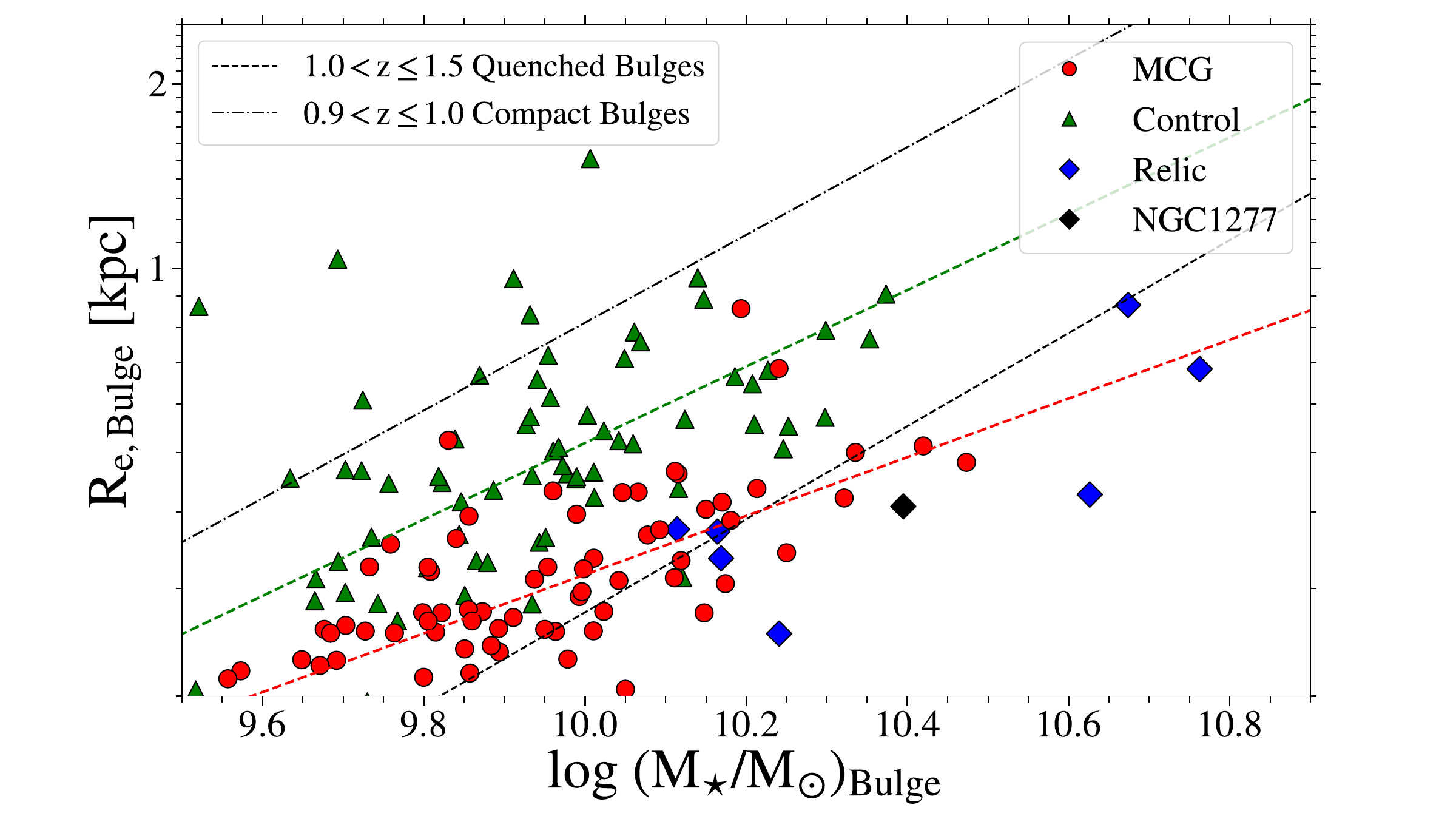}
\includegraphics[width=\linewidth, trim=45 8 90 0, clip]{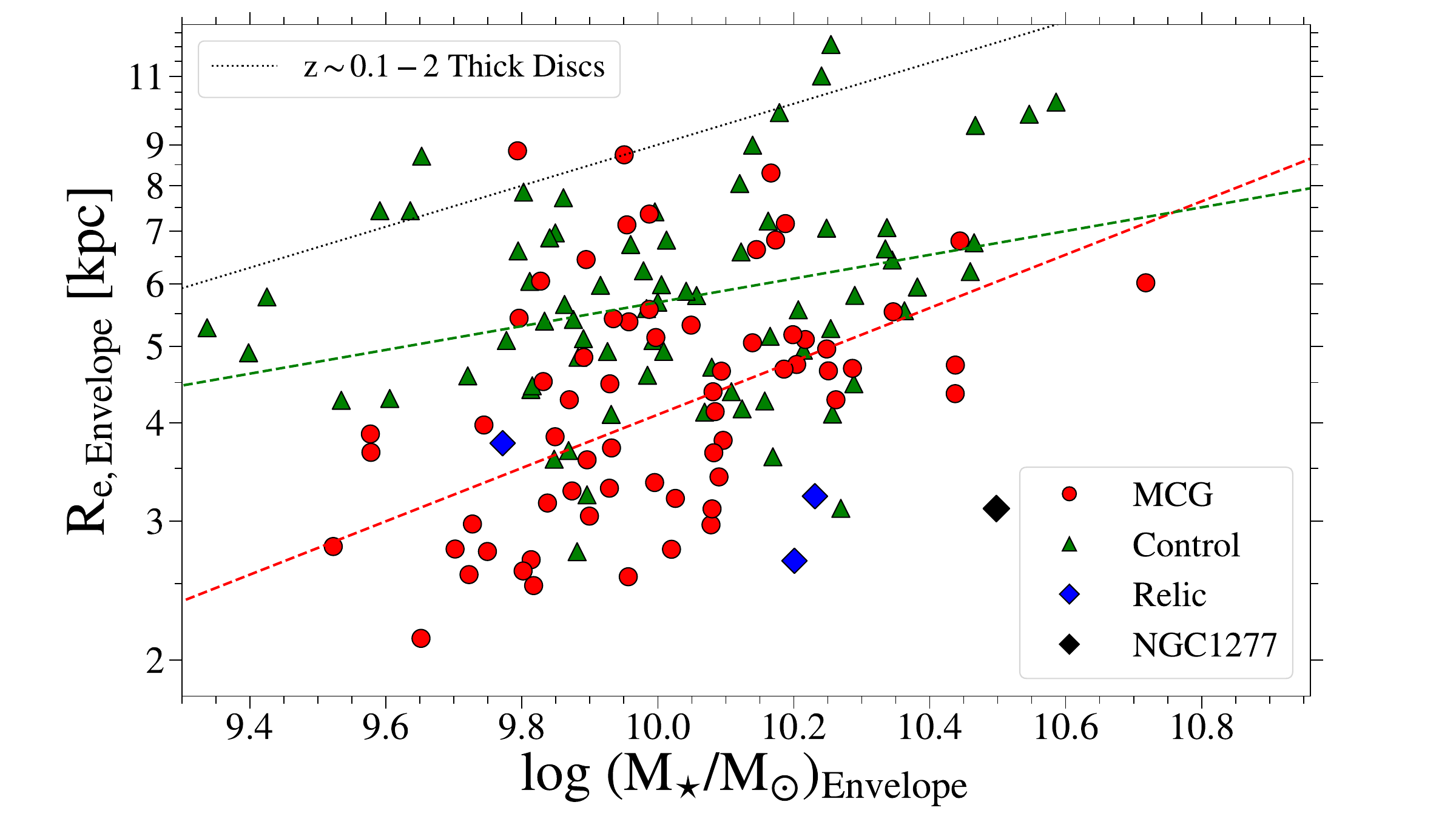}
\end{minipage}
\begin{minipage}{0.445\linewidth}
\includegraphics[width=\linewidth, trim=45 8 90 0, clip]{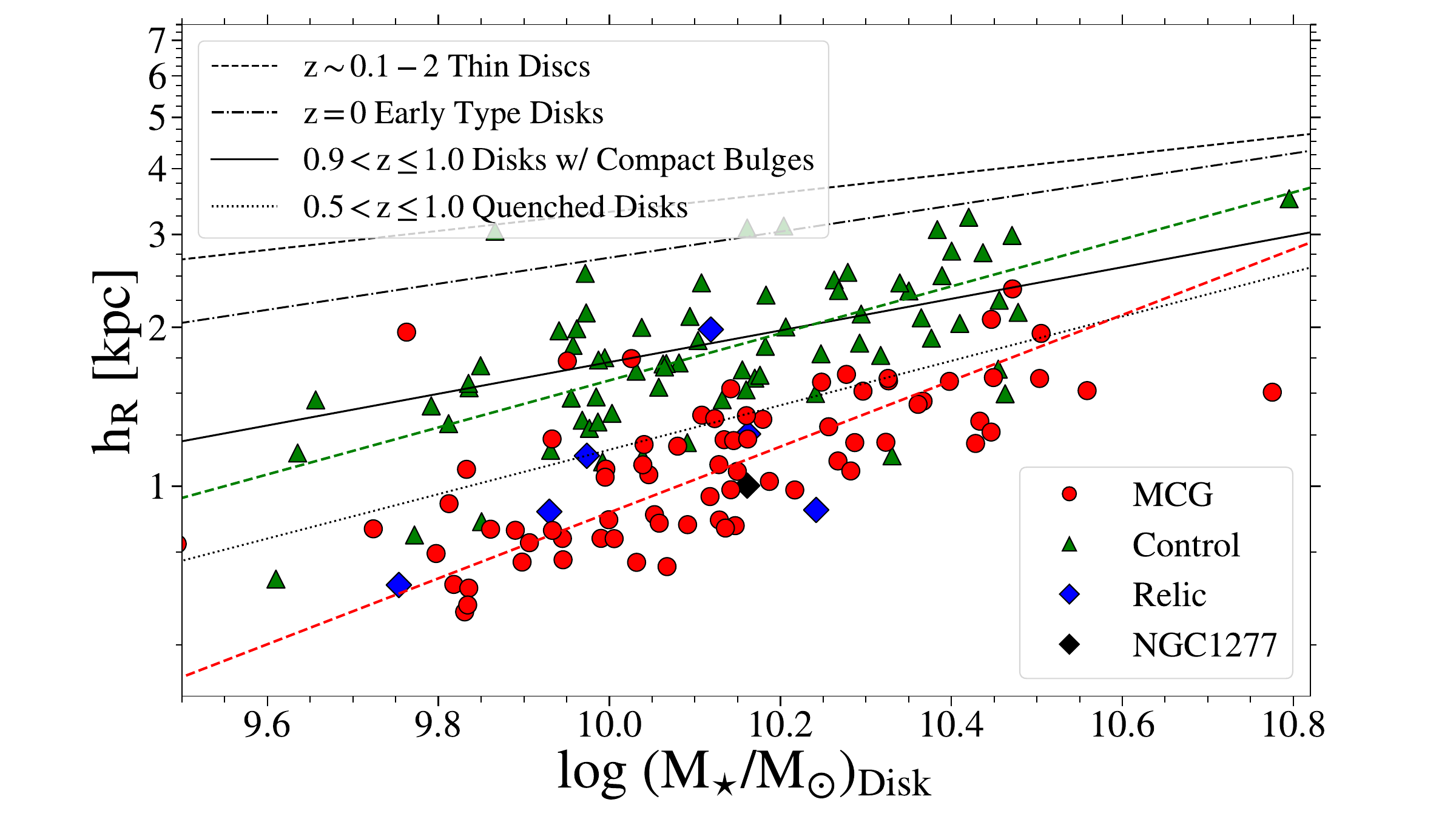}
\includegraphics[width=\linewidth, trim=45 8 90 0, clip]{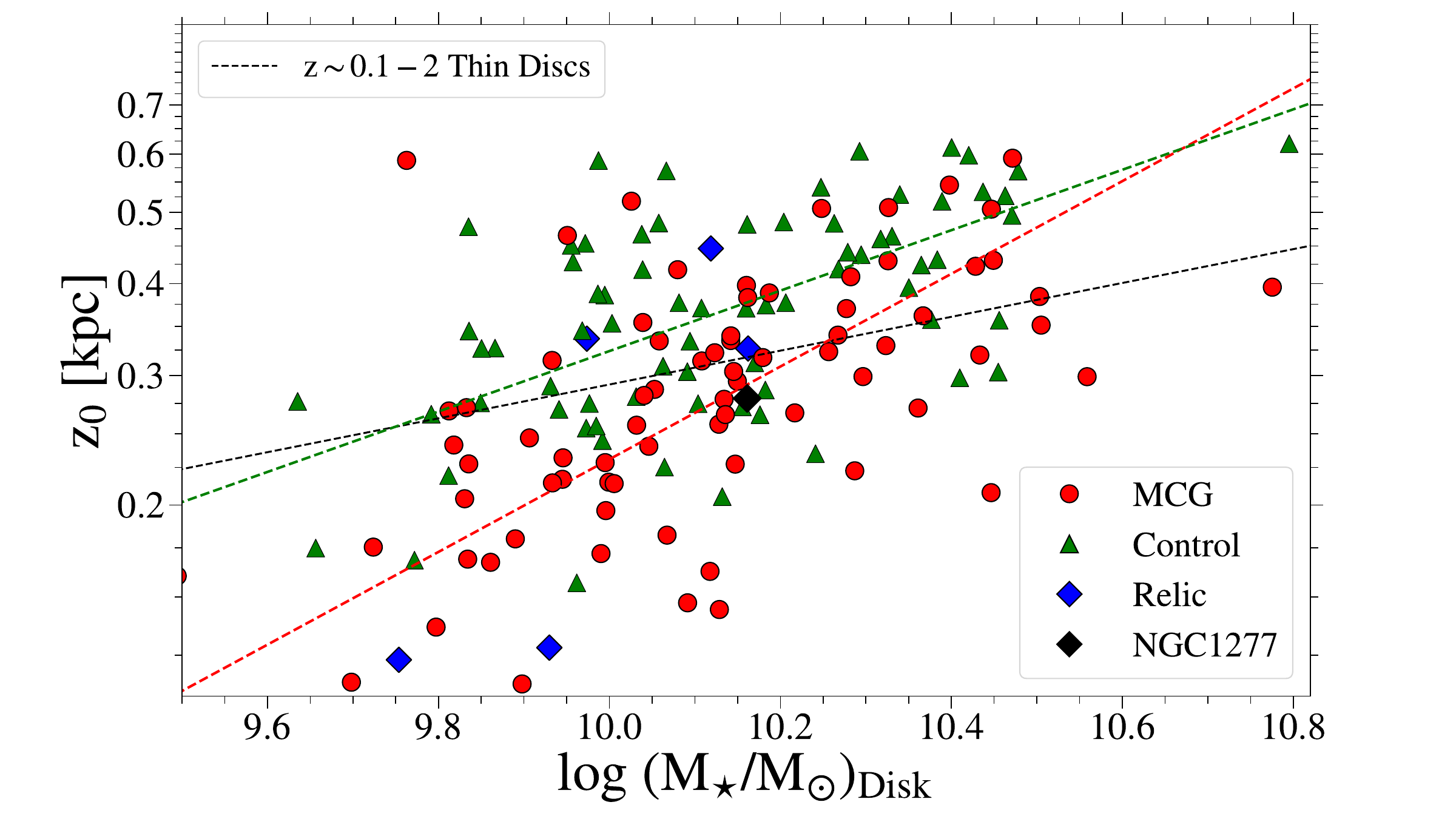}
\end{minipage}

\caption{Size--mass relations for the bulge (top left), disk (top right), and envelope (bottom left), together with the scale height--disk mass relation (bottom right). MCGs are shown as red circles, CSGs as green triangles, and relic galaxies as blue diamonds, with the prototypical relic galaxy NGC\,1277 highlighted as a black diamond. Best-fit relations from the literature for local and high-redshift galaxies are overplotted in each panel for comparison. The dashed red line shows a linear fit to the combined MCG and relic galaxy distribution, while the dashed green line indicates the best fit to the CSG distribution. Galaxies best fitted with a four-component model are excluded from the envelope size--mass relation.} 
\label{fig:bulge_disk_env_relics}
\end{figure*}

In Fig.\,\ref{fig:bulge_disk_env_relics}, we show the mass–size relations for bulges, disks, and envelopes, along with the disk mass–scale height relation. MCGs, relics, and CSGs are shown as red circles, blue diamonds, and green triangles, respectively. Given the similar distributions of MCGs and relics, we perform a combined fit to these two populations, indicated by the dashed red line, while the dashed green line shows the best fit to the CSG distribution. Reference relations from local and high-redshift samples are also included.

The top-left panel shows the bulge mass–size relation. Compared to MCGs, CSGs are shifted to larger bulge sizes at fixed stellar mass, although there is some overlap between the samples at the low-mass end. Relics mainly occupy the high-mass end, extending the MCG trend toward larger masses and sizes. Compared to best-fitting relations from the literature, the MCG fit lies close to the quenched bulge relation at $1.0 < z \leq 1.5$ \citep{nedkova24}, but due to its different slope, it lies above that relation at $\log(M_\star/M_\odot) \sim 10.3$ and below it at higher masses. Notably, relic galaxies tend to fall on or below the \citet{nedkova24} relation, whereas MCGs exhibit a broader spread, likely reflecting the resolution limits of the HSC data. The CSG fit has a similar slope to the \citet{nedkova24} and \citet{cook25} relations, lying between them. 

The top-right panel shows the disk mass–scale length relation. The best-fitting relations for MCGs and CSGs have similar slopes. When compared to size–mass relations from the literature (converted to $h_R$ using $R_e = 1.678\,h_R$ for an exponential disk), both MCG and CSG disks appear more compact than the thin disks of $z \sim 0.1-2$ star-forming galaxies \citep{tsukui25} and the disks of local early-type galaxies \citep{lange16}. Moreover, MCG disks are also shorter than those of $z \sim 1$ quenched galaxies \citep{nedkova24} and galaxies hosting classical bulges \citep{cook25}. Notably, the MCG relation differs from these not only in zero point but also in slope, showing a stronger mass dependence. Relic disks have comparable sizes and stellar masses to MCG disks. However, as relics tend to host more massive bulges at fixed total stellar mass, their disks and envelopes carry a smaller fraction of the total mass. Since component  sizes scale with component mass, this naturally explains why relics are more compact than MCGs at fixed total stellar mass: higher $B/T$ ratios reduce galaxy sizes not only through the larger bulge contribution, but also because less massive disks and  envelopes are intrinsically smaller.

The bottom-left panel shows the envelope mass–size relation. Both samples display considerable scatter, particularly the CSGs, but MCG envelopes are typically more compact. The MCG relation shows a stronger mass dependence and converges toward the CSG trend at higher masses. As one possible interpretation is that the envelope corresponds to a thick disk, we compare its mass–size relation to that of thick disks in star-forming galaxies \citep{tsukui25}, finding that most envelopes lie below this relation.

The bottom-right panel shows the disk scale height as a function of mass. In contrast to the disk scale length, MCGs and CSGs do not show a clear separation in scale height, with the two samples largely overlapping. Consequently, given their shorter scale lengths, MCG disks are proportionally thicker. Compared to the scale height–stellar mass relation of star-forming galaxies \citep{tsukui25}, the disk scale heights of MCGs show a stronger mass dependence.

\subsection{Coupling of Component Lengths}

\begin{figure}
\centering
\includegraphics[width=0.95\linewidth, trim=0 0 0 0, clip]{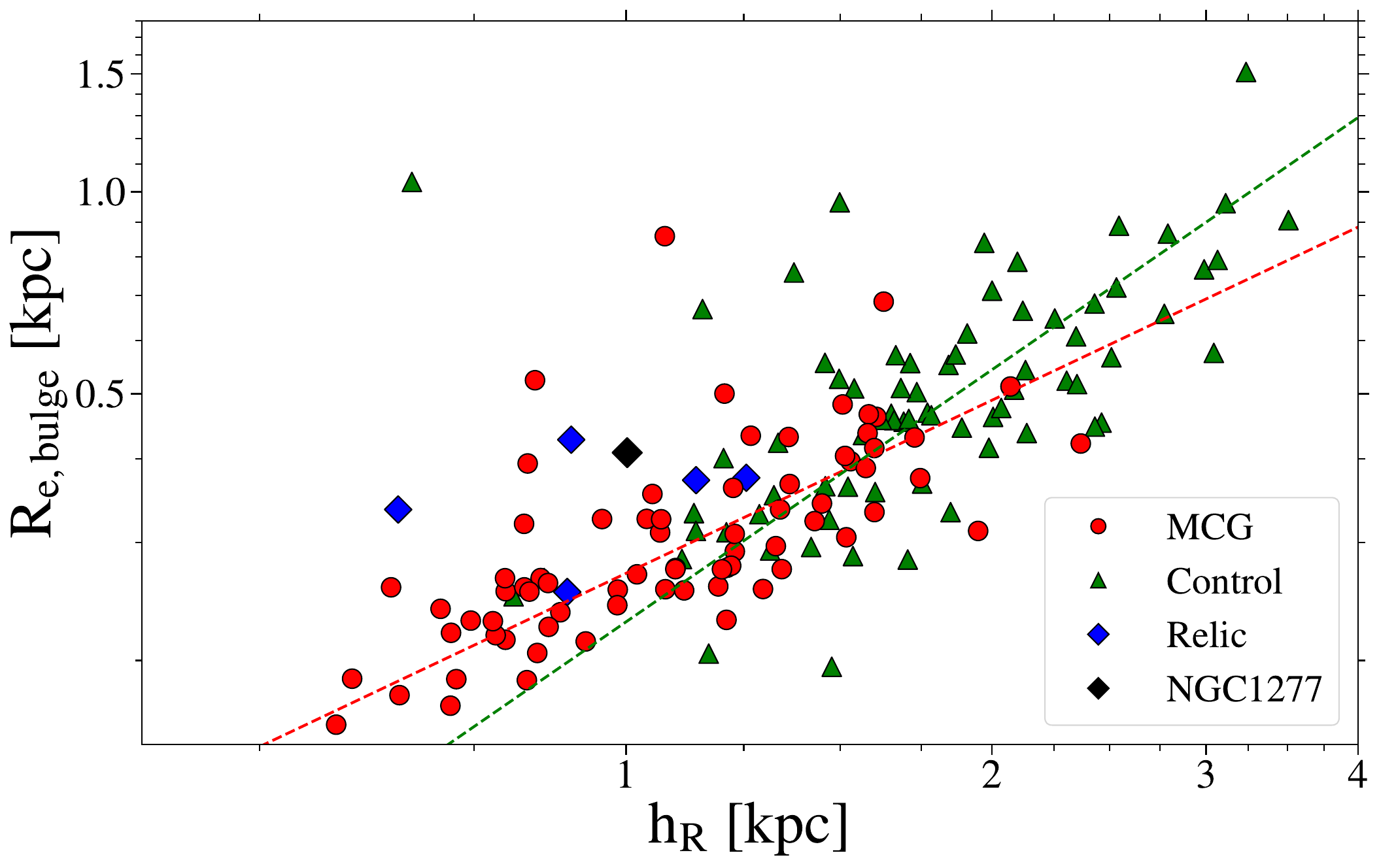}
\includegraphics[width=0.95\linewidth, trim=0 0 0 0, clip]{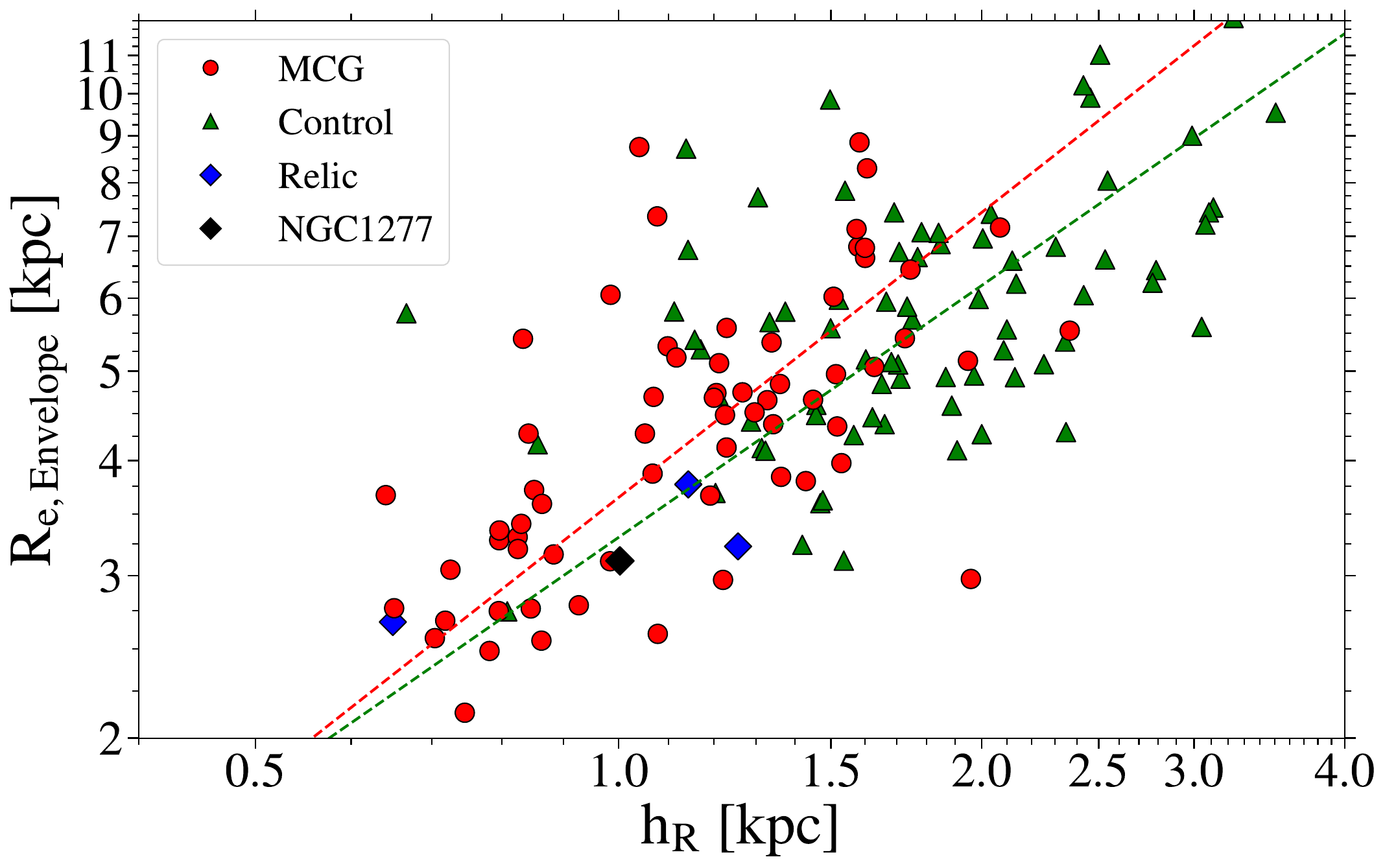}
\caption{Coupling between structural components. Disk scale length versus bulge effective radius is shown in the top panel, and disk scale length versus envelope effective radius in the bottom panel. MCGs are shown as red circles, CSGs as green triangles, and relics as blue diamonds with the prototypical relic galaxy NGC\,1277 highlighted as a black diamond. A linear fit to the combined MCG and relic distribution is indicated by the dashed red line, while the dashed green line shows the best fit to the CSG distribution. Galaxies best fitted with a four-component model are excluded from the bottom panel.}
\label{fig:coupling}
\end{figure}

S0 and spiral galaxies display a coupling between $R_{e,\mathrm{bulge}}$ and $h_R$, which has been interpreted as evidence that the formation processes of these components are linked \citep{aguerri05,carollo07,nordemeer07,laurikainen09,laurikainen10}. To test whether a similar coupling is present in MCGs, we show in the top panel of Fig.\,\ref{fig:coupling} $R_{e,\mathrm{bulge}}$ as a function of $h_R$. The dashed lines indicate linear fits to the CSG (green) and the combined MCG+relic (red) distributions. A coupling is present, with Spearman correlation coefficients of $r = 0.72$ for MCGs and $r = 0.62$ for CSGs.

To test whether a similar coupling is present between the disk and the envelope, we show in the bottom panel of Fig.\,\ref{fig:coupling} $h_R$ as a function of $R_{e,\mathrm{Envelope}}$. A correlation is clearly present, with Spearman coefficients of $r = 0.69$ for MCGs and $r = 0.48$ for CSGs. The presence of this coupling suggests a relatively quiet post-quenching accretion history in MCGs, as repeated dry minor mergers are expected to disrupt both the disk and the envelope, thereby weakening or erasing this relation, and only gas-rich mergers could plausibly re-establish it \citep{querejeta15}.

\section{Discussion\label{subsec:discuss}}

\subsection{Insights into the Formation of MCGs from the Disk Structure}

The comparison between the disk scale lengths and scale heights of MCGs and CSGs shown in Figs.\,\ref{fig:morph} and \ref{fig:bulge_disk_env_relics} offers valuable insight into the conditions under which the disks of MCGs formed. The scale height of a self-gravitating isothermal sheet is given by $z_0 = \sigma_\star^2/(2\pi G \Sigma_\star)$ \citep{binney08}, where $\sigma_\star$ and $\Sigma_\star$ are the stellar velocity dispersion and stellar surface density, respectively. Since the $z_0$ distributions of MCGs and CSGs largely overlap, while MCG disks are systematically shorter at fixed stellar mass (and therefore denser), the disks of MCGs are expected to have higher $\sigma_\star$. If stellar disks preserve, at least partially, the kinematic properties of the gas disks in which they formed, this suggests that the star-forming progenitors of MCGs hosted more turbulent gas disks than the progenitors of CSGs. An evolutionary connection between MCGs and compact high-redshift starbursts provides a possible explanation for this scenario, since intensely star-forming dust-obscured galaxies at $z \sim 1.3$--$2.6$ have been found to exhibit elevated H$\alpha$ velocity dispersions compared to less actively star-forming galaxies at similar redshifts \citep{birkin24}. It is worth noting that the stellar population properties of MCGs, namely their old ages, supersolar metallicities, and enhanced [$\alpha$/Fe] reported by \citet{slodkowski_clerici24}, are also consistent with such an evolutionary connection.

\subsection{The Nature of the Envelope}

An extended stellar envelope surrounds the disks of MCGs, containing $\sim 25\%$ of the total flux. The envelope ellipticity ranges from $0.0$ to $0.6$, with a median value of $\epsilon_{\mathrm{env}} \sim 0.4$. This suggests that in some cases the envelope corresponds to a stellar halo, while in others it is more consistent with a thick disk. Although the nature of the envelope cannot be determined unambiguously without spatially resolved kinematics, examining MCGs at the extremes of the ellipticity distribution provides valuable insight into its origin.

\begin{figure}
\centering
\hspace*{-0.85cm}
\includegraphics[width=\linewidth, trim=0 0 0 0, clip]{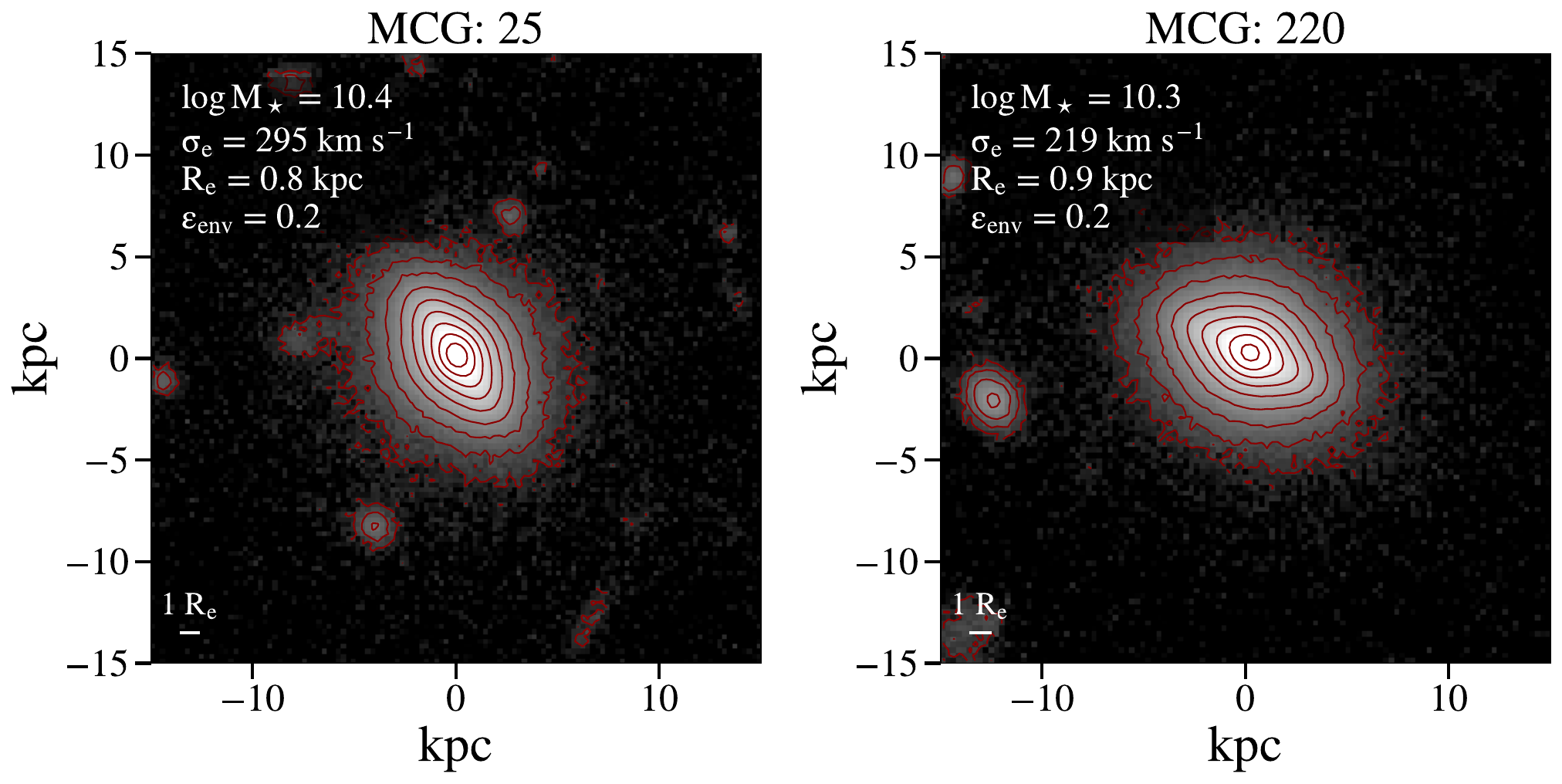}
\hspace*{-0.85cm}
\includegraphics[width=\linewidth, trim=0 0 0 0, clip]{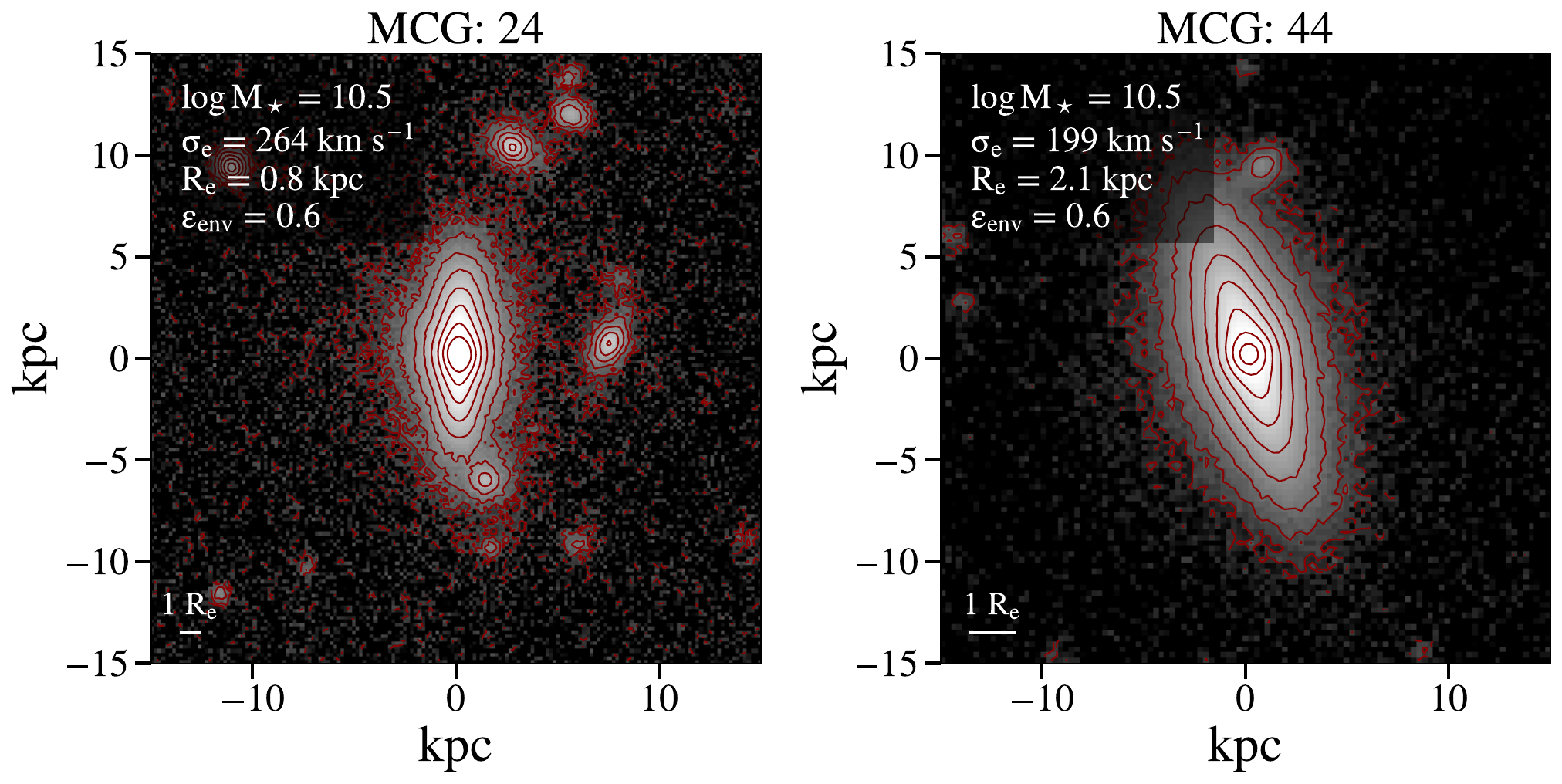}
\caption{Top panels: example \textit{r}-band images of MCGs with outer ellipticity $\epsilon_{\mathrm{env}} < 0.3$. 
Bottom panels: example \textit{r}-band images of MCGs with outer ellipticity $\epsilon_{\mathrm{env}} > 0.45$.
Overlaid are isophotal contours spanning surface brightness levels from $\mu = 18.5$ to $26\,\mathrm{mag\,arcsec^{-2}}$, highlighting the galaxies’ morphological structure.}
\label{fig:env_02_04}
\end{figure}

We first consider MCGs with low envelope ellipticity ($\epsilon_{\mathrm{env}} < 0.3$), which account for $17\%$ of the sample. Two representative examples, MCGs ID 25 and 220, are shown in the top panels of Fig.\,\ref{fig:env_02_04}. Galaxies in this group exhibit compact disks embedded within round stellar envelopes. They resemble compact analogs of the halo-embedded disk galaxies identified in IllustrisTNG50 by \citet{du21} and \citet{he25}, which consist of intermediate-scale disks surrounded by diffuse stellar halos. \citet{du21} proposed an evolutionary scenario in which a major merger disrupts an early disk and builds a stellar halo, followed by disk regrowth through residual star formation fueled by leftover gas after coalescence. A similar mechanism may apply to some MCGs, provided the merger occurred at $z \gtrsim 2$, consistent with their old stellar populations (see \citealt{slodkowski_clerici24}). In this scenario, the disk-dominated structure of low-$\epsilon_{\mathrm{env}}$ MCGs ($\mathrm{D/T} \sim 0.4$, $\mathrm{Env/T} \sim 0.2$) may reflect the high gas fractions expected at early cosmic times, which favor efficient disk regrowth post-merger.

A similar conclusion can be drawn by comparing MCGs and relic galaxies. In a recent study, \citet{zhu25} analyzed the dynamical properties of compact early-type galaxies in IllustrisTNG50 and classified them as ``merger-free'' or not based on whether the fraction of accreted stars is below 10\%. They then applied this classification to a sample of nearby compact elliptical galaxies (which includes all galaxies in our relic sample except for PGC12557) by comparing their observed dynamics to those of the simulated systems. \citet{zhu25} found that the relic galaxies Mrk\,1216, NGC\,1270, and NGC\,1271—all of which in our analysis require four structural components—and NGC\,1281, which has a relatively low envelope ellipticity ($\epsilon_{\mathrm{env}} = 0.2$), are likely to have experienced mergers that significantly altered their internal structure. The remaining relic galaxies were classified as ``merger-free''.

We now turn to MCGs with high envelope ellipticity ($\epsilon_{\mathrm{env}} > 0.45$), comprising $29\%$ of the sample. Two representative examples, MCGs ID 24 and 44, are shown in the bottom panels of Fig.\,\ref{fig:env_02_04}. This group is morphologically homogeneous, with members resembling compact, thicker versions of the S0s in the control sample. Given their high ellipticity, we interpret their envelopes as thick disks.

Simulations suggest that thick disks form through a combination of early in situ star formation and later growth via accretion and dynamical heating of thin disk stars \citep{park21,yu21,pinna24,yi24}, with the relative contributions varying between galaxies. While thin disks can survive minor mergers \citep{bournaud07,villalobos08,qu11}, mergers with stellar mass ratios of $0.1$--$0.2$ often leave only $15$--$25\%$ of the stellar mass in a surviving thin disk \citep{villalobos08}. This is at odds with the observed $\mathrm{D/T} \sim 0.4$ in MCGs, moreover only $4\%$ of the sample have $\mathrm{D/T} < 0.25$. Therefore, if high-$\epsilon_{\mathrm{env}}$ envelopes formed via minor mergers, these events must have been gas-rich, allowing for substantial thin-disk regrowth.  Alternatively, a predominantly in-situ formation of high-ellipticity envelopes is also plausible.  We note that the relics classified as merger-free by \citet{zhu25} almost invariably have $\epsilon_{\mathrm{env}} \geq 0.5$, supporting this scenario.

This discussion raises the broader question of the role mergers play in the formation of MCGs. Simulations suggest that this role is limited: in \citet{lohmann23}, we analyzed simulated MCG analogs at $z = 0$ in Illustris TNG100 and found a median ex situ stellar mass fraction of only $\sim 7\%$, although this can reach $\sim 20\%$ in some cases. Similarly, \citet{kurinchi24} find that mergers play a minimal role in the quenching of star formation in quiescent galaxies at $z > 3$. In contrast, \citet{pathak21} report that the majority of young quiescent galaxies at $z \sim 2$ experienced significant mergers, although this fraction is much smaller in older quiescent systems. An important caveat, however, is whether these simulated analogs adequately capture the formation physics of such rare, extreme galaxies.

An observational perspective is provided by morphological studies of $z > 2$ starbursts and sub-millimeter galaxies, the likely progenitors of high-redshift compact quiescent galaxies. Studies with the Atacama Large Millimeter/submillimeter Array (ALMA) and the James Webb Space Telescope (JWST) consistently find that the majority of these galaxies show signs of interactions, with major merger fractions of $\sim 20-25\%$ and minor merger fractions of $\sim 40-50\%$ \citep{guijarro18,colina23,gillman24,polletta24,ren25,hodge25}. However, galaxies with undisturbed morphologies make up $20-40\%$ of the population, implying secular dynamical processes can also play a role in shaping their morphology \citep{ren25,hodge25,stacey25}.

Taken together, these results suggest that early gas-rich mergers are likely to play an important role in the formation of many MCGs, although a non-negligible fraction of merger-free MCGs is expected.

\subsection{Are MCGs Relics of the High Redshift Universe?}

There is no consensus definition of relic galaxies, with different studies adopting varying criteria, but the following conditions are usually required to be satisfied \citep{trujillo14,ferre-mateu15,ferre_mateu17,tortora25}: I) the galaxy must host predominantly old ($\gtrsim 10\,\mathrm{Gyr}$) stellar populations out to a few $R_e$, with the fraction of younger stellar populations below $\sim 10\%$; II) it should show a regular velocity field and exhibit an undisturbed morphology, as these are features consistent with a passive post-quenching evolution; III) the galaxy should be massive and compact. This last point is particularly ill-defined, as the mass and size thresholds vary significantly across studies. While a quantitative estimate of the fraction of MCGs satisfying the relic galaxy criteria can not be obtained with the currently available data, we believe a qualitative discussion on this topic is pertinent given the structural similarities between MCGs and relics highlighted earlier in this paper.

We begin by assessing criterion II). Based on a visual inspection of the HSC images, we identify only ten MCGs with minor asymmetries or disturbances that could plausibly be associated with either minor mergers or tidal stripping. It is important to bear in mind that such signatures typically remain visible for only a few Gyr \citep{mancillas19}, and therefore trace only recent merger activity. However, it is unlikely that more than a small fraction of MCGs experienced significant dry merging at earlier times, given that their high $D/T$ ratios are inconsistent with the expected for remnants of dry mergers with mass ratios $> 0.1$, as discussed in the previous section. We therefore expect that a significant fraction of MCGs satisfy criterion II).

\begin{figure}
\centering
\includegraphics[width=\linewidth, trim=0 0 0 0, clip]{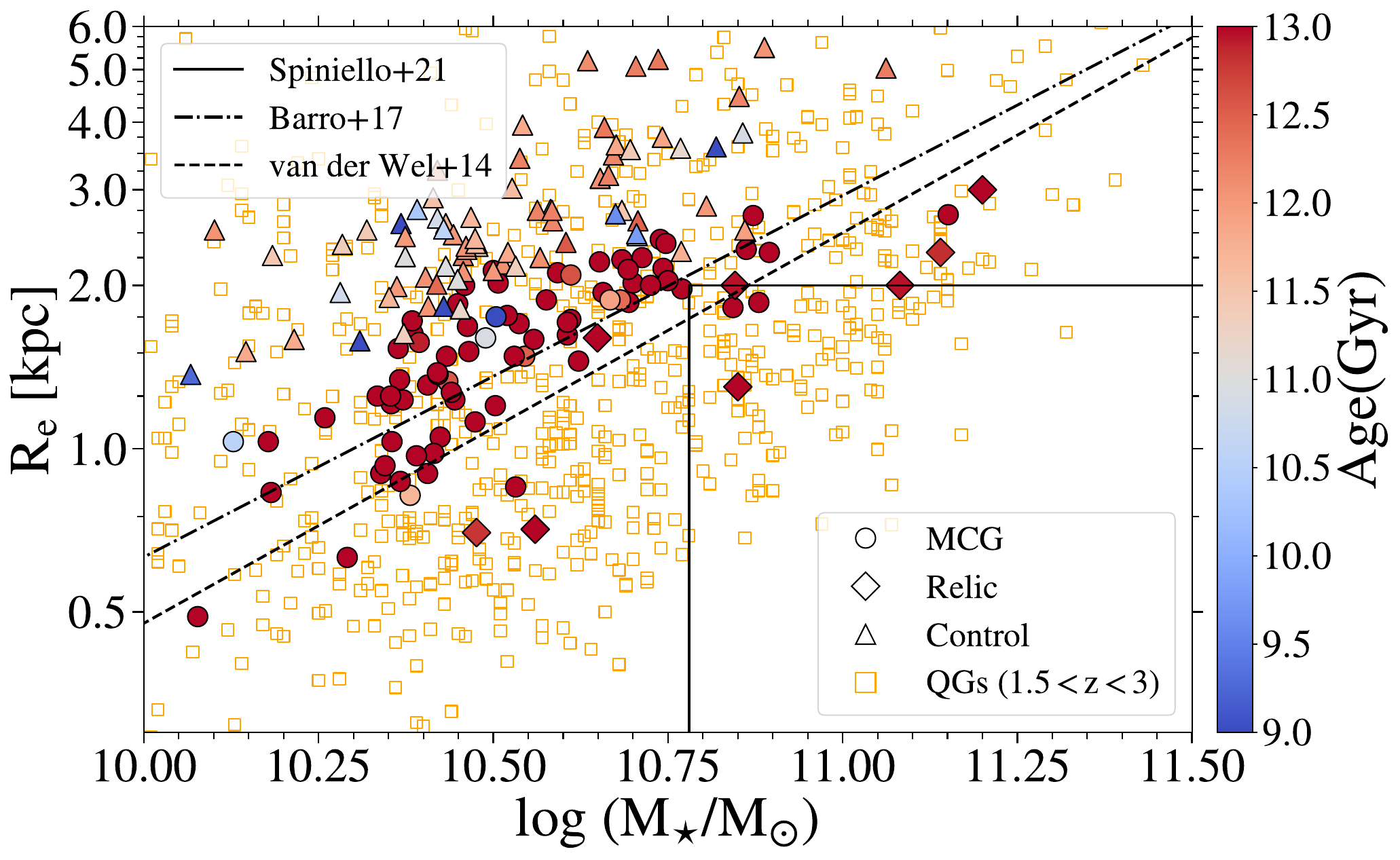}
\caption{Distribution of MCGs, CSGs, relics, and $1.5 < z < 3$ quiescent galaxies in the $M_\star$–$R_e$ plane. MCGs, CSGs, and relics are color-coded by stellar population age. The dashed, dot-dashed, and solid lines represent the compactness thresholds proposed by \citet{vanderwel14}, \citet{barro17}, and \citet{spiniello21}, respectively. Despite their differing sizes, relics and MCGs exhibit similarly old stellar populations. In contrast, CSGs are significantly younger, even though some have sizes comparable to those of MCGs.}
\label{fig:age}
\end{figure}

To assess criteria I) and III), we present in Fig.\,\ref{fig:age} the size–mass distribution of MCGs and relic galaxies, color-coded by mass-weighted stellar population age. CSGs are also shown for comparison. The dashed, dot-dashed, and solid lines mark the compactness thresholds proposed by \citet{vanderwel14}, \citet{barro17}, and \citet{spiniello21}, respectively. Stellar ages for MCGs were taken from \citet{slodkowski_clerici24}. Ages for CSGs and relic galaxies (except Mrk\,1266) were derived from their SDSS spectra following the methodology of \citet{slodkowski_clerici24}, while the age of Mrk\,1266, which has no SDSS spectrum available, was taken from \citet{ferre_mateu17}. Within the region covered by the SDSS fiber ($1.4 \pm 0.5\,R_e$ on average in our sample), MCGs typically exhibit ages comparable to those of relics and older than those of CSGs, with only one galaxy younger than $10$\,Gyr. Given the arguments presented above against significant dry merging, and considering the envelope-origin scenarios discussed earlier, which suggest that the envelope was largely in place by $z \gtrsim 2$, it is plausible that MCGs generally host similarly old stellar populations out to several $R_e$, therefore we expect a significant fraction fulfills criterion I, although spatially resolved spectroscopy is required to confirm this. Regarding compactness (criterion III), 2 MCGs satisfy the \citet{spiniello21} threshold, 7 meet the \citet{vanderwel14} threshold, and 23 fulfill the \citet{barro17} criterion. Since all three relic criteria (I–III) must be satisfied simultaneously, only a few MCGs in our sample qualify as relic galaxy candidates, despite the large fraction that show undisturbed morphologies and host very old stellar populations within the fiber aperture.

To explore this further, we also show in Fig.\,\ref{fig:age} the distribution in the size–mass plane of $1.5 < z < 3$ quiescent galaxies extracted from the 3D-HST survey \citep{brammer12,skelton14,momcheva16}. Following \citet{wuyts07} and \citet{williams09}, 3D-HST galaxies were classified as star-forming or quiescent based on their rest-frame colors. We extracted their sizes from \citet{vanderwel14}. Some 3D-HST quiescent galaxies occupy the same region as MCGs, and some are even larger, meaning that many high-z quiescent galaxies, if they evolved passively, would not satisfy the relic criteria at $z \sim 0$. This reflects the fact that the compactness thresholds were designed to select local analogues with sizes at the lower end of the high-z size distribution. Therefore, a passively evolving galaxy with a relatively large size may fulfill criteria I) and II) but fail to meet the compactness requirement (criterion III), and thus would not be classified as a relic under current definitions. This limited overlap with compactness thresholds raises the question of how well local relic definitions capture the properties of the broader high-redshift quiescent population. 

High-redshift quiescent galaxies differ from their local counterparts not only in the size–mass relation but also in the stellar mass–velocity dispersion ($M_\star$–$\sigma_e$) relation. As illustrated in Fig.\,\ref{fig:MCGs_CSGs_SDSS}, the $M_\star$–$\sigma_e$ relation has undergone significant evolution since $z = 2$. In \citet{slodkowski_clerici24}, we demonstrated that the stellar population properties of quiescent galaxies systematically vary with their location in the $M_\star$–$\sigma_e$ plane. Massive compact galaxies, $+2\sigma$ outliers above the local $M_\star$–$\sigma_e$ relation by definition, are older and more $\alpha$-enhanced than quiescent galaxies with similar $\sigma_e$ along the central trend. These differences diminish with increasing $\sigma_e$ and largely disappear at $\sigma_e \gtrsim 225$\,km\,s$^{-1}$, where galaxies on the local relation exhibit comparable ages and $\alpha$-enhancements. A similar trend was reported by \citet{graves09a} for elliptical galaxies. Furthermore, \citet{graves09b} found that, at fixed $\sigma_e$, stellar population properties of ellipticals show no dependence on $R_e$ (and thus on compactness), a result we also confirm for MCGs. Consequently, quiescent galaxy samples selected solely on the basis of compactness—without considering the $M_\star$–$\sigma_e$ relation—include systems spanning a wide range of stellar ages, from ancient galaxies to young post-starburst objects \citep{ferre-mateu12,damjanov13,zahid16,schnorr21,spiniello21}. An age diversity is also visible in Fig.\,\ref{fig:age}, where CSGs consistently exhibit younger ages than MCGs regardless of their size, although to a smaller degree than in the aforementioned studies.

The results presented in Sec.\,\ref{sec:size_mass} hint at a possible additional drawback of strict compactness thresholds: they tend to select galaxies with more massive bulges, thereby preferentially excluding disk-dominated systems. However, our strict selection based on the $M_\star-\sigma_e$ relation is also subject to biases. Among the 225 HSC-observed MCGs presented in Paper I, 75 were classified as edge-on systems, representing a disproportionate number of highly inclined galaxies. This reflects a selection bias introduced by the contribution of rotational motion to the integrated velocity dispersion, which leads to systematically higher $\sigma_e$ measurements in highly inclined systems. A similar inclination bias is evident in the \citet{yildirim17} compact elliptical galaxy sample, where high velocity dispersion was a selection criterion: all 17 galaxies have inclinations $\gtrsim 70^\circ$. A possible way to mitigate these biases may be to combine constraints from both the $M_\star-R_e$ and $M_\star-\sigma_e$ relations, using thresholds less stringent than those adopted here and in previous studies. We leave the exploration of such combined selection strategies for future work.

\subsection{Structural Similarities between MCGs and High-z Quiescent Galaxies}

In this study, we have shown that the morphological structure of MCGs is well described by a combination of compact bulges and disks embedded within a stellar envelope of varying intrinsic ellipticity, with the disk generally being the dominant component. Given that the stellar population properties of MCGs indicate a formation epoch at $z \gtrsim 2$, one would expect a significant fraction of high-$z$ quiescent galaxies to exhibit similar structures if MCGs indeed underwent passive evolution. Observations provide suggestive evidence supporting this expectation. 

\citet{davari17} performed bulge–disk decompositions of massive ($\log(M_\star/M_\odot) > 10.7$) quiescent galaxies at $0.5 < z < 2.5$ in the CANDELS fields, finding that at $z \sim 2$, more than 50\% of galaxies have $B/T \leq 0.5$. Furthermore, spatially resolved stellar kinematics of massive quiescent galaxies at $z = 1{-}3$ reveals the majority are fast rotators, consistent with the presence of stellar disks. However, the measured values of the spin parameter $\lambda_e$ show considerable variation, suggesting differences in the degree of rotational support among galaxies \citep{deugenio24,pascalau25,forrest25,slob25}.

A few $z \gtrsim 2$ quiescent galaxies have been the subject of detailed studies, allowing for a deeper comparison. A double Sérsic decomposition of galaxies MRG-2129 and MRG-M0138 by \citet{newman18a} reveals edge-on disks embedded in mildly flattened envelopes ($\epsilon \approx 0.2$ and $\epsilon \approx 0.3$, respectively) with $n \sim 1$ and no evidence of $n = 4$ classical bulges, sharing notable similarities with MCGs. Similarly, a bulge+disk decomposition of the $z \sim 3$ quiescent galaxy ZF-UDS-7329 by \citet{turner25} shows an edge-on disk and an $n \sim 2$ bulge, further reinforcing the disky nature of high-$z$ quiescent galaxies. Taken together, these results suggest that the structural properties of MCGs are consistent with those of massive quiescent galaxies at $z \sim 2{-}3$, supporting an evolutionary link between the two populations.

\section{Summary and Conclusions} \label{sec:conclusions}

In this work, we presented the results of a multi-component morphological decomposition of $r$-band HSC images for a sample of 75 edge-on massive compact galaxies, alongside an equal-sized control sample of edge-on S0 galaxies matched in stellar mass and redshift, as well as a sample of 8 relic galaxies with HST imaging. Galaxies were modeled with three structural components: a Sérsic bulge, an edge-on disk, and an exponential stellar envelope.

We find that:
\begin{itemize}
    \item The smaller sizes of MCGs compared to CSGs are not driven preferentially by a single component. Instead, all structural components are more compact. MCGs exhibit more compact bulges (median $R_\mathrm{e,bulge} \sim 0.3$\,kpc vs.\ $0.5$\,kpc) and envelopes ($R_\mathrm{e,env} \sim 4.4$\,kpc vs.\ $5.7$\,kpc), as well as shorter disks ($h_R \sim 1.1$\,kpc vs.\ $1.7$\,kpc);
    \item MCGs host thicker disks ($h_R/z_0 \sim 3.9$ vs.\ $5.3$) and rounder envelopes (median $\epsilon_\mathrm{env} \sim 0.40$ vs.\ $0.56$) than CSGs;
    \item The flux fractions of the structural components are broadly consistent between the two samples, with median $B/T \sim 0.3$ and $D/T \sim 0.4$ in both;
    \item Considering the $z_0$ distributions of MCGs and CSGs largely overlap while MCG disks are systematically shorter at fixed stellar mass, we argue that MCG disks likely have higher stellar velocity dispersions. We suggest this difference could, at least in part, be due to the star-forming progenitors of MCGs hosting more turbulent gas disks;
    \item We find that the sizes of the structural components are coupled in MCGs, suggesting that the formation and growth of the disk, bulge, and envelope are linked. This coupling, together with the survival of a disk component with $D/T \sim 0.4$, argues against dry minor mergers as the dominant mechanism responsible for building the envelope, at least for the MCG population as a whole: repeated minor mergers would heat and disrupt the thin disk, and deposit material preferentially at large radii without producing the observed coupling between component sizes;
    \item A comparison with 8 relic galaxies reveals that MCGs and relics share many structural properties, although MCGs are typically somewhat larger. Relics also exhibit compact bulges and disks embedded within stellar envelopes. The structural components of MCGs and relics are consistent with following the same size--mass relations, distinct from those followed by the components of CSGs. This supports the idea that relics and MCGs belong to the same structural family;
    \item Despite these similarities, relics tend to have higher $B/T$ ratios (median 0.37 vs. 0.26). This is likely to be, at least in part, a contributing factor to the smaller sizes of relics compared to MCGs. At fixed total stellar mass, higher $B/T$ ratios are associated with smaller galaxy  sizes not only because of the larger bulge contribution, but also because the disk  and envelope carry less of the total mass --- and since component sizes scale with  component mass, this makes all components smaller simultaneously;
    \item Given that higher $B/T$ ratios lead to smaller galaxy sizes, the strict compactness criteria frequently employed in relic searches are likely to bias the selected samples toward bulge-dominated galaxies;
    \item The envelope ellipticities of MCGs and relics span a wide range, $\epsilon_\mathrm{Env} \sim 0$--$0.7$. This suggests that envelopes correspond  to different physical structures in different galaxies. Based on comparisons with simulations and $z > 2$ starburst observations, we propose that low-ellipticity envelopes ($\epsilon_{Env} \lesssim 0.2$) are stellar haloes formed in gas-rich major mergers at $z \gtrsim 2$,  while the most flattened ones ($\epsilon_{Env} \gtrsim 0.5$) are thick disks, formed either in-situ or through gas-rich minor mergers. The nature of envelopes with intermediate ellipticities remains ambiguous from photometry alone;
    \item The majority of MCGs show undisturbed morphologies, and within the region probed by SDSS spectra ($1.4 \pm 0.5\,R_e$ on average) they exhibit old stellar populations ($\gtrsim 10$\,Gyr), characteristics comparable to those of confirmed relics. However, most MCGs  fail to meet the compactness thresholds commonly adopted in relic searches: only 2  satisfy the \citet{spiniello21} criterion, 7 meet the \citet{vanderwel14} threshold, and 23 fulfill the \citet{barro17} criterion. Since spatially resolved spectroscopy  is required to confirm that old stellar populations extend to several $R_e$, a definitive assessment on the fraction of passively evolving MCGs is not yet possible.  Nonetheless, the structural, morphological, and stellar population similarities between MCGs and relics suggest that a significant fraction of MCGs are members of a population of passively evolved compact galaxies that current relic search criteria fail to capture.
\end{itemize}

\begin{acknowledgments}

ASM acknowledges the financial support from the Brazilian National Council for Scientific and Technological Development (CNPq) and from the Fundação de Amparo à Pesquisa do Estado do Rio Grande do Sul (FAPERGS). KSC acknowledges the Coordination for the Improvement of Higher Education Personnel (CAPES) for the financial support
(88887.629089/2021-00). FP acknowledges support from the Horizon Europe research and innovation programme under the Maria Skłodowska-Curie grant “TraNSLate” No 101108180. ACSM acknowledges support from the European Southern Observatory (ESO) as an SCV visitor at the ESO Science Office in Vitacura, as well as financial support from CAPES (process no. 88887.001289/2024-00). MT thanks the support from CNPq (process 307111/2025-3) and from FAPERGS.
\end{acknowledgments}



\bibliography{sample701}{}

@ARTICLE{dominguez-sanchez18,
       author = {{Dom{\'\i}nguez S{\'a}nchez}, H. and {Huertas-Company}, M. and {Bernardi}, M. and {Tuccillo}, D. and {Fischer}, J.~L.},
        title = "{Improving galaxy morphologies for SDSS with Deep Learning}",
      journal = {\mnras},
         year = 2018,
        month = feb,
       volume = {476},
       number = {3},
        pages = {3661-3676},
          doi = {10.1093/mnras/sty338},
archivePrefix = {arXiv},
       eprint = {1711.05744},
 primaryClass = {astro-ph.GA},
       adsurl = {https://ui.adsabs.harvard.edu/abs/2018MNRAS.476.3661D}
}

@ARTICLE{fangzhou12,
       author = {{Jiang}, Fangzhou and {van Dokkum}, Pieter and {Bezanson}, Rachel and {Franx}, Marijn},
        title = "{A Nearby Analog of z \raisebox{-0.5ex}\textasciitilde 2 Compact Quiescent Galaxies with a Rotating Disk}",
      journal = {\apjl},
         year = 2012,
        month = apr,
       volume = {749},
       number = {1},
          eid = {L10},
        pages = {L10},
          doi = {10.1088/2041-8205/749/1/L10},
archivePrefix = {arXiv},
       eprint = {1203.1317},
 primaryClass = {astro-ph.CO},
       adsurl = {https://ui.adsabs.harvard.edu/abs/2012ApJ...749L..10J}
}

@ARTICLE{llasker13,
       author = {{Lasker}, R. and {van den Bosch}, R.~C.~E. and {van de Ven}, G. and {Ferreras}, I. and {La Barbera}, F. and {Vazdekis}, A. and {Falcon-Barroso}, J.},
        title = "{Bottom-heavy initial mass function in a nearby compact l\{star\} galaxy.}",
      journal = {\mnras},
         year = 2013,
        month = jul,
       volume = {434},
        pages = {L31-L35},
          doi = {10.1093/mnrasl/slt070},
archivePrefix = {arXiv},
       eprint = {1305.5542},
 primaryClass = {astro-ph.CO},
       adsurl = {https://ui.adsabs.harvard.edu/abs/2013MNRAS.434L..31L}
}

@ARTICLE{mancillas19,
       author = {{Mancillas}, Brisa and {Duc}, Pierre-Alain and {Combes}, Fran{\c{c}}oise and {Bournaud}, Fr{\'e}d{\'e}ric and {Emsellem}, Eric and {Martig}, Marie and {Michel-Dansac}, Leo},
        title = "{Probing the merger history of red early-type galaxies with their faint stellar substructures}",
      journal = {\aap},
         year = 2019,
        month = dec,
       volume = {632},
          eid = {A122},
        pages = {A122},
          doi = {10.1051/0004-6361/201936320},
archivePrefix = {arXiv},
       eprint = {1909.07500},
 primaryClass = {astro-ph.GA},
       adsurl = {https://ui.adsabs.harvard.edu/abs/2019A&A...632A.122M}
}

@ARTICLE{trujillo14,
       author = {{Trujillo}, Ignacio and {Ferr{\'e}-Mateu}, Anna and {Balcells}, Marc and {Vazdekis}, Alexandre and {S{\'a}nchez-Bl{\'a}zquez}, Patricia},
        title = "{NGC 1277: A Massive Compact Relic Galaxy in the Nearby Universe}",
      journal = {\apjl},
         year = 2014,
        month = jan,
       volume = {780},
       number = {2},
          eid = {L20},
        pages = {L20},
          doi = {10.1088/2041-8205/780/2/L20},
archivePrefix = {arXiv},
       eprint = {1310.6367},
 primaryClass = {astro-ph.CO},
       adsurl = {https://ui.adsabs.harvard.edu/abs/2014ApJ...780L..20T}
}

@ARTICLE{tortora25,
       author = {{Tortora}, C. and {Tozzi}, G. and {Agapito}, G. and {Barbera}, F. La and {Spiniello}, C. and {Li}, R. and {Carl{\`a}}, G. and {D'Ago}, G. and {Ghose}, E. and {Mannucci}, F. and {Napolitano}, N.~R. and {Pinna}, E. and {Arnaboldi}, M. and {Bevacqua}, D. and {Ferr{\'e}-Mateu}, A. and {Gallazzi}, A. and {Hartke}, J. and {Hunt}, L.~K. and {Maksymowicz-Maciata}, M. and {Pulsoni}, C. and {Saracco}, P. and {Scognamiglio}, D. and {Spavone}, M.},
        title = "{INSPIRE: INvestigating Stellar Populations In RElics - IX. KiDS J0842 + 0059: the first fully confirmed relic beyond the local Universe}",
      journal = {\mnras},
         year = 2025,
        month = jul,
       volume = {540},
       number = {3},
        pages = {2555-2565},
          doi = {10.1093/mnras/staf831},
archivePrefix = {arXiv},
       eprint = {2505.13611},
 primaryClass = {astro-ph.GA},
       adsurl = {https://ui.adsabs.harvard.edu/abs/2025MNRAS.540.2555T}
}

@ARTICLE{ferre-mateu15,
       author = {{Ferr{\'e}-Mateu}, Anna and {Mezcua}, Mar and {Trujillo}, Ignacio and {Balcells}, Marc and {van den Bosch}, Remco C.~E.},
        title = "{Massive Relic Galaxies Challenge the Co-evolution of Super-massive Black Holes and Their Host Galaxies}",
      journal = {\apj},
         year = 2015,
        month = jul,
       volume = {808},
       number = {1},
          eid = {79},
        pages = {79},
          doi = {10.1088/0004-637X/808/1/79},
archivePrefix = {arXiv},
       eprint = {1506.02663},
 primaryClass = {astro-ph.GA},
       adsurl = {https://ui.adsabs.harvard.edu/abs/2015ApJ...808...79F}
}

@ARTICLE{hodge25,
       author = {{Hodge}, J.~A. and {da Cunha}, E. and {Kendrew}, S. and {Li}, J. and {Smail}, I. and {Westoby}, B.~A. and {Nayak}, O. and {Swinbank}, A.~M. and {Chen}, C. -C. and {Walter}, F. and {van der Werf}, P. and {Cracraft}, M. and {Battisti}, A. and {Brandt}, W.~N. and {Calistro Rivera}, G. and {Chapman}, S.~C. and {Cox}, P. and {Dannerbauer}, H. and {Decarli}, R. and {Frias Castillo}, M. and {Greve}, T.~R. and {Knudsen}, K.~K. and {Leslie}, S. and {Menten}, K.~M. and {Rybak}, M. and {Schinnerer}, E. and {Wardlow}, J.~L. and {Weiss}, A.},
        title = "{ALESS-JWST: Joint (Sub)kiloparsec JWST and ALMA Imaging of z \raisebox{-0.5ex}\textasciitilde 3 Submillimeter Galaxies Reveals Heavily Obscured Bulge Formation Events}",
      journal = {\apj},
         year = 2025,
        month = jan,
       volume = {978},
       number = {2},
          eid = {165},
        pages = {165},
          doi = {10.3847/1538-4357/ad9a52},
archivePrefix = {arXiv},
       eprint = {2407.15846},
 primaryClass = {astro-ph.GA},
       adsurl = {https://ui.adsabs.harvard.edu/abs/2025ApJ...978..165H}
}

@ARTICLE{querejeta15,
       author = {{Querejeta}, M. and {Eliche-Moral}, M.~C. and {Tapia}, T. and {Borlaff}, A. and {Rodr{\'\i}guez-P{\'e}rez}, C. and {Zamorano}, J. and {Gallego}, J.},
        title = "{Formation of S0 galaxies through mergers. Bulge-disc structural coupling resulting from major mergers}",
      journal = {\aap},
         year = 2015,
        month = jan,
       volume = {573},
          eid = {A78},
        pages = {A78},
          doi = {10.1051/0004-6361/201424303},
archivePrefix = {arXiv},
       eprint = {1409.5126},
 primaryClass = {astro-ph.GA},
       adsurl = {https://ui.adsabs.harvard.edu/abs/2015A&A...573A..78Q}
}

@ARTICLE{colina23,
       author = {{Colina}, L. and {Crespo G{\'o}mez}, A. and {{\'A}lvarez-M{\'a}rquez}, J. and {Bik}, A. and {Walter}, F. and {Boogaard}, L. and {Labiano}, A. and {Peissker}, F. and {P{\'e}rez-Gonz{\'a}lez}, P. and {{\"O}stlin}, G. and {Greve}, T.~R. and {N{\o}rgaard-Nielsen}, H.~U. and {Wright}, G. and {Alonso-Herrero}, A. and {Azollini}, R. and {Caputi}, K.~I. and {Dicken}, D. and {Garc{\'\i}a-Mar{\'\i}n}, M. and {Hjorth}, J. and {Ilbert}, O. and {Kendrew}, S. and {Pye}, J.~P. and {Tikkanen}, T. and {van der Werf}, P. and {Costantin}, L. and {Iani}, E. and {Gillman}, S. and {Jermann}, I. and {Langeroodi}, D. and {Moutard}, T. and {Rinaldi}, P. and {Topinka}, M. and {van Dishoeck}, E.~F. and {G{\"u}del}, M. and {Henning}, Th. and {Lagage}, P.~O. and {Ray}, T. and {Vandenbussche}, B.},
        title = "{Uncovering the stellar structure of the dusty star-forming galaxy GN20 at z = 4.055 with MIRI/JWST}",
      journal = {\aap},
         year = 2023,
        month = may,
       volume = {673},
          eid = {L6},
        pages = {L6},
          doi = {10.1051/0004-6361/202346535},
archivePrefix = {arXiv},
       eprint = {2304.13529},
 primaryClass = {astro-ph.GA},
       adsurl = {https://ui.adsabs.harvard.edu/abs/2023A&A...673L...6C}
}

@ARTICLE{polletta24,
       author = {{Polletta}, M. and {Frye}, B.~L. and {Garuda}, N. and {Willner}, S.~P. and {Berta}, S. and {Kneissl}, R. and {Dole}, H. and {Jansen}, R.~A. and {Lehnert}, M.~D. and {Cohen}, S.~H. and {Summers}, J. and {Windhorst}, R.~A. and {D'Silva}, J.~C.~J. and {Koekemoer}, A.~M. and {Coe}, D. and {Conselice}, C.~J. and {Driver}, S.~P. and {Grogin}, N.~A. and {Marshall}, M.~A. and {Nonino}, M. and {Ortiz}, III, R. and {Pirzkal}, N. and {Robotham}, A. and {Ryan}, R.~E. and {Willmer}, C.~N.~A. and {Yan}, H. and {Arumugam}, V. and {Cheng}, C. and {Gim}, H.~B. and {Hathi}, N.~P. and {Holwerda}, B. and {Kamieneski}, P. and {Keel}, W.~C. and {Li}, J. and {Pascale}, M. and {Rottgering}, H. and {Smith}, B.~M. and {Yun}, M.~S.},
        title = "{JWST's PEARLS: Resolved study of the stellar and dust components in starburst galaxies at cosmic noon}",
      journal = {\aap},
         year = 2024,
        month = oct,
       volume = {690},
          eid = {A285},
        pages = {A285},
          doi = {10.1051/0004-6361/202450671},
archivePrefix = {arXiv},
       eprint = {2405.07986},
 primaryClass = {astro-ph.GA},
       adsurl = {https://ui.adsabs.harvard.edu/abs/2024A&A...690A.285P}
}

@ARTICLE{stacey25,
       author = {{Stacey}, H.~R. and {Kaasinen}, M. and {O'Riordan}, C.~M. and {McKean}, J.~P. and {Powell}, D.~M. and {Rizzo}, F.},
        title = "{A nuclear spiral in a dusty star-forming galaxy at z = 2.78}",
      journal = {\aap},
         year = 2025,
        month = jan,
       volume = {693},
          eid = {L17},
        pages = {L17},
          doi = {10.1051/0004-6361/202452518},
archivePrefix = {arXiv},
       eprint = {2412.03644},
 primaryClass = {astro-ph.GA},
       adsurl = {https://ui.adsabs.harvard.edu/abs/2025A&A...693L..17S}
}

@ARTICLE{gillman24,
       author = {{Gillman}, Steven and {Smail}, Ian and {Gullberg}, Bitten and {Swinbank}, A.~M. and {Vijayan}, Aswin P. and {Lee}, Minju and {Brammer}, Gabe and {Dudzevi{\v{c}}i{\={u}}t{\.{e}}}, Ugn{\.{e}} and {Greve}, Thomas R. and {Almaini}, Omar and {Brinch}, Malte and {Chapman}, Scott C. and {Chen}, Chian-Chou and {Ikarashi}, Soh and {Matsuda}, Yuichi and {Wang}, Wei-Hao and {Walter}, Fabian and {van der Werf}, Paul P.},
        title = "{The structure of massive star-forming galaxies from JWST and ALMA: Dusty, high-redshift disc galaxies}",
      journal = {\aap},
         year = 2024,
        month = nov,
       volume = {691},
          eid = {A299},
        pages = {A299},
          doi = {10.1051/0004-6361/202451006},
archivePrefix = {arXiv},
       eprint = {2406.03544},
 primaryClass = {astro-ph.GA},
       adsurl = {https://ui.adsabs.harvard.edu/abs/2024A&A...691A.299G}
}

@ARTICLE{guijarro18,
       author = {{G{\'o}mez-Guijarro}, C. and {Toft}, S. and {Karim}, A. and {Magnelli}, B. and {Magdis}, G.~E. and {Jim{\'e}nez-Andrade}, E.~F. and {Capak}, P.~L. and {Fraternali}, F. and {Fujimoto}, S. and {Riechers}, D.~A. and {Schinnerer}, E. and {Smol{\v{c}}i{\'c}}, V. and {Aravena}, M. and {Bertoldi}, F. and {Cortzen}, I. and {Hasinger}, G. and {Hu}, E.~M. and {Jones}, G.~C. and {Koekemoer}, A.~M. and {Lee}, N. and {McCracken}, H.~J. and {Micha{\l}owski}, M.~J. and {Navarrete}, F. and {Povi{\'c}}, M. and {Puglisi}, A. and {Romano-D{\'\i}az}, E. and {Sheth}, K. and {Silverman}, J.~D. and {Staguhn}, J. and {Steinhardt}, C.~L. and {Stockmann}, M. and {Tanaka}, M. and {Valentino}, F. and {van Kampen}, E. and {Zirm}, A.},
        title = "{Starburst to Quiescent from HST/ALMA: Stars and Dust Unveil Minor Mergers in Submillimeter Galaxies at z {\ensuremath{\sim}} 4.5}",
      journal = {\apj},
         year = 2018,
        month = apr,
       volume = {856},
       number = {2},
          eid = {121},
        pages = {121},
          doi = {10.3847/1538-4357/aab206},
archivePrefix = {arXiv},
       eprint = {1802.07751},
 primaryClass = {astro-ph.GA},
       adsurl = {https://ui.adsabs.harvard.edu/abs/2018ApJ...856..121G}
}

@ARTICLE{ren25,
       author = {{Ren}, Jian and {Liu}, F.~S. and {Li}, Nan and {Zhao}, Pinsong and {Cui}, Qifan and {Song}, Qi and {Li}, Yubin and {Mo}, Hao and {Yesuf}, Hassen M. and {Wang}, Weichen and {An}, Fangxia and {Zheng}, Xian Zhong},
        title = "{The Evolution of the Size and Merger Fraction of Submillimeter Galaxies across 1 < z {\ensuremath{\lesssim}} 6 as Observed by JWST}",
      journal = {\apj},
         year = 2025,
        month = apr,
       volume = {982},
       number = {2},
          eid = {200},
        pages = {200},
          doi = {10.3847/1538-4357/adb961},
archivePrefix = {arXiv},
       eprint = {2502.15569},
 primaryClass = {astro-ph.GA},
       adsurl = {https://ui.adsabs.harvard.edu/abs/2025ApJ...982..200R}
}

@ARTICLE{ferre_mateu17,
       author = {{Ferr{\'e}-Mateu}, Anna and {Trujillo}, Ignacio and {Mart{\'\i}n-Navarro}, Ignacio and {Vazdekis}, Alexandre and {Mezcua}, Mar and {Balcells}, Marc and {Dom{\'\i}nguez}, Lilian},
        title = "{Two new confirmed massive relic galaxies: red nuggets in the present-day Universe}",
      journal = {\mnras},
         year = 2017,
        month = may,
       volume = {467},
       number = {2},
        pages = {1929-1939},
          doi = {10.1093/mnras/stx171},
archivePrefix = {arXiv},
       eprint = {1701.05197},
 primaryClass = {astro-ph.GA},
       adsurl = {https://ui.adsabs.harvard.edu/abs/2017MNRAS.467.1929F}
}

@ARTICLE{brammer12,
       author = {{Brammer}, Gabriel B. and {van Dokkum}, Pieter G. and {Franx}, Marijn and {Fumagalli}, Mattia and {Patel}, Shannon and {Rix}, Hans-Walter and {Skelton}, Rosalind E. and {Kriek}, Mariska and {Nelson}, Erica and {Schmidt}, Kasper B. and {Bezanson}, Rachel and {da Cunha}, Elisabete and {Erb}, Dawn K. and {Fan}, Xiaohui and {F{\"o}rster Schreiber}, Natascha and {Illingworth}, Garth D. and {Labb{\'e}}, Ivo and {Leja}, Joel and {Lundgren}, Britt and {Magee}, Dan and {Marchesini}, Danilo and {McCarthy}, Patrick and {Momcheva}, Ivelina and {Muzzin}, Adam and {Quadri}, Ryan and {Steidel}, Charles C. and {Tal}, Tomer and {Wake}, David and {Whitaker}, Katherine E. and {Williams}, Anna},
        title = "{3D-HST: A Wide-field Grism Spectroscopic Survey with the Hubble Space Telescope}",
      journal = {\apjs},
         year = 2012,
        month = jun,
       volume = {200},
       number = {2},
          eid = {13},
        pages = {13},
          doi = {10.1088/0067-0049/200/2/13},
archivePrefix = {arXiv},
       eprint = {1204.2829},
 primaryClass = {astro-ph.CO},
       adsurl = {https://ui.adsabs.harvard.edu/abs/2012ApJS..200...13B}
}

@ARTICLE{skelton14,
       author = {{Skelton}, Rosalind E. and {Whitaker}, Katherine E. and {Momcheva}, Ivelina G. and {Brammer}, Gabriel B. and {van Dokkum}, Pieter G. and {Labb{\'e}}, Ivo and {Franx}, Marijn and {van der Wel}, Arjen and {Bezanson}, Rachel and {Da Cunha}, Elisabete and {Fumagalli}, Mattia and {F{\"o}rster Schreiber}, Natascha and {Kriek}, Mariska and {Leja}, Joel and {Lundgren}, Britt F. and {Magee}, Daniel and {Marchesini}, Danilo and {Maseda}, Michael V. and {Nelson}, Erica J. and {Oesch}, Pascal and {Pacifici}, Camilla and {Patel}, Shannon G. and {Price}, Sedona and {Rix}, Hans-Walter and {Tal}, Tomer and {Wake}, David A. and {Wuyts}, Stijn},
        title = "{3D-HST WFC3-selected Photometric Catalogs in the Five CANDELS/3D-HST Fields: Photometry, Photometric Redshifts, and Stellar Masses}",
      journal = {\apjs},
         year = 2014,
        month = oct,
       volume = {214},
       number = {2},
          eid = {24},
        pages = {24},
          doi = {10.1088/0067-0049/214/2/24},
archivePrefix = {arXiv},
       eprint = {1403.3689},
 primaryClass = {astro-ph.GA},
       adsurl = {https://ui.adsabs.harvard.edu/abs/2014ApJS..214...24S}
}

@ARTICLE{graves09a,
       author = {{Graves}, Genevieve J. and {Faber}, S.~M. and {Schiavon}, Ricardo P.},
        title = "{Dissecting the Red Sequence. I. Star-Formation Histories of Quiescent Galaxies: The Color-Magnitude versus the Color-{\ensuremath{\sigma}} Relation}",
      journal = {\apj},
         year = 2009,
        month = mar,
       volume = {693},
       number = {1},
        pages = {486-506},
          doi = {10.1088/0004-637X/693/1/486},
archivePrefix = {arXiv},
       eprint = {0810.4334},
 primaryClass = {astro-ph},
       adsurl = {https://ui.adsabs.harvard.edu/abs/2009ApJ...693..486G}
}

@ARTICLE{graves09b,
       author = {{Graves}, Genevieve J. and {Faber}, S.~M. and {Schiavon}, Ricardo P.},
        title = "{Dissecting the Red Sequence. II. Star Formation Histories of Early-Type Galaxies Throughout the Fundamental Plane}",
      journal = {\apj},
         year = 2009,
        month = jun,
       volume = {698},
       number = {2},
        pages = {1590-1608},
          doi = {10.1088/0004-637X/698/2/1590},
archivePrefix = {arXiv},
       eprint = {0903.3603},
 primaryClass = {astro-ph.CO},
       adsurl = {https://ui.adsabs.harvard.edu/abs/2009ApJ...698.1590G}
}

@ARTICLE{buitrago18,
       author = {{Buitrago}, F. and {Ferreras}, I. and {Kelvin}, L.~S. and {Baldry}, I.~K. and {Davies}, L. and {Angthopo}, J. and {Khochfar}, S. and {Hopkins}, A.~M. and {Driver}, S.~P. and {Brough}, S. and {Sabater}, J. and {Conselice}, C.~J. and {Liske}, J. and {Holwerda}, B.~W. and {Bremer}, M.~N. and {Phillipps}, S. and {L{\'o}pez-S{\'a}nchez}, {\'A}. R. and {Graham}, A.~W.},
        title = "{Galaxy and Mass Assembly (GAMA): Accurate number densities and environments of massive ultra-compact galaxies at 0.02 < z < 0.3}",
      journal = {\aap},
         year = 2018,
        month = nov,
       volume = {619},
          eid = {A137},
        pages = {A137},
          doi = {10.1051/0004-6361/201833785},
archivePrefix = {arXiv},
       eprint = {1807.02534},
 primaryClass = {astro-ph.GA},
       adsurl = {https://ui.adsabs.harvard.edu/abs/2018A&A...619A.137B}
}

@ARTICLE{carollo13,
       author = {{Carollo}, C.~M. and {Bschorr}, T.~J. and {Renzini}, A. and {Lilly}, S.~J. and {Capak}, P. and {Cibinel}, A. and {Ilbert}, O. and {Onodera}, M. and {Scoville}, N. and {Cameron}, E. and {Mobasher}, B. and {Sanders}, D. and {Taniguchi}, Y.},
        title = "{Newly Quenched Galaxies as the Cause for the Apparent Evolution in Average Size of the Population}",
      journal = {\apj},
         year = 2013,
        month = aug,
       volume = {773},
       number = {2},
          eid = {112},
        pages = {112},
          doi = {10.1088/0004-637X/773/2/112},
archivePrefix = {arXiv},
       eprint = {1302.5115},
 primaryClass = {astro-ph.CO},
       adsurl = {https://ui.adsabs.harvard.edu/abs/2013ApJ...773..112C}
}

@ARTICLE{williams17,
       author = {{Williams}, Christina C. and {Giavalisco}, Mauro and {Bezanson}, Rachel and {Cappelluti}, Nico and {Cassata}, Paolo and {Liu}, Teng and {Lee}, Bomee and {Tundo}, Elena and {Vanzella}, Eros},
        title = "{Morphology Dependence of Stellar Age in Quenched Galaxies at Redshift {\ensuremath{\sim}}1.2: Massive Compact Galaxies Are Older than More Extended Ones}",
      journal = {\apj},
         year = 2017,
        month = apr,
       volume = {838},
       number = {2},
          eid = {94},
        pages = {94},
          doi = {10.3847/1538-4357/aa662f},
archivePrefix = {arXiv},
       eprint = {1607.06089},
 primaryClass = {astro-ph.GA},
       adsurl = {https://ui.adsabs.harvard.edu/abs/2017ApJ...838...94W}
}

@ARTICLE{naab09,
       author = {{Naab}, Thorsten and {Ostriker}, Jeremiah P.},
        title = "{Are Disk Galaxies the Progenitors of Giant Ellipticals?}",
      journal = {\apj},
         year = 2009,
        month = jan,
       volume = {690},
       number = {2},
        pages = {1452-1462},
          doi = {10.1088/0004-637X/690/2/1452},
archivePrefix = {arXiv},
       eprint = {astro-ph/0702535},
 primaryClass = {astro-ph},
       adsurl = {https://ui.adsabs.harvard.edu/abs/2009ApJ...690.1452N}
}

@ARTICLE{barro13,
       author = {{Barro}, Guillermo and {Faber}, S.~M. and {P{\'e}rez-Gonz{\'a}lez}, Pablo G. and {Koo}, David C. and {Williams}, Christina C. and {Kocevski}, Dale D. and {Trump}, Jonathan R. and {Mozena}, Mark and {McGrath}, Elizabeth and {van der Wel}, Arjen and {Wuyts}, Stijn and {Bell}, Eric F. and {Croton}, Darren J. and {Ceverino}, Daniel and {Dekel}, Avishai and {Ashby}, M.~L.~N. and {Cheung}, Edmond and {Ferguson}, Henry C. and {Fontana}, Adriano and {Fang}, Jerome and {Giavalisco}, Mauro and {Grogin}, Norman A. and {Guo}, Yicheng and {Hathi}, Nimish P. and {Hopkins}, Philip F. and {Huang}, Kuang-Han and {Koekemoer}, Anton M. and {Kartaltepe}, Jeyhan S. and {Lee}, Kyoung-Soo and {Newman}, Jeffrey A. and {Porter}, Lauren A. and {Primack}, Joel R. and {Ryan}, Russell E. and {Rosario}, David and {Somerville}, Rachel S. and {Salvato}, Mara and {Hsu}, Li-Ting},
        title = "{CANDELS: The Progenitors of Compact Quiescent Galaxies at z \raisebox{-0.5ex}\textasciitilde 2}",
      journal = {\apj},
         year = 2013,
        month = mar,
       volume = {765},
       number = {2},
          eid = {104},
        pages = {104},
          doi = {10.1088/0004-637X/765/2/104},
archivePrefix = {arXiv},
       eprint = {1206.5000},
 primaryClass = {astro-ph.CO},
       adsurl = {https://ui.adsabs.harvard.edu/abs/2013ApJ...765..104B}
}

@ARTICLE{davidzon17,
       author = {{Davidzon}, I. and {Ilbert}, O. and {Laigle}, C. and {Coupon}, J. and {McCracken}, H.~J. and {Delvecchio}, I. and {Masters}, D. and {Capak}, P. and {Hsieh}, B.~C. and {Le F{\`e}vre}, O. and {Tresse}, L. and {Bethermin}, M. and {Chang}, Y. -Y. and {Faisst}, A.~L. and {Le Floc'h}, E. and {Steinhardt}, C. and {Toft}, S. and {Aussel}, H. and {Dubois}, C. and {Hasinger}, G. and {Salvato}, M. and {Sanders}, D.~B. and {Scoville}, N. and {Silverman}, J.~D.},
        title = "{The COSMOS2015 galaxy stellar mass function . Thirteen billion years of stellar mass assembly in ten snapshots}",
      journal = {\aap},
         year = 2017,
        month = sep,
       volume = {605},
          eid = {A70},
        pages = {A70},
          doi = {10.1051/0004-6361/201730419},
archivePrefix = {arXiv},
       eprint = {1701.02734},
 primaryClass = {astro-ph.GA},
       adsurl = {https://ui.adsabs.harvard.edu/abs/2017A&A...605A..70D}
}

@ARTICLE{oser10,
       author = {{Oser}, Ludwig and {Ostriker}, Jeremiah P. and {Naab}, Thorsten and {Johansson}, Peter H. and {Burkert}, Andreas},
        title = "{The Two Phases of Galaxy Formation}",
      journal = {\apj},
         year = 2010,
        month = dec,
       volume = {725},
       number = {2},
        pages = {2312-2323},
          doi = {10.1088/0004-637X/725/2/2312},
archivePrefix = {arXiv},
       eprint = {1010.1381},
 primaryClass = {astro-ph.CO},
       adsurl = {https://ui.adsabs.harvard.edu/abs/2010ApJ...725.2312O}
}

@ARTICLE{brinchmann04,
       author = {{Brinchmann}, J. and {Charlot}, S. and {White}, S.~D.~M. and
         {Tremonti}, C. and {Kauffmann}, G. and {Heckman}, T. and
         {Brinkmann}, J.},
        title = "{The physical properties of star-forming galaxies in the low-redshift Universe}",
      journal = {\mnras},
         year = 2004,
        month = jul,
       volume = {351},
       number = {4},
        pages = {1151-1179},
          doi = {10.1111/j.1365-2966.2004.07881.x},
archivePrefix = {arXiv},
       eprint = {astro-ph/0311060},
 primaryClass = {astro-ph},
       adsurl = {https://ui.adsabs.harvard.edu/abs/2004MNRAS.351.1151B}
}

@BOOK{binney08,
       author = {{Binney}, James and {Tremaine}, Scott},
        title = "{Galactic Dynamics: Second Edition}",
         year = 2008,
       adsurl = {https://ui.adsabs.harvard.edu/abs/2008gady.book.....B}
}

@ARTICLE{birkin24,
       author = {{Birkin}, Jack E. and {Puglisi}, A. and {Swinbank}, A.~M. and {Smail}, Ian and {An}, Fang Xia and {Chapman}, S.~C. and {Chen}, Chian-Chou and {Conselice}, C.~J. and {Dudzevi{\v{c}}i{\={u}}t{\.{e}}}, U. and {Farrah}, D. and {Gullberg}, B. and {Matsuda}, Y. and {Schinnerer}, E. and {Scott}, D. and {Wardlow}, J.~L. and {van der Werf}, P.},
        title = "{KAOSS: turbulent, but disc-like kinematics in dust-obscured star-forming galaxies at z   1.3-2.6}",
      journal = {\mnras},
         year = 2024,
        month = jun,
       volume = {531},
       number = {1},
        pages = {61-83},
          doi = {10.1093/mnras/stae1089},
archivePrefix = {arXiv},
       eprint = {2301.05720},
 primaryClass = {astro-ph.GA},
       adsurl = {https://ui.adsabs.harvard.edu/abs/2024MNRAS.531...61B}
}

@ARTICLE{clerici26,
       author = {{Slodkowski Clerici}, K. and {Schnorr-M{\"u}ller}, A. and {Santiago-Menezes}, A.~C. and {Trevisan}, M. and {Ricci}, T.~V. and {Merib-Dias}, R. and {Palacios}, F. and {Becker}, W.~L. and {Ferrari}, F.},
        title = "{Small and Complex I: The Three Component Structure of $z \sim 0$ Massive Compact Quiescent Galaxies}",
      journal = {arXiv e-prints},
         year = 2026,
        month = may,
          eid = {arXiv:2605.09733},
        pages = {arXiv:2605.09733},
archivePrefix = {arXiv},
       eprint = {2605.09733},
 primaryClass = {astro-ph.GA},
       adsurl = {https://ui.adsabs.harvard.edu/abs/2026arXiv260509733S}
}

@ARTICLE{gadotti26,
       author = {{Gadotti}, Dimitri A.},
        title = "{Robust galaxy image decompositions with differential evolution optimization and the problem of classical bulges in and beyond the nearby Universe}",
      journal = {\mnras},
         year = 2026,
        month = feb,
       volume = {545},
       number = {4},
          eid = {staf2072},
        pages = {staf2072},
          doi = {10.1093/mnras/staf2072},
archivePrefix = {arXiv},
       eprint = {2511.13823},
 primaryClass = {astro-ph.GA},
       adsurl = {https://ui.adsabs.harvard.edu/abs/2026MNRAS.545f2072G}
}

@ARTICLE{deugenio24,
       author = {{D'Eugenio}, Francesco and {P{\'e}rez-Gonz{\'a}lez}, Pablo G. and {Maiolino}, Roberto and {Scholtz}, Jan and {Perna}, Michele and {Circosta}, Chiara and {{\"U}bler}, Hannah and {Arribas}, Santiago and {B{\"o}ker}, Torsten and {Bunker}, Andrew J. and {Carniani}, Stefano and {Charlot}, Stephane and {Chevallard}, Jacopo and {Cresci}, Giovanni and {Curtis-Lake}, Emma and {Jones}, Gareth C. and {Kumari}, Nimisha and {Lamperti}, Isabella and {Looser}, Tobias J. and {Parlanti}, Eleonora and {Rix}, Hans-Walter and {Robertson}, Brant and {Rodr{\'\i}guez Del Pino}, Bruno and {Tacchella}, Sandro and {Venturi}, Giacomo and {Willott}, Chris J.},
        title = "{A fast-rotator post-starburst galaxy quenched by supermassive black-hole feedback at z = 3}",
      journal = {Nature Astronomy},
         year = 2024,
        month = nov,
       volume = {8},
        pages = {1443-1456},
          doi = {10.1038/s41550-024-02345-1},
archivePrefix = {arXiv},
       eprint = {2308.06317},
 primaryClass = {astro-ph.GA},
       adsurl = {https://ui.adsabs.harvard.edu/abs/2024NatAs...8.1443D}
}

@ARTICLE{newman18a,
       author = {{Newman}, Andrew B. and {Belli}, Sirio and {Ellis}, Richard S. and {Patel}, Shannon G.},
        title = "{Resolving Quiescent Galaxies at z {\ensuremath{\gtrsim}} 2. I. Search for Gravitationally Lensed Sources and Characterization of Their Structure, Stellar Populations, and Line Emission}",
      journal = {\apj},
         year = 2018,
        month = aug,
       volume = {862},
       number = {2},
          eid = {125},
        pages = {125},
          doi = {10.3847/1538-4357/aacd4d},
archivePrefix = {arXiv},
       eprint = {1806.06814},
 primaryClass = {astro-ph.GA},
       adsurl = {https://ui.adsabs.harvard.edu/abs/2018ApJ...862..125N}
}

@ARTICLE{pathak21,
       author = {{Pathak}, Debosmita and {Belli}, Sirio and {Weinberger}, Rainer},
        title = "{Quenching, Mergers, and Age Profiles for z = 2 Galaxies in IllustrisTNG}",
      journal = {\apjl},
         year = 2021,
        month = aug,
       volume = {916},
       number = {2},
          eid = {L23},
        pages = {L23},
          doi = {10.3847/2041-8213/ac13a7},
archivePrefix = {arXiv},
       eprint = {2105.02234},
 primaryClass = {astro-ph.GA},
       adsurl = {https://ui.adsabs.harvard.edu/abs/2021ApJ...916L..23P}
}

@ARTICLE{kurinchi24,
       author = {{Kurinchi-Vendhan}, Shalini and {Farcy}, Marion and {Hirschmann}, Michaela and {Valentino}, Francesco},
        title = "{On the origin of star formation quenching in massive galaxies at z {\ensuremath{\gtrsim}} 3 in the cosmological simulations IllustrisTNG}",
      journal = {\mnras},
         year = 2024,
        month = nov,
       volume = {534},
       number = {4},
        pages = {3974-3988},
          doi = {10.1093/mnras/stae2297},
archivePrefix = {arXiv},
       eprint = {2310.03083},
 primaryClass = {astro-ph.GA},
       adsurl = {https://ui.adsabs.harvard.edu/abs/2024MNRAS.534.3974K}
}

@ARTICLE{forrest25,
       author = {{Forrest}, Ben and {Muzzin}, Adam and {Marchesini}, Danilo and {Pan}, Richard and {Ozden}, Nehir and {Antwi-Danso}, Jacqueline and {Chang}, Wenjun and {Cooper}, M.~C. and {Edward}, Adit H. and {Gomez}, Percy and {Kimmig}, Lucas and {Lemaux}, Brian C. and {McConachie}, Ian and {Noble}, Allison and {Remus}, Rhea-Silvia and {Urbano Stawinski}, Stephanie M. and {Wilson}, Gillian and {Wisz}, M.~E.},
        title = "{A massive, evolved slow-rotating galaxy in the early Universe}",
      journal = {arXiv e-prints},
         year = 2025,
        month = aug,
          eid = {arXiv:2508.10987},
        pages = {arXiv:2508.10987},
          doi = {10.48550/arXiv.2508.10987},
archivePrefix = {arXiv},
       eprint = {2508.10987},
 primaryClass = {astro-ph.GA},
       adsurl = {https://ui.adsabs.harvard.edu/abs/2025arXiv250810987F}
}

@ARTICLE{turner25,
       author = {{Turner}, Crispin and {Tacchella}, Sandro and {D'Eugenio}, Francesco and {Carniani}, Stefano and {Curti}, Mirko and {Glazebrook}, Karl and {Johnson}, Benjamin D. and {Lim}, Seunghwan and {Looser}, Tobias and {Maiolino}, Roberto and {Nanayakkara}, Themiya and {Wan}, Jenny},
        title = "{Age-dating early quiescent galaxies: high star formation efficiency, but consistent with direct, higher-redshift observations}",
      journal = {\mnras},
         year = 2025,
        month = feb,
       volume = {537},
       number = {2},
        pages = {1826-1848},
          doi = {10.1093/mnras/staf128},
archivePrefix = {arXiv},
       eprint = {2410.05377},
 primaryClass = {astro-ph.GA},
       adsurl = {https://ui.adsabs.harvard.edu/abs/2025MNRAS.537.1826T}
}

@ARTICLE{pascalau25,
       author = {{Pascalau}, Robert G. and {D'Eugenio}, Francesco and {Tacchella}, Sandro and {Maiolino}, Roberto and {Cappellari}, Michele and {Lagos}, Claudia del P. and {Bunker}, Andrew J. and {Jones}, Gareth C. and {Scholtz}, Jan and {{\"U}bler}, Hannah and {Cresci}, Giovanni and {Arribas}, Santiago and {Perna}, Michele and {van der Wel}, Arjen and {Danhaive}, A. Lola and {McClymont}, William and {Vani}, Akash and {Maseda}, Michael V. and {Carnall}, Adam C. and {Charlot}, St{\'e}phane and {Carniani}, Stefano and {Duan}, Qiao and {Goh}, Tze P. and {de Graaff}, Anna and {Ji}, Zhiyuan and {P{\'e}rez-Gonz{\'a}lez}, Pablo},
        title = "{When relics were made: vigorous stellar rotation and low dark matter content in the massive ultra-compact galaxy GS-9209 at z=4.66}",
      journal = {arXiv e-prints},
         year = 2025,
        month = may,
          eid = {arXiv:2505.06349},
        pages = {arXiv:2505.06349},
          doi = {10.48550/arXiv.2505.06349},
archivePrefix = {arXiv},
       eprint = {2505.06349},
 primaryClass = {astro-ph.GA},
       adsurl = {https://ui.adsabs.harvard.edu/abs/2025arXiv250506349P}
}

@ARTICLE{slob25,
       author = {{Slob}, Martje and {Kriek}, Mariska and {de Graaff}, Anna and {Cheng}, Chloe M. and {Beverage}, Aliza G. and {Bezanson}, Rachel and {F{\"o}rster Schreiber}, Natascha M. and {Lorenz}, Brian and {Mancera Pi{\~n}a}, Pavel E. and {Marchesini}, Danilo and {Muzzin}, Adam and {Newman}, Andrew B. and {Price}, Sedona H. and {Suess}, Katherine A. and {van de Sande}, Jesse and {van Dokkum}, Pieter and {Weisz}, Daniel R.},
        title = "{Fast rotators at cosmic noon: Stellar kinematics for 15 quiescent galaxies from JWST-SUSPENSE}",
      journal = {\aap},
         year = 2025,
        month = oct,
       volume = {702},
          eid = {A110},
        pages = {A110},
          doi = {10.1051/0004-6361/202555812},
archivePrefix = {arXiv},
       eprint = {2506.04310},
 primaryClass = {astro-ph.GA},
       adsurl = {https://ui.adsabs.harvard.edu/abs/2025A&A...702A.110S}
}

@ARTICLE{laurikainen10,
       author = {{Laurikainen}, E. and {Salo}, H. and {Buta}, R. and {Knapen}, J.~H. and {Comer{\'o}n}, S.},
        title = "{Photometric scaling relations of lenticular and spiral galaxies}",
      journal = {\mnras},
         year = 2010,
        month = jun,
       volume = {405},
       number = {2},
        pages = {1089-1118},
          doi = {10.1111/j.1365-2966.2010.16521.x},
archivePrefix = {arXiv},
       eprint = {1002.4370},
 primaryClass = {astro-ph.CO},
       adsurl = {https://ui.adsabs.harvard.edu/abs/2010MNRAS.405.1089L}
}

@ARTICLE{damjanov13,
       author = {{Damjanov}, Ivana and {Chilingarian}, Igor and {Hwang}, Ho Seong and {Geller}, Margaret J.},
        title = "{Discovery of Nine Intermediate-redshift Compact Quiescent Galaxies in the Sloan Digital Sky Survey}",
      journal = {\apjl},
         year = 2013,
        month = oct,
       volume = {775},
       number = {2},
          eid = {L48},
        pages = {L48},
          doi = {10.1088/2041-8205/775/2/L48},
archivePrefix = {arXiv},
       eprint = {1309.2948},
 primaryClass = {astro-ph.CO},
       adsurl = {https://ui.adsabs.harvard.edu/abs/2013ApJ...775L..48D}
}

@ARTICLE{ferre-mateu12,
       author = {{Ferr{\'e}-Mateu}, A. and {Vazdekis}, A. and {Trujillo}, I. and {S{\'a}nchez-Bl{\'a}zquez}, P. and {Ricciardelli}, E. and {de la Rosa}, I.~G.},
        title = "{Young ages and other intriguing properties of massive compact galaxies in the local Universe}",
      journal = {\mnras},
         year = 2012,
        month = jun,
       volume = {423},
       number = {1},
        pages = {632-646},
          doi = {10.1111/j.1365-2966.2012.20897.x},
archivePrefix = {arXiv},
       eprint = {1203.2623},
 primaryClass = {astro-ph.CO},
       adsurl = {https://ui.adsabs.harvard.edu/abs/2012MNRAS.423..632F}
}

@ARTICLE{momcheva16,
       author = {{Momcheva}, Ivelina G. and {Brammer}, Gabriel B. and {van Dokkum}, Pieter G. and {Skelton}, Rosalind E. and {Whitaker}, Katherine E. and {Nelson}, Erica J. and {Fumagalli}, Mattia and {Maseda}, Michael V. and {Leja}, Joel and {Franx}, Marijn and {Rix}, Hans-Walter and {Bezanson}, Rachel and {Da Cunha}, Elisabete and {Dickey}, Claire and {F{\"o}rster Schreiber}, Natascha M. and {Illingworth}, Garth and {Kriek}, Mariska and {Labb{\'e}}, Ivo and {Ulf Lange}, Johannes and {Lundgren}, Britt F. and {Magee}, Daniel and {Marchesini}, Danilo and {Oesch}, Pascal and {Pacifici}, Camilla and {Patel}, Shannon G. and {Price}, Sedona and {Tal}, Tomer and {Wake}, David A. and {van der Wel}, Arjen and {Wuyts}, Stijn},
        title = "{The 3D-HST Survey: Hubble Space Telescope WFC3/G141 Grism Spectra, Redshifts, and Emission Line Measurements for \raisebox{-0.5ex}\textasciitilde 100,000 Galaxies}",
      journal = {\apjs},
         year = 2016,
        month = aug,
       volume = {225},
       number = {2},
          eid = {27},
        pages = {27},
          doi = {10.3847/0067-0049/225/2/27},
archivePrefix = {arXiv},
       eprint = {1510.02106},
 primaryClass = {astro-ph.GA},
       adsurl = {https://ui.adsabs.harvard.edu/abs/2016ApJS..225...27M}
}

@ARTICLE{bundy10,
       author = {{Bundy}, Kevin and {Scarlata}, Claudia and {Carollo}, C.~M. and
         {Ellis}, Richard S. and {Drory}, Niv and {Hopkins}, Philip and
         {Salvato}, Mara and {Leauthaud}, Alexie and {Koekemoer}, Anton M. and
         {Murray}, Norman and {Ilbert}, Olivier and {Oesch}, Pascal and
         {Ma}, Chung-Pei and {Capak}, Peter and {Pozzetti}, Lucia and
         {Scoville}, Nick},
        title = "{The Rise and Fall of Passive Disk Galaxies: Morphological Evolution Along the Red Sequence Revealed by COSMOS}",
      journal = {\apj},
         year = 2010,
        month = aug,
       volume = {719},
       number = {2},
        pages = {1969-1983},
          doi = {10.1088/0004-637X/719/2/1969},
archivePrefix = {arXiv},
       eprint = {0912.1077},
 primaryClass = {astro-ph.CO},
       adsurl = {https://ui.adsabs.harvard.edu/abs/2010ApJ...719.1969B}
}

@ARTICLE{davari17,
       author = {{Davari}, Roozbeh H. and {Ho}, Luis C. and {Mobasher}, Bahram and
         {Canalizo}, Gabriela},
        title = "{Detection of Prominent Stellar Disks in the Progenitors of Present-day Massive Elliptical Galaxies}",
      journal = {\apj},
         year = 2017,
        month = feb,
       volume = {836},
       number = {1},
          eid = {75},
        pages = {75},
          doi = {10.3847/1538-4357/836/1/75},
archivePrefix = {arXiv},
       eprint = {1606.07571},
 primaryClass = {astro-ph.GA},
       adsurl = {https://ui.adsabs.harvard.edu/abs/2017ApJ...836...75D}
}

@ARTICLE{anderson00,
       author = {{Anderson}, Jay and {King}, Ivan R.},
        title = "{Toward High-Precision Astrometry with WFPC2. I. Deriving an Accurate Point-Spread Function}",
      journal = {\pasp},
         year = 2000,
        month = oct,
       volume = {112},
       number = {776},
        pages = {1360-1382},
          doi = {10.1086/316632},
archivePrefix = {arXiv},
       eprint = {astro-ph/0006325},
 primaryClass = {astro-ph},
       adsurl = {https://ui.adsabs.harvard.edu/abs/2000PASP..112.1360A}
}

@MISC{anderson16,
       author = {{Anderson}, Jay},
        title = "{Empirical Models for the WFC3/IR PSF}",
 howpublished = {Instrument Science Report WFC3 2016-12, 42 pages},
         year = 2016,
        month = mar,
        pages = {12},
       adsurl = {https://ui.adsabs.harvard.edu/abs/2016wfc..rept...12A}
}

@ARTICLE{williams09,
       author = {{Williams}, Rik J. and {Quadri}, Ryan F. and {Franx}, Marijn and {van Dokkum}, Pieter and {Labb{\'e}}, Ivo},
        title = "{Detection of Quiescent Galaxies in a Bicolor Sequence from Z = 0-2}",
      journal = {\apj},
         year = 2009,
        month = feb,
       volume = {691},
       number = {2},
        pages = {1879-1895},
          doi = {10.1088/0004-637X/691/2/1879},
archivePrefix = {arXiv},
       eprint = {0806.0625},
 primaryClass = {astro-ph},
       adsurl = {https://ui.adsabs.harvard.edu/abs/2009ApJ...691.1879W}
}

@ARTICLE{damjanov19,
       author = {{Damjanov}, Ivana and {Zahid}, H. Jabran and {Geller}, Margaret J. and
         {Utsumi}, Yousuke and {Sohn}, Jubee and {Souchereau}, Harrison},
        title = "{Quiescent Galaxy Size and Spectroscopic Evolution: Combining HSC Imaging and Hectospec Spectroscopy}",
      journal = {\apj},
         year = 2019,
        month = feb,
       volume = {872},
       number = {1},
          eid = {91},
        pages = {91},
          doi = {10.3847/1538-4357/aaf97d},
archivePrefix = {arXiv},
       eprint = {1809.10705},
 primaryClass = {astro-ph.GA},
       adsurl = {https://ui.adsabs.harvard.edu/abs/2019ApJ...872...91D}
}

@ARTICLE{fagioli16,
       author = {{Fagioli}, Martina and {Carollo}, C. Marcella and {Renzini}, Alvio and {Lilly}, Simon J. and {Onodera}, Masato and {Tacchella}, Sandro},
        title = "{Minor Mergers or Progenitor Bias? The Stellar Ages of Small and Large Quenched Galaxies}",
      journal = {\apj},
         year = 2016,
        month = nov,
       volume = {831},
       number = {2},
          eid = {173},
        pages = {173},
          doi = {10.3847/0004-637X/831/2/173},
archivePrefix = {arXiv},
       eprint = {1607.03493},
 primaryClass = {astro-ph.GA},
       adsurl = {https://ui.adsabs.harvard.edu/abs/2016ApJ...831..173F}
}

@ARTICLE{wuyts07,
       author = {{Wuyts}, Stijn and {Labb{\'e}}, Ivo and {Franx}, Marijn and {Rudnick}, Gregory and {van Dokkum}, Pieter G. and {Fazio}, Giovanni G. and {F{\"o}rster Schreiber}, Natascha M. and {Huang}, Jiasheng and {Moorwood}, Alan F.~M. and {Rix}, Hans-Walter and {R{\"o}ttgering}, Huub and {van der Werf}, Paul},
        title = "{What Do We Learn from IRAC Observations of Galaxies at 2 < z < 3.5?}",
      journal = {\apj},
         year = 2007,
        month = jan,
       volume = {655},
       number = {1},
        pages = {51-65},
          doi = {10.1086/509708},
archivePrefix = {arXiv},
       eprint = {astro-ph/0609548},
 primaryClass = {astro-ph},
       adsurl = {https://ui.adsabs.harvard.edu/abs/2007ApJ...655...51W}
}

@ARTICLE{lange16,
       author = {{Lange}, Rebecca and {Moffett}, Amanda J. and {Driver}, Simon P. and {Robotham}, Aaron S.~G. and {Lagos}, Claudia del P. and {Kelvin}, Lee S. and {Conselice}, Christopher and {Margalef-Bentabol}, Berta and {Alpaslan}, Mehmet and {Baldry}, Ivan and {Bland-Hawthorn}, Joss and {Bremer}, Malcolm and {Brough}, Sarah and {Cluver}, Michelle and {Colless}, Matthew and {Davies}, Luke J.~M. and {H{\"a}u{\ss}ler}, Boris and {Holwerda}, Benne W. and {Hopkins}, Andrew M. and {Kafle}, Prajwal R. and {Kennedy}, Rebecca and {Liske}, Jochen and {Phillipps}, Steven and {Popescu}, Cristina C. and {Taylor}, Edward N. and {Tuffs}, Richard and {van Kampen}, Eelco and {Wright}, Angus H.},
        title = "{Galaxy And Mass Assembly (GAMA): M\_star - R\_e relations of z = 0 bulges, discs and spheroids}",
      journal = {\mnras},
         year = 2016,
        month = oct,
       volume = {462},
       number = {2},
        pages = {1470-1500},
          doi = {10.1093/mnras/stw1495},
archivePrefix = {arXiv},
       eprint = {1607.01096},
 primaryClass = {astro-ph.GA},
       adsurl = {https://ui.adsabs.harvard.edu/abs/2016MNRAS.462.1470L}
}

@ARTICLE{spitzer42,
       author = {{Spitzer}, Jr., Lyman},
        title = "{The Dynamics of the Interstellar Medium. III. Galactic Distribution.}",
      journal = {\apj},
         year = 1942,
        month = may,
       volume = {95},
        pages = {329},
          doi = {10.1086/144407},
       adsurl = {https://ui.adsabs.harvard.edu/abs/1942ApJ....95..329S}
}

@ARTICLE{zahid16,
       author = {{Zahid}, H. Jabran and {Baeza Hochmuth}, Nicholas and {Geller}, Margaret J. and {Damjanov}, Ivana and {Chilingarian}, Igor V. and {Sohn}, Jubee and {Salmi}, Fadia and {Hwang}, Ho Seong},
        title = "{Compact E+A Galaxies as a Progenitor of Massive Compact Quiescent Galaxies at 0.2 < z < 0.8}",
      journal = {\apj},
         year = 2016,
        month = nov,
       volume = {831},
       number = {2},
          eid = {146},
        pages = {146},
          doi = {10.3847/0004-637X/831/2/146},
archivePrefix = {arXiv},
       eprint = {1605.09734},
 primaryClass = {astro-ph.GA},
       adsurl = {https://ui.adsabs.harvard.edu/abs/2016ApJ...831..146Z}
}

@ARTICLE{pinna24,
       author = {{Pinna}, Francesca and {Walo-Mart{\'\i}n}, Daniel and {Grand}, Robert J.~J. and {Martig}, Marie and {Fragkoudi}, Francesca and {G{\'o}mez}, Facundo A. and {Marinacci}, Federico and {Pakmor}, R{\"u}diger},
        title = "{Stellar populations and the origin of thick disks in AURIGA simulations}",
      journal = {\aap},
         year = 2024,
        month = mar,
       volume = {683},
          eid = {A236},
        pages = {A236},
          doi = {10.1051/0004-6361/202347388},
archivePrefix = {arXiv},
       eprint = {2311.13700},
 primaryClass = {astro-ph.GA},
       adsurl = {https://ui.adsabs.harvard.edu/abs/2024A&A...683A.236P}
}

@ARTICLE{yildirim15,
       author = {{Y{\i}ld{\i}r{\i}m}, Ak{\i}n and {van den Bosch}, Remco C.~E. and {van de Ven}, Glenn and {Husemann}, Bernd and {Lyubenova}, Mariya and {Walsh}, Jonelle L. and {Gebhardt}, Karl and {G{\"u}ltekin}, Kayhan},
        title = "{MRK 1216 and NGC 1277 - an orbit-based dynamical analysis of compact, high-velocity dispersion galaxies}",
      journal = {\mnras},
         year = 2015,
        month = sep,
       volume = {452},
       number = {2},
        pages = {1792-1816},
          doi = {10.1093/mnras/stv1381},
archivePrefix = {arXiv},
       eprint = {1506.06762},
 primaryClass = {astro-ph.GA},
       adsurl = {https://ui.adsabs.harvard.edu/abs/2015MNRAS.452.1792Y}
}

@ARTICLE{qu11,
       author = {{Qu}, Y. and {Di Matteo}, P. and {Lehnert}, M.~D. and {van Driel}, W.},
        title = "{Characteristics of thick disks formed through minor mergers: stellar excesses and scale lengths}",
      journal = {\aap},
         year = 2011,
        month = jun,
       volume = {530},
          eid = {A10},
        pages = {A10},
          doi = {10.1051/0004-6361/201015224},
archivePrefix = {arXiv},
       eprint = {1102.1879},
 primaryClass = {astro-ph.GA},
       adsurl = {https://ui.adsabs.harvard.edu/abs/2011A&A...530A..10Q}
}

@ARTICLE{villalobos08,
       author = {{Villalobos}, {\'A}lvaro and {Helmi}, Amina},
        title = "{Simulations of minor mergers - I. General properties of thick discs}",
      journal = {\mnras},
         year = 2008,
        month = dec,
       volume = {391},
       number = {4},
        pages = {1806-1827},
          doi = {10.1111/j.1365-2966.2008.13979.x},
archivePrefix = {arXiv},
       eprint = {0803.2323},
 primaryClass = {astro-ph},
       adsurl = {https://ui.adsabs.harvard.edu/abs/2008MNRAS.391.1806V}
}

@ARTICLE{bournaud07,
       author = {{Bournaud}, F. and {Jog}, C.~J. and {Combes}, F.},
        title = "{Multiple minor mergers: formation of elliptical galaxies and constraints for the growth of spiral disks}",
      journal = {\aap},
         year = 2007,
        month = dec,
       volume = {476},
       number = {3},
        pages = {1179-1190},
          doi = {10.1051/0004-6361:20078010},
archivePrefix = {arXiv},
       eprint = {0709.3439},
 primaryClass = {astro-ph},
       adsurl = {https://ui.adsabs.harvard.edu/abs/2007A&A...476.1179B}
}

@ARTICLE{yi24,
       author = {{Yi}, Sukyoung K. and {Jang}, J.~K. and {Devriendt}, Julien and {Dubois}, Yohan and {Han}, San and {Kimm}, Taysun and {Kraljic}, Katarina and {Park}, Minjung and {Peirani}, Sebastien and {Pichon}, Christophe and {Rhee}, Jinsu},
        title = "{On the Significance of the Thick Disks of Disk Galaxies}",
      journal = {\apjs},
         year = 2024,
        month = mar,
       volume = {271},
       number = {1},
          eid = {1},
        pages = {1},
          doi = {10.3847/1538-4365/ad0e71},
archivePrefix = {arXiv},
       eprint = {2308.03566},
 primaryClass = {astro-ph.GA},
       adsurl = {https://ui.adsabs.harvard.edu/abs/2024ApJS..271....1Y}
}

@ARTICLE{yu21,
       author = {{Yu}, Sijie and {Bullock}, James S. and {Klein}, Courtney and {Stern}, Jonathan and {Wetzel}, Andrew and {Ma}, Xiangcheng and {Moreno}, Jorge and {Hafen}, Zachary and {Gurvich}, Alexander B. and {Hopkins}, Philip F. and {Kere{\v{s}}}, Du{\v{s}}an and {Faucher-Gigu{\`e}re}, Claude-Andr{\'e} and {Feldmann}, Robert and {Quataert}, Eliot},
        title = "{The bursty origin of the Milky Way thick disc}",
      journal = {\mnras},
         year = 2021,
        month = jul,
       volume = {505},
       number = {1},
        pages = {889-902},
          doi = {10.1093/mnras/stab1339},
archivePrefix = {arXiv},
       eprint = {2103.03888},
 primaryClass = {astro-ph.GA},
       adsurl = {https://ui.adsabs.harvard.edu/abs/2021MNRAS.505..889Y}
}

@ARTICLE{park21,
       author = {{Park}, Minjung J. and {Yi}, Sukyoung K. and {Peirani}, Sebastien and {Pichon}, Christophe and {Dubois}, Yohan and {Choi}, Hoseung and {Devriendt}, Julien and {Kaviraj}, Sugata and {Kimm}, Taysun and {Kraljic}, Katarina and {Volonteri}, Marta},
        title = "{Exploring the Origin of Thick Disks Using the NewHorizon and Galactica Simulations}",
      journal = {\apjs},
         year = 2021,
        month = may,
       volume = {254},
       number = {1},
          eid = {2},
        pages = {2},
          doi = {10.3847/1538-4365/abe937},
archivePrefix = {arXiv},
       eprint = {2009.12373},
 primaryClass = {astro-ph.GA},
       adsurl = {https://ui.adsabs.harvard.edu/abs/2021ApJS..254....2P}
}

@ARTICLE{zhu25,
       author = {{Zhu}, Ling and {Chies-Santos}, Ana L. and {Moura}, Micheli T. and {Shi}, Hanjing},
        title = "{Distinguishing the formation paths of massive compact early-type galaxies through their internal dynamical structures}",
      journal = {\aap},
         year = 2025,
        month = jun,
       volume = {698},
          eid = {A195},
        pages = {A195},
          doi = {10.1051/0004-6361/202453083},
archivePrefix = {arXiv},
       eprint = {2504.09677},
 primaryClass = {astro-ph.GA},
       adsurl = {https://ui.adsabs.harvard.edu/abs/2025A&A...698A.195Z}
}

@ARTICLE{du21,
       author = {{Du}, Min and {Ho}, Luis C. and {Debattista}, Victor P. and {Pillepich}, Annalisa and {Nelson}, Dylan and {Hernquist}, Lars and {Weinberger}, Rainer},
        title = "{The Evolutionary Pathways of Disk-, Bulge-, and Halo-dominated Galaxies}",
      journal = {\apj},
         year = 2021,
        month = oct,
       volume = {919},
       number = {2},
          eid = {135},
        pages = {135},
          doi = {10.3847/1538-4357/ac0e98},
archivePrefix = {arXiv},
       eprint = {2101.12373},
 primaryClass = {astro-ph.GA},
       adsurl = {https://ui.adsabs.harvard.edu/abs/2021ApJ...919..135D}
}

@ARTICLE{he25,
       author = {{He}, Wu-Tao and {Du}, Min and {Li}, Zhao-Yu and {Li}, Yuan},
        title = "{Beyond Morphology: Challenges in Decomposing Massive Stellar Halos in Sombrero-like, Halo-Embedded Disk Galaxies}",
      journal = {\aap},
         year = 2025,
        month = jul,
       volume = {699},
          eid = {A99},
        pages = {A99},
          doi = {10.1051/0004-6361/202554551},
archivePrefix = {arXiv},
       eprint = {2505.04133},
 primaryClass = {astro-ph.GA},
       adsurl = {https://ui.adsabs.harvard.edu/abs/2025A&A...699A..99H}
}

@ARTICLE{bosch18,
       author = {{Bosch}, James and {Armstrong}, Robert and {Bickerton}, Steven and {Furusawa}, Hisanori and {Ikeda}, Hiroyuki and {Koike}, Michitaro and {Lupton}, Robert and {Mineo}, Sogo and {Price}, Paul and {Takata}, Tadafumi and {Tanaka}, Masayuki and {Yasuda}, Naoki and {AlSayyad}, Yusra and {Becker}, Andrew C. and {Coulton}, William and {Coupon}, Jean and {Garmilla}, Jose and {Huang}, Song and {Krughoff}, K. Simon and {Lang}, Dustin and {Leauthaud}, Alexie and {Lim}, Kian-Tat and {Lust}, Nate B. and {MacArthur}, Lauren A. and {Mandelbaum}, Rachel and {Miyatake}, Hironao and {Miyazaki}, Satoshi and {Murata}, Ryoma and {More}, Surhud and {Okura}, Yuki and {Owen}, Russell and {Swinbank}, John D. and {Strauss}, Michael A. and {Yamada}, Yoshihiko and {Yamanoi}, Hitomi},
        title = "{The Hyper Suprime-Cam software pipeline}",
      journal = {\pasj},
         year = 2018,
        month = jan,
       volume = {70},
          eid = {S5},
        pages = {S5},
          doi = {10.1093/pasj/psx080},
archivePrefix = {arXiv},
       eprint = {1705.06766},
 primaryClass = {astro-ph.IM},
       adsurl = {https://ui.adsabs.harvard.edu/abs/2018PASJ...70S...5B}
}

@ARTICLE{bertin96,
       author = {{Bertin}, E. and {Arnouts}, S.},
        title = "{SExtractor: Software for source extraction.}",
      journal = {\aaps},
         year = 1996,
        month = jun,
       volume = {117},
        pages = {393-404},
          doi = {10.1051/aas:1996164},
       adsurl = {https://ui.adsabs.harvard.edu/abs/1996A&AS..117..393B}
}

@ARTICLE{vanderkruit81,
       author = {{van der Kruit}, P.~C. and {Searle}, L.},
        title = "{Surface photometry of edge-on spiral galaxies. I - A model for the three-dimensional distribution of light in galactic disks.}",
      journal = {\aap},
         year = 1981,
        month = feb,
       volume = {95},
        pages = {105-115},
       adsurl = {https://ui.adsabs.harvard.edu/abs/1981A&A....95..105V}
}

@ARTICLE{erwin15,
       author = {{Erwin}, Peter},
        title = "{IMFIT: A Fast, Flexible New Program for Astronomical Image Fitting}",
      journal = {\apj},
         year = 2015,
        month = feb,
       volume = {799},
       number = {2},
          eid = {226},
        pages = {226},
          doi = {10.1088/0004-637X/799/2/226},
archivePrefix = {arXiv},
       eprint = {1408.1097},
 primaryClass = {astro-ph.IM},
       adsurl = {https://ui.adsabs.harvard.edu/abs/2015ApJ...799..226E}
}

@ARTICLE{spiniello21,
       author = {{Spiniello}, C. and {Tortora}, C. and {D'Ago}, G. and {Coccato}, L. and {La Barbera}, F. and {Ferr{\'e}-Mateu}, A. and {Napolitano}, N.~R. and {Spavone}, M. and {Scognamiglio}, D. and {Arnaboldi}, M. and {Gallazzi}, A. and {Hunt}, L. and {Moehler}, S. and {Radovich}, M. and {Zibetti}, S.},
        title = "{INSPIRE: INvestigating Stellar Population In RElics. I. Survey presentation and pilot study}",
      journal = {\aap},
         year = 2021,
        month = feb,
       volume = {646},
          eid = {A28},
        pages = {A28},
          doi = {10.1051/0004-6361/202038936},
archivePrefix = {arXiv},
       eprint = {2011.05347},
 primaryClass = {astro-ph.GA},
       adsurl = {https://ui.adsabs.harvard.edu/abs/2021A&A...646A..28S}
}

@ARTICLE{barro17,
       author = {{Barro}, Guillermo and {Faber}, S.~M. and {Koo}, David C. and {Dekel}, Avishai and {Fang}, Jerome J. and {Trump}, Jonathan R. and {P{\'e}rez-Gonz{\'a}lez}, Pablo G. and {Pacifici}, Camilla and {Primack}, Joel R. and {Somerville}, Rachel S. and {Yan}, Haojing and {Guo}, Yicheng and {Liu}, Fengshan and {Ceverino}, Daniel and {Kocevski}, Dale D. and {McGrath}, Elizabeth},
        title = "{Structural and Star-forming Relations since z {\ensuremath{\sim}} 3: Connecting Compact Star-forming and Quiescent Galaxies}",
      journal = {\apj},
         year = 2017,
        month = may,
       volume = {840},
       number = {1},
          eid = {47},
        pages = {47},
          doi = {10.3847/1538-4357/aa6b05},
archivePrefix = {arXiv},
       eprint = {1509.00469},
 primaryClass = {astro-ph.GA},
       adsurl = {https://ui.adsabs.harvard.edu/abs/2017ApJ...840...47B}
}

@ARTICLE{cook25,
       author = {{Cook}, Robin H.~W. and {Davies}, Luke J.~M. and {Bellstedt}, Sabine and {Robotham}, Aaron S.~G. and {Driver}, Simon P. and {Siudek}, Malgorzata and {Wolf}, Christian},
        title = "{Deep Extragalactic VIsible Legacy Survey (DEVILS): the size{\textendash}mass relation of galaxies and their components in HST-COSMOS over the last 8 billion years}",
      journal = {\mnras},
         year = 2025,
        month = jun,
       volume = {539},
       number = {4},
        pages = {2829-2854},
          doi = {10.1093/mnras/staf575},
       adsurl = {https://ui.adsabs.harvard.edu/abs/2025MNRAS.539.2829C}
}

@ARTICLE{tsukui25,
       author = {{Tsukui}, Takafumi and {Wisnioski}, Emily and {Bland-Hawthorn}, Joss and {Freeman}, Ken},
        title = "{The emergence of galactic thin and thick discs across cosmic history}",
      journal = {\mnras},
         year = 2025,
        month = jul,
       volume = {540},
       number = {4},
        pages = {3493-3522},
          doi = {10.1093/mnras/staf604},
archivePrefix = {arXiv},
       eprint = {2409.15909},
 primaryClass = {astro-ph.GA},
       adsurl = {https://ui.adsabs.harvard.edu/abs/2025MNRAS.540.3493T}
}

@ARTICLE{nedkova24,
       author = {{Nedkova}, Kalina V. and {H{\"a}u{\ss}ler}, Boris and {Marchesini}, Danilo and {Brammer}, Gabriel B. and {Feinstein}, Adina D. and {Johnston}, Evelyn J. and {Kartaltepe}, Jeyhan S. and {Koekemoer}, Anton M. and {Martis}, Nicholas S. and {Muzzin}, Adam and {Rafelski}, Marc and {Shipley}, Heath V. and {Skelton}, Rosalind E. and {Stefanon}, Mauro and {van der Wel}, Arjen and {Whitaker}, Katherine E.},
        title = "{Bulge+disc decomposition of HFF and CANDELS galaxies: UVJ diagrams and stellar mass-size relations of galaxy components at 0.2 {\ensuremath{\leq}} z {\ensuremath{\leq}} 1.5}",
      journal = {\mnras},
         year = 2024,
        month = aug,
       volume = {532},
       number = {4},
        pages = {3747-3777},
          doi = {10.1093/mnras/stae1702},
archivePrefix = {arXiv},
       eprint = {2406.14613},
 primaryClass = {astro-ph.GA},
       adsurl = {https://ui.adsabs.harvard.edu/abs/2024MNRAS.532.3747N}
}

@ARTICLE{lim17,
       author = {{Lim}, S.~H. and {Mo}, H.~J. and {Lu}, Yi and {Wang}, Huiyuan and {Yang}, Xiaohu},
        title = "{Galaxy groups in the low-redshift Universe}",
      journal = {\mnras},
         year = 2017,
        month = sep,
       volume = {470},
       number = {3},
        pages = {2982-3005},
          doi = {10.1093/mnras/stx1462},
archivePrefix = {arXiv},
       eprint = {1706.02307},
 primaryClass = {astro-ph.GA},
       adsurl = {https://ui.adsabs.harvard.edu/abs/2017MNRAS.470.2982L}
}

@Article{rosenbaum83,
	author = "P. R. {Rosenbaum} and D. B. {Rubin}",
	title = "{The Central Role of the Propensity Score in Observational Studies for Causal Effects}",
	journal = "Biometrika",
	year = 1983,
	volume = 70,
	pages = "41--55",
	doi = "10.1093/biomet/70.1.41"
}

@Article{ho11,
  title = {{MatchIt}: Nonparametric Preprocessing for Parametric Causal Inference},
  author = {Daniel E. Ho and Kosuke Imai and Gary King and Elizabeth A. Stuart},
  journal = {Journal of Statistical Software},
  year = {2011},
  volume = {42},
  number = {8},
  pages = {1--28},
  url = {http://www.jstatsoft.org/v42/i08/},
}

@ARTICLE{cappellari06,
       author = {{Cappellari}, Michele and {Bacon}, R. and {Bureau}, M. and {Damen}, M.~C. and {Davies}, Roger L. and {de Zeeuw}, P.~T. and {Emsellem}, Eric and {Falc{\'o}n-Barroso}, Jes{\'u}s and {Krajnovi{\'c}}, Davor and {Kuntschner}, Harald and {McDermid}, Richard M. and {Peletier}, Reynier F. and {Sarzi}, Marc and {van den Bosch}, Remco C.~E. and {van de Ven}, Glenn},
        title = "{The SAURON project - IV. The mass-to-light ratio, the virial mass estimator and the Fundamental Plane of elliptical and lenticular galaxies}",
      journal = {\mnras},
         year = 2006,
        month = mar,
       volume = {366},
       number = {4},
        pages = {1126-1150},
          doi = {10.1111/j.1365-2966.2005.09981.x},
archivePrefix = {arXiv},
       eprint = {astro-ph/0505042},
 primaryClass = {astro-ph},
       adsurl = {https://ui.adsabs.harvard.edu/abs/2006MNRAS.366.1126C}
}

@ARTICLE{simard11,
       author = {{Simard}, Luc and {Mendel}, J. Trevor and {Patton}, David R. and {Ellison}, Sara L. and {McConnachie}, Alan W.},
        title = "{A Catalog of Bulge+disk Decompositions and Updated Photometry for 1.12 Million Galaxies in the Sloan Digital Sky Survey}",
      journal = {\apjs},
         year = 2011,
        month = sep,
       volume = {196},
       number = {1},
          eid = {11},
        pages = {11},
          doi = {10.1088/0067-0049/196/1/11},
archivePrefix = {arXiv},
       eprint = {1107.1518},
 primaryClass = {astro-ph.CO},
       adsurl = {https://ui.adsabs.harvard.edu/abs/2011ApJS..196...11S}
}

@ARTICLE{salim18,
       author = {{Salim}, Samir and {Boquien}, M{\'e}d{\'e}ric and {Lee}, Janice C.},
        title = "{Dust Attenuation Curves in the Local Universe: Demographics and New Laws for Star-forming Galaxies and High-redshift Analogs}",
      journal = {\apj},
         year = 2018,
        month = may,
       volume = {859},
       number = {1},
          eid = {11},
        pages = {11},
          doi = {10.3847/1538-4357/aabf3c},
archivePrefix = {arXiv},
       eprint = {1804.05850},
 primaryClass = {astro-ph.GA},
       adsurl = {https://ui.adsabs.harvard.edu/abs/2018ApJ...859...11S}
}

@ARTICLE{carollo07,
       author = {{Carollo}, C.~M. and {Scarlata}, C. and {Stiavelli}, M. and {Wyse}, R.~F.~G. and {Mayer}, L.},
        title = "{Old and Young Bulges in Late-Type Disk Galaxies}",
      journal = {\apj},
         year = 2007,
        month = apr,
       volume = {658},
       number = {2},
        pages = {960-979},
          doi = {10.1086/511125},
archivePrefix = {arXiv},
       eprint = {astro-ph/0610638},
 primaryClass = {astro-ph},
       adsurl = {https://ui.adsabs.harvard.edu/abs/2007ApJ...658..960C}
}

@ARTICLE{laurikainen09,
       author = {{Laurikainen}, E. and {Salo}, H. and {Buta}, R. and {Knapen}, J.~H.},
        title = "{Bars, Ovals, and Lenses in Early-Type Disk Galaxies: Probes of Galaxy Evolution}",
      journal = {\apjl},
         year = 2009,
        month = feb,
       volume = {692},
       number = {1},
        pages = {L34-L39},
          doi = {10.1088/0004-637X/692/1/L34},
archivePrefix = {arXiv},
       eprint = {0901.0641},
 primaryClass = {astro-ph.GA},
       adsurl = {https://ui.adsabs.harvard.edu/abs/2009ApJ...692L..34L}
}

@ARTICLE{aguerri05,
       author = {{Aguerri}, J.~A.~L. and {Elias-Rosa}, N. and {Corsini}, E.~M. and {Mu{\~n}oz-Tu{\~n}{\'o}n}, C.},
        title = "{Photometric properties and origin of bulges in SB0 galaxies}",
      journal = {\aap},
         year = 2005,
        month = apr,
       volume = {434},
       number = {1},
        pages = {109-122},
          doi = {10.1051/0004-6361:20041743},
archivePrefix = {arXiv},
       eprint = {astro-ph/0501163},
 primaryClass = {astro-ph},
       adsurl = {https://ui.adsabs.harvard.edu/abs/2005A&A...434..109A}
}

@ARTICLE{nordemeer07,
       author = {{Noordermeer}, E. and {van der Hulst}, J.~M.},
        title = "{The stellar mass distribution in early-type disc galaxies: surface photometry and bulge-disc decompositions}",
      journal = {\mnras},
         year = 2007,
        month = apr,
       volume = {376},
       number = {4},
        pages = {1480-1512},
          doi = {10.1111/j.1365-2966.2007.11532.x},
archivePrefix = {arXiv},
       eprint = {astro-ph/0701730},
 primaryClass = {astro-ph},
       adsurl = {https://ui.adsabs.harvard.edu/abs/2007MNRAS.376.1480N}
}

@ARTICLE{abolfathi18,
       author = {{Abolfathi}, Bela and {Aguado}, D.~S. and {Aguilar}, Gabriela and {Allende Prieto}, Carlos and {Almeida}, Andres and {Ananna}, Tonima Tasnim and {Anders}, Friedrich and {Anderson}, Scott F. and {Andrews}, Brett H. and {Anguiano}, Borja and {Arag{\'o}n-Salamanca}, Alfonso and {Argudo-Fern{\'a}ndez}, Maria and {Armengaud}, Eric and {Ata}, Metin and {Aubourg}, Eric and {Avila-Reese}, Vladimir and {Badenes}, Carles and {Bailey}, Stephen and {Balland}, Christophe and {Barger}, Kathleen A. and {Barrera-Ballesteros}, Jorge and {Bartosz}, Curtis and {Bastien}, Fabienne and {Bates}, Dominic and {Baumgarten}, Falk and {Bautista}, Julian and {Beaton}, Rachael and {Beers}, Timothy C. and {Belfiore}, Francesco and {Bender}, Chad F. and {Bernardi}, Mariangela and {Bershady}, Matthew A. and {Beutler}, Florian and {Bird}, Jonathan C. and {Bizyaev}, Dmitry and {Blanc}, Guillermo A. and {Blanton}, Michael R. and {Blomqvist}, Michael and {Bolton}, Adam S. and {Boquien}, M{\'e}d{\'e}ric and {Borissova}, Jura and {Bovy}, Jo and {Bradna Diaz}, Christian Andres and {Brandt}, William Nielsen and {Brinkmann}, Jonathan and {Brownstein}, Joel R. and {Bundy}, Kevin and {Burgasser}, Adam J. and {Burtin}, Etienne and {Busca}, Nicol{\'a}s G. and {Ca{\~n}as}, Caleb I. and {Cano-D{\'\i}az}, Mariana and {Cappellari}, Michele and {Carrera}, Ricardo and {Casey}, Andrew R. and {Cervantes Sodi}, Bernardo and {Chen}, Yanping and {Cherinka}, Brian and {Chiappini}, Cristina and {Choi}, Peter Doohyun and {Chojnowski}, Drew and {Chuang}, Chia-Hsun and {Chung}, Haeun and {Clerc}, Nicolas and {Cohen}, Roger E. and {Comerford}, Julia M. and {Comparat}, Johan and {Correa do Nascimento}, Janaina and {da Costa}, Luiz and {Cousinou}, Marie-Claude and {Covey}, Kevin and {Crane}, Jeffrey D. and {Cruz-Gonzalez}, Irene and {Cunha}, Katia and {da Silva Ilha}, Gabriele and {Damke}, Guillermo J. and {Darling}, Jeremy and {Davidson}, Jr., James W. and {Dawson}, Kyle and {de Icaza Lizaola}, Miguel Angel C. and {de la Macorra}, Axel and {de la Torre}, Sylvain and {De Lee}, Nathan and {de Sainte Agathe}, Victoria and {Deconto Machado}, Alice and {Dell'Agli}, Flavia and {Delubac}, Timoth{\'e}e and {Diamond-Stanic}, Aleksandar M. and {Donor}, John and {Downes}, Juan Jos{\'e} and {Drory}, Niv and {du Mas des Bourboux}, H{\'e}lion and {Duckworth}, Christopher J. and {Dwelly}, Tom and {Dyer}, Jamie and {Ebelke}, Garrett and {Davis Eigenbrot}, Arthur and {Eisenstein}, Daniel J. and {Elsworth}, Yvonne P. and {Emsellem}, Eric and {Eracleous}, Michael and {Erfanianfar}, Ghazaleh and {Escoffier}, Stephanie and {Fan}, Xiaohui and {Fern{\'a}ndez Alvar}, Emma and {Fernandez-Trincado}, J.~G. and {Fernando Cirolini}, Rafael and {Feuillet}, Diane and {Finoguenov}, Alexis and {Fleming}, Scott W. and {Font-Ribera}, Andreu and {Freischlad}, Gordon and {Frinchaboy}, Peter and {Fu}, Hai and {G{\'o}mez Maqueo Chew}, Yilen and {Galbany}, Llu{\'\i}s and {Garc{\'\i}a P{\'e}rez}, Ana E. and {Garcia-Dias}, R. and {Garc{\'\i}a-Hern{\'a}ndez}, D.~A. and {Garma Oehmichen}, Luis Alberto and {Gaulme}, Patrick and {Gelfand}, Joseph and {Gil-Mar{\'\i}n}, H{\'e}ctor and {Gillespie}, Bruce A. and {Goddard}, Daniel and {Gonz{\'a}lez Hern{\'a}ndez}, Jonay I. and {Gonzalez-Perez}, Violeta and {Grabowski}, Kathleen and {Green}, Paul J. and {Grier}, Catherine J. and {Gueguen}, Alain and {Guo}, Hong and {Guy}, Julien and {Hagen}, Alex and {Hall}, Patrick and {Harding}, Paul and {Hasselquist}, Sten and {Hawley}, Suzanne and {Hayes}, Christian R. and {Hearty}, Fred and {Hekker}, Saskia and {Hernandez}, Jesus and {Hernandez Toledo}, Hector and {Hogg}, David W. and {Holley-Bockelmann}, Kelly and {Holtzman}, Jon A. and {Hou}, Jiamin and {Hsieh}, Bau-Ching and {Hunt}, Jason A.~S. and {Hutchinson}, Timothy A. and {Hwang}, Ho Seong and {Jimenez Angel}, Camilo Eduardo and {Johnson}, Jennifer A. and {Jones}, Amy and {J{\"o}nsson}, Henrik and {Jullo}, Eric and {Khan}, Fahim Sakil and {Kinemuchi}, Karen and {Kirkby}, David and {Kirkpatrick}, IV, Charles C. and {Kitaura}, Francisco-Shu and {Knapp}, Gillian R. and {Kneib}, Jean-Paul and {Kollmeier}, Juna A. and {Lacerna}, Ivan and {Lane}, Richard R. and {Lang}, Dustin and {Law}, David R. and {Le Goff}, Jean-Marc and {Lee}, Young-Bae and {Li}, Hongyu and {Li}, Cheng and {Lian}, Jianhui and {Liang}, Yu and {Lima}, Marcos and {Lin}, Lihwai and {Long}, Dan and {Lucatello}, Sara and {Lundgren}, Britt and {Mackereth}, J. Ted and {MacLeod}, Chelsea L. and {Mahadevan}, Suvrath and {Maia}, Marcio Antonio Geimba and {Majewski}, Steven and {Manchado}, Arturo and {Maraston}, Claudia and {Mariappan}, Vivek and {Marques-Chaves}, Rui and {Masseron}, Thomas and {Masters}, Karen L. and {McDermid}, Richard M. and {McGreer}, Ian D. and {Melendez}, Matthew and {Meneses-Goytia}, Sofia and {Merloni}, Andrea and {Merrifield}, Michael R. and {Meszaros}, Szabolcs and {Meza}, Andres and {Minchev}, Ivan and {Minniti}, Dante},
        title = "{The Fourteenth Data Release of the Sloan Digital Sky Survey: First Spectroscopic Data from the Extended Baryon Oscillation Spectroscopic Survey and from the Second Phase of the Apache Point Observatory Galactic Evolution Experiment}",
      journal = {\apjs},
         year = 2018,
        month = apr,
       volume = {235},
       number = {2},
          eid = {42},
        pages = {42},
          doi = {10.3847/1538-4365/aa9e8a},
archivePrefix = {arXiv},
       eprint = {1707.09322},
 primaryClass = {astro-ph.GA},
       adsurl = {https://ui.adsabs.harvard.edu/abs/2018ApJS..235...42A}
}

@ARTICLE{yildirim17,
       author = {{Y{\i}ld{\i}r{\i}m}, Ak{\i}n and {van den Bosch}, Remco C.~E. and {van de Ven}, Glenn and {Mart{\'\i}n-Navarro}, Ignacio and {Walsh}, Jonelle L. and {Husemann}, Bernd and {G{\"u}ltekin}, Kayhan and {Gebhardt}, Karl},
        title = "{The structural and dynamical properties of compact elliptical galaxies}",
      journal = {\mnras},
         year = 2017,
        month = jul,
       volume = {468},
       number = {4},
        pages = {4216-4245},
          doi = {10.1093/mnras/stx732},
archivePrefix = {arXiv},
       eprint = {1701.05898},
 primaryClass = {astro-ph.GA},
       adsurl = {https://ui.adsabs.harvard.edu/abs/2017MNRAS.468.4216Y}
}

@ARTICLE{vanderwel14,
       author = {{van der Wel}, A. and {Franx}, M. and {van Dokkum}, P.~G. and {Skelton}, R.~E. and {Momcheva}, I.~G. and {Whitaker}, K.~E. and {Brammer}, G.~B. and {Bell}, E.~F. and {Rix}, H. -W. and {Wuyts}, S. and {Ferguson}, H.~C. and {Holden}, B.~P. and {Barro}, G. and {Koekemoer}, A.~M. and {Chang}, Yu-Yen and {McGrath}, E.~J. and {H{\"a}ussler}, B. and {Dekel}, A. and {Behroozi}, P. and {Fumagalli}, M. and {Leja}, J. and {Lundgren}, B.~F. and {Maseda}, M.~V. and {Nelson}, E.~J. and {Wake}, D.~A. and {Patel}, S.~G. and {Labb{\'e}}, I. and {Faber}, S.~M. and {Grogin}, N.~A. and {Kocevski}, D.~D.},
        title = "{3D-HST+CANDELS: The Evolution of the Galaxy Size-Mass Distribution since z = 3}",
      journal = {\apj},
         year = 2014,
        month = jun,
       volume = {788},
       number = {1},
          eid = {28},
        pages = {28},
          doi = {10.1088/0004-637X/788/1/28},
archivePrefix = {arXiv},
       eprint = {1404.2844},
 primaryClass = {astro-ph.GA},
       adsurl = {https://ui.adsabs.harvard.edu/abs/2014ApJ...788...28V}
}

@ARTICLE{slodkowski_clerici24,
       author = {{Clerici}, K. Slodkowski and {Schnorr-M{\"u}ller}, A. and {Trevisan}, M. and {Ricci}, T.~V.},
        title = "{Massive compact quiescent galaxies in the M$_{{\ensuremath{\star}}}$ versus {\ensuremath{\sigma}}$_{e}$ plane: insights from stellar population properties}",
      journal = {\mnras},
         year = 2024,
        month = jun,
       volume = {531},
       number = {1},
        pages = {1034-1055},
          doi = {10.1093/mnras/stae1213},
archivePrefix = {arXiv},
       eprint = {2405.02348},
 primaryClass = {astro-ph.GA},
       adsurl = {https://ui.adsabs.harvard.edu/abs/2024MNRAS.531.1034C}
}

@ARTICLE{aihara18,
       author = {{Aihara}, Hiroaki and {Armstrong}, Robert and {Bickerton}, Steven and {Bosch}, James and {Coupon}, Jean and {Furusawa}, Hisanori and {Hayashi}, Yusuke and {Ikeda}, Hiroyuki and {Kamata}, Yukiko and {Karoji}, Hiroshi and {Kawanomoto}, Satoshi and {Koike}, Michitaro and {Komiyama}, Yutaka and {Lang}, Dustin and {Lupton}, Robert H. and {Mineo}, Sogo and {Miyatake}, Hironao and {Miyazaki}, Satoshi and {Morokuma}, Tomoki and {Obuchi}, Yoshiyuki and {Oishi}, Yukie and {Okura}, Yuki and {Price}, Paul A. and {Takata}, Tadafumi and {Tanaka}, Manobu M. and {Tanaka}, Masayuki and {Tanaka}, Yoko and {Uchida}, Tomohisa and {Uraguchi}, Fumihiro and {Utsumi}, Yousuke and {Wang}, Shiang-Yu and {Yamada}, Yoshihiko and {Yamanoi}, Hitomi and {Yasuda}, Naoki and {Arimoto}, Nobuo and {Chiba}, Masashi and {Finet}, Francois and {Fujimori}, Hiroki and {Fujimoto}, Seiji and {Furusawa}, Junko and {Goto}, Tomotsugu and {Goulding}, Andy and {Gunn}, James E. and {Harikane}, Yuichi and {Hattori}, Takashi and {Hayashi}, Masao and {He{\l}miniak}, Krzysztof G. and {Higuchi}, Ryo and {Hikage}, Chiaki and {Ho}, Paul T.~P. and {Hsieh}, Bau-Ching and {Huang}, Kuiyun and {Huang}, Song and {Imanishi}, Masatoshi and {Iwata}, Ikuru and {Jaelani}, Anton T. and {Jian}, Hung-Yu and {Kashikawa}, Nobunari and {Katayama}, Nobuhiko and {Kojima}, Takashi and {Konno}, Akira and {Koshida}, Shintaro and {Kusakabe}, Haruka and {Leauthaud}, Alexie and {Lee}, Chien-Hsiu and {Lin}, Lihwai and {Lin}, Yen-Ting and {Mandelbaum}, Rachel and {Matsuoka}, Yoshiki and {Medezinski}, Elinor and {Miyama}, Shoken and {Momose}, Rieko and {More}, Anupreeta and {More}, Surhud and {Mukae}, Shiro and {Murata}, Ryoma and {Murayama}, Hitoshi and {Nagao}, Tohru and {Nakata}, Fumiaki and {Niida}, Mana and {Niikura}, Hiroko and {Nishizawa}, Atsushi J. and {Oguri}, Masamune and {Okabe}, Nobuhiro and {Ono}, Yoshiaki and {Onodera}, Masato and {Onoue}, Masafusa and {Ouchi}, Masami and {Pyo}, Tae-Soo and {Shibuya}, Takatoshi and {Shimasaku}, Kazuhiro and {Simet}, Melanie and {Speagle}, Joshua and {Spergel}, David N. and {Strauss}, Michael A. and {Sugahara}, Yuma and {Sugiyama}, Naoshi and {Suto}, Yasushi and {Suzuki}, Nao and {Tait}, Philip J. and {Takada}, Masahiro and {Terai}, Tsuyoshi and {Toba}, Yoshiki and {Turner}, Edwin L. and {Uchiyama}, Hisakazu and {Umetsu}, Keiichi and {Urata}, Yuji and {Usuda}, Tomonori and {Yeh}, Sherry and {Yuma}, Suraphong},
        title = "{First data release of the Hyper Suprime-Cam Subaru Strategic Program}",
      journal = {\pasj},
         year = 2018,
        month = jan,
       volume = {70},
          eid = {S8},
        pages = {S8},
          doi = {10.1093/pasj/psx081},
archivePrefix = {arXiv},
       eprint = {1702.08449},
 primaryClass = {astro-ph.IM},
       adsurl = {https://ui.adsabs.harvard.edu/abs/2018PASJ...70S...8A}
}

@ARTICLE{lohmann23,
       author = {{Lohmann}, Felipe S. and {Schnorr-M{\"u}ller}, Allan and {Trevisan}, Marina and {Ricci}, T.~V. and {Clerici}, K. Slodkowski},
        title = "{The origin of massive compact galaxies: lessons from IllustrisTNG}",
      journal = {\mnras},
         year = 2023,
        month = oct,
       volume = {524},
       number = {4},
        pages = {5266-5279},
          doi = {10.1093/mnras/stad2176},
archivePrefix = {arXiv},
       eprint = {2307.08911},
 primaryClass = {astro-ph.GA},
       adsurl = {https://ui.adsabs.harvard.edu/abs/2023MNRAS.524.5266L}
}

@ARTICLE{cannarozzo20,
       author = {{Cannarozzo}, Carlo and {Sonnenfeld}, Alessandro and {Nipoti}, Carlo},
        title = "{The cosmic evolution of the stellar mass-velocity dispersion relation of early-type galaxies}",
      journal = {\mnras},
         year = 2020,
        month = oct,
       volume = {498},
       number = {1},
        pages = {1101-1120},
          doi = {10.1093/mnras/staa2147},
archivePrefix = {arXiv},
       eprint = {1910.06987},
 primaryClass = {astro-ph.GA},
       adsurl = {https://ui.adsabs.harvard.edu/abs/2020MNRAS.498.1101C}
}

@ARTICLE{schnorr21,
       author = {{Schnorr-M{\"u}ller}, A. and {Trevisan}, M. and {Riffel}, R. and {Chies-Santos}, A.~L. and {Furlanetto}, C. and {Ricci}, T.~V. and {Lohmann}, F.~S. and {Flores-Freitas}, R. and {Mallmann}, N.~D. and {Alamo-Mart{\'\i}nez}, K.~A.},
        title = "{The puzzling origin of massive compact galaxies in MaNGA}",
      journal = {\mnras},
         year = 2021,
        month = oct,
       volume = {507},
       number = {1},
        pages = {300-317},
          doi = {10.1093/mnras/stab2116},
archivePrefix = {arXiv},
       eprint = {2104.12737},
 primaryClass = {astro-ph.GA},
       adsurl = {https://ui.adsabs.harvard.edu/abs/2021MNRAS.507..300S}
}
\bibliographystyle{aasjournal}

\appendix

\section{MCGs with HST imaging}\label{app:mcg_hst}

\begin{figure*}
\centering
\includegraphics[width=0.98\linewidth, trim=0 0 0 0,clip]{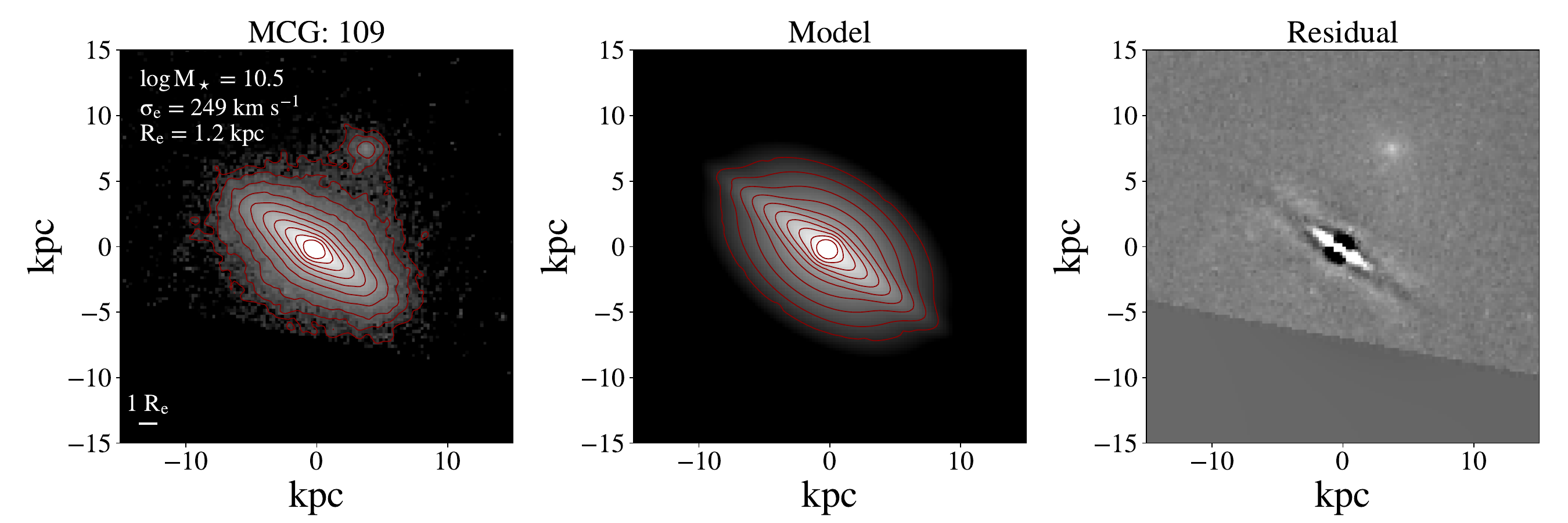}
\caption{HST WFPC2/F606W image, best-fitting model, and residuals for MCG ID: 109. The bulge of this galaxy was fitted with a point source, which gives an upper limit of $\sim 0.2$\,kpc for $R_{e,\mathrm{bulge}}$.}
\label{fig:mcg_hst}
\end{figure*}

Of the 1858 SDSS MCGs in the \citet{slodkowski_clerici24} sample, four have HST archival imaging available. In table\,\ref{tab:mcgs_hst_1} we list the Proposal ID, PI, instrument and filter of these observations. 
We performed multi-component decomposition of these galaxies following the same strategy used for the HSC images. PSFs were built for each observation following the same strategy outlined in Sec.\,\ref{sec:methods}. The bulge of MCG ID: 109 was best fitted by a point source, so an upper limit for $R_{e,\mathrm{bulge}}$ is given in table\,\ref{tab:mcgs_hst_2}. In Fig.\,\ref{fig:mcg_hst} we show as an example the HST image, best fitting model and residuals for MCG ID 109.

\begin{table}
\centering
\caption{Summary of HST observations of MCGs. The table lists the MCG identifier, right ascension, declination, the associated HST proposal ID, the Principal Investigator (PI) of the program, and the instrument/filter combination used for the observations.}
\begin{tabular}{cccccc}
\hline
MCG ID & RA (deg) & Dec (deg) & Proposal ID & PI & Instrument / Filter \\
\hline
101 & 194.9083 & 27.8404  & 10861 & D. Carter   & ACS / F814W  \\
108 & 144.0975 & 59.4158  & 12903 & L. C. Ho    & WFC3 / F105W \\
109 & 168.2287 & 13.4059  & 8719  & A. Edge     & WFPC2 / F606W \\
110 & 202.6897 & -1.8116  & 11202 & L. Koopmans & WFPC2 / F606W\\
\hline
\label{tab:mcgs_hst_1}
\end{tabular}
\end{table}

\begin{table*}
\centering
\hspace*{-1.95cm}
\scriptsize
\caption{Structural properties of MCGs with available HST imaging. Reported values include stellar mass ($M_\star$), effective velocity dispersion ($\sigma_e$), global effective radius ($R_e$), bulge ellipticity ($\epsilon_{\mathrm{bulge}}$), Sérsic index of the bulge ($n_{\mathrm{bulge}}$), bulge effective radius ($R_{e,\mathrm{bulge}}$), bulge-to-total light ratio (B/T), disk scale length ($h_R$), disk scale height ($z_0$), disk flattening ($h_R/z_0$), disk-to-total light ratio (D/T), envelope ellipticity ($\epsilon_{\mathrm{env}}$), envelope effective radius ($R_{e,\mathrm{env}}$), and envelope-to-total light ratio (Env/T). All radii are given in kpc. }

\begin{tabular}{lcccccccccccccc}
\hline
ID & $\log(M_\star/M_\odot)$ & $\sigma_e$ & $R_e$ & $\epsilon_{\mathrm{bulge}}$ & $n_{\mathrm{bulge}}$ & $R_{e,\mathrm{bulge}}$ & B/T & $h_R$ & $z_0$ & $h_R/z_0$ & D/T & $\epsilon_{\mathrm{env}}$ & $R_{e,\mathrm{env}}$ & Env/T \\
 &  &  [km\,s$^{-1}$] & [kpc] & &  & [kpc] &  & [kpc] & [kpc] &  &  &  & [kpc] &  \\
\hline
101   & 10.5 & 227 & 1.9 & 0.3 & 1.9   & 0.26 & 0.26 & 0.7 & 0.21 & 3.2 & 0.20 & 0.6 & 2.8 & 0.54 \\
108   & 10.5 & 207 & 1.5 & 0.2 & 2.0  & 0.16 & 0.31 & 1.1 & 0.30 &  3.6 & 0.43 & 0.5 & 2.5 & 0.26 \\
109   & 10.5 & 249 & 1.2 & -- & -- & $<0.2$ & 0.22 & 0.7 & 0.12 &  5.8 & 0.43 & 0.4 & 1.9 & 0.35 \\
110   & 10.5 & 184 & 2.0 & 0.5 & 2.1 & 0.34 & 0.37 & 1.0 & 0.24 &  4.0 & 0.21 & 0.6 & 3.0 & 0.42 \\
\hline
\end{tabular}
\label{tab:mcgs_hst_2}
\end{table*}

\end{document}